\documentclass[11pt,a4paper]{article}
\usepackage{jheppub_mod}
\usepackage[utf8]{inputenc}
\usepackage{amsmath,amsfonts,amssymb,physics,subcaption}
\usepackage[usenames,dvipsnames]{xcolor}
\usepackage{setspace}
\usepackage{verbatim}
\usepackage{graphicx}
\usepackage{bbm}
\usepackage{cancel}

\definecolor{dgreen}{rgb}{0.,0.6,0.}

\newcommand{\rme}{\mathrm e}

\newcommand{\be}{\begin{equation}}
\newcommand{\ee}{\end{equation}}
\newcommand{\bea}{\begin{eqnarray}}
\newcommand{\eea}{\end{eqnarray}}
\newcommand{\bi}{\begin{itemize}}
\newcommand{\ei}{\end{itemize}}
\newcommand{\ben}{\begin{enumerate}}
\newcommand{\een}{\end{enumerate}}

\newcommand{\ie}{{\it i.e.,}\ }

\newcommand{\mL}{\mathcal{L}}
\newcommand{\pd}{\partial}

\begin{document}
\title{Holographic complexity: braneworld gravity versus the Lloyd bound}

\author[1]{Sergio E. Aguilar-Gutierrez,}
\author[2]{Ben Craps,}
\author[2]{Juan Hernandez,}
\author[2,3]{Mikhail Khramtsov,}
\author[2]{Maria Knysh,}
\author[4]{and Ashish Shukla}

\affiliation[1]{Instituut voor Theoretische Fysica, KU Leuven, Celestijnenlaan 200D, B-3001 Leuven, Belgium}
\affiliation[2]{Theoretische Natuurkunde, Vrije Universiteit Brussel (VUB) and The International Solvay Institutes, Pleinlaan 2, B-1050 Brussels, Belgium}
\affiliation[3]{David Rittenhouse Laboratory, University of Pennsylvania, 209 S. 33rd Street, Philadelphia, PA 19104, USA}
\affiliation[4]{CPHT, CNRS, \'{E}cole polytechnique, Institut Polytechnique de Paris, 91120 Palaiseau, France}

\emailAdd{sergio.ernesto.aguilar@gmail.com}
\emailAdd{ben.craps@vub.be}
\emailAdd{juan.hernandez@vub.be}
\emailAdd{mikhail.khramtsov@vub.be}
\emailAdd{maria.knysh@vub.be} 
\emailAdd{ashish.shukla@polytechnique.edu}

\preprint{CPHT-RR068.112023}

\abstract{
We explore the complexity equals volume proposal for planar black holes in anti-de Sitter (AdS) spacetime in 2+1 dimensions, with an end of the world (ETW) brane behind the horizon. We allow for the possibility of intrinsic gravitational dynamics in the form of Jackiw-Teitelboim (JT) gravity to be localized on the brane. We compute the asymptotic rate of change of volume complexity analytically and obtain the full time dependence using numerical techniques. We find that the inclusion of JT gravity on the brane leads to interesting effects on time dependence of holographic complexity. We identify the region in parameter space (the brane location and the JT coupling) for which the rate of change of complexity violates the Lloyd bound. In an equivalent description of the model in terms of an asymptotically AdS wormhole, we connect the violation of the Lloyd bound to the violation of a suitable energy condition in the bulk that we introduce. We also compare the Lloyd bound constraints to previously derived constraints on the bulk parameters in this model that are based on bounds on entanglement growth in the dual CFT state.
} 


\maketitle


\section{Introduction}
\label{sec:intro}
Over the years, many important lessons we have learned about the nature of quantum gravity have been catalyzed by the use of information-theoretic concepts in the setting of the Anti-de Sitter/Conformal Field Theory (AdS/CFT) correspondence \cite{Maldacena:1997re, Gubser:1998bc, Witten:1998qj}. Ongoing efforts in this direction include the study of quantum computational complexity for field theory states, which is a measure of the number of steps needed to reach the said state starting from a simple reference state, using a set of ``simple'' unitary operators (see \cite{Chapman:2021jbh} for a review). For states in holographic CFTs, this is complemented by various holographic proposals that aim to compute the complexity of the state in terms of a bulk observable. The earliest proposals for holographic complexity are the \emph{complexity equals volume} (CV) \cite{Susskind:2014rva,Stanford:2014jda} and the \emph{complexity equals action} (CA) \cite{Brown:2015bva, Brown:2015lvg} proposals, which were quickly followed by the \emph{complexity equals volume 2.0} (CV2.0) proposal \cite{Couch:2016exn}. More recently, it has been observed that there exist an infinite number of gravitational observables that can all serve as holographic measures of complexity \cite{Belin:2021bga, Belin:2022xmt}, since they display the telltale features expected of computational complexity, e.g.~a late time linear growth\footnote{We are ignoring saturation of complexity due to finite system sizes, which is not captured by the classical holographic complexity proposals we focus on in the present work.} as well as the switchback effect in the presence of shock waves. This broad variety of candidates for holographic complexity is reminiscent of the broad variety of possible complexity frameworks in quantum systems.

An important ingredient in the study of complexity is the notion of the Lloyd bound~\cite{lloyd2000ultimate}, originally proposed as a bound on the maximum rate of computation achievable by a quantum system. Interpreting complexity growth as computation, it was initially argued that the Lloyd bound translates to an upper bound on the rate of change of complexity~\cite{Brown:2015bva}. However, further studies into the time dependence of holographic complexity found several instances where the CA proposal violates the Lloyd bound. These include eternal AdS black holes \cite{Carmi:2017jqz, Yang:2017czx}, rotating Ba\~{n}ados-Teitelboim-Zanelli (BTZ) black holes \cite{Bernamonti:2021jyu}, warped AdS$_3$ black holes \cite{Auzzi:2018pbc}, the four-dimensional non-commutative $\mathcal{N} = 4$ super Yang-Mills theory \cite{Couch:2017yil}, as well as several bulk geometries that appear as solutions to Einstein-Maxwell-dilaton gravity \cite{Swingle:2017zcd, An:2018xhv, Alishahiha:2018tep, Wang:2023ipy}. A rigorous study on a class of spacetimes where the Lloyd bound holds for the CA proposal has been done in \cite{Yang:2016awy}.
For the CV proposal, it was argued recently in \cite{Engelhardt:2021mju} that by assuming the weak energy condition (WEC), a version of the Lloyd bound always holds true in asymptotically AdS$_{D}$ geometries, with spacetime dimensionality $D \ge 4$, in minimally coupled Einstein-Maxwell-scalar theories. It thus seems that CV is more robust in meeting an upper bound on the rate of complexity growth compared to the CA proposal. In $D = 3$, to the best of our knowledge, a violation of the Lloyd bound for the CV proposal has only been found for asymptotically AdS$_3$ geometries with a de Sitter bubble in their interior \cite{Auzzi:2023qbm}, for multi-boundary AdS$_3$ wormhole geometries \cite{Zolfi:2023bdp}, and in the holographic local quench scenario \cite{Ageev1,Ageev2}. These examples involve bulk geometries that are either exotic or they have less symmetry compared to the bulk duals of generic holographic states. This signifies the importance of further exploring the CV proposal with reference to meeting or violating the Lloyd bound in less exotic setups, that can perhaps be arrived at by using simple bottom-up constructions.

In the present work, we study the time dependence of the CV proposal for planar AdS black holes with an end of the world (ETW) brane embedded in the geometry, with an emphasis on the late time behaviour of complexity growth and its relation to the Lloyd bound. The ETW brane cuts off the second asymptotic region of the maximally extended spacetime. 
From a top-down perspective, the ETW brane may correspond to branes in string theory, or a region of large backreaction such that the geometry caps off \cite{DHoker:2007zhm, DHoker:2007hhe, Chiodaroli:2012vc, Bak:2020enw, Uhlemann:2021nhu, VanRaamsdonk:2021duo, Sugimoto:2023oul}. Such geometries also arise as the gravitational dual description for pure states in boundary conformal field theories (BCFT), which are CFTs defined on manifolds with boundaries, with conformally invariant boundary conditions \cite{Karch:2000gx, Takayanagi:2011zk, Fujita:2011fp, Kourkoulou:2017zaj, Almheiri:2018ijj, Miyaji:2021ktr, Chandra:2022fwi}; see also \cite{Cooper:2019rwk, Reeves:2021sab, Belin:2021nck, Kusuki:2021gpt, Kawamoto:2022etl, Izumi:2022opi, Anous:2022wqh, Kusuki:2022ozk, Kanda:2023zse, Neuenfeld:2023svs}. More recently, they have played an important role in holographic constructions providing a resolution to the black hole information loss problem via the quantum extremal island prescription \cite{Almheiri:2019hni, Rozali:2019day, Almheiri:2019psy, Balasubramanian:2020hfs, Sully:2020pza, Geng:2020qvw, Chen:2020uac, Chen:2020hmv, Grimaldi:2022suv, Krishnan:2020fer, Deng:2020ent, May:2021zyu, Fallows:2021sge, Neuenfeld:2021wbl, Geng:2021iyq, Chu:2021gdb, Miyaji:2021lcq, Verheijden:2021yrb, Geng:2021mic, Suzuki:2022xwv, Bianchi:2022ulu, Geng:2022slq, Geng:2022dua, Jeong:2023hrb}, as well as in attempts to embed cosmology in a holographic perspective \cite{Cooper:2018cmb, Antonini:2019qkt, Chen:2020tes, VanRaamsdonk:2020tlr, Wang:2021xih, Fallows:2022ioc, Waddell:2022fbn, Antonini:2022blk,Yadav:2023qfg,Aguilar-Gutierrez:2023zoi}. Holographic complexity for asymptotically AdS spacetimes with an ETW brane has previously been explored in \cite{Chapman:2018bqj, Ross:2019rtu, Sato:2019kik, Braccia:2019xxi, Hernandez:2020nem, Omidi:2020oit, Bhattacharya:2021jrn, Auzzi:2021ozb,Craps:2022ahp,Aguilar-Gutierrez:2023tic}. Braneworld theories have proven to be a useful framework in which to investigate aspects of entanglement and complexity. Well-known examples include studying the role of the graviton mass in the formation of entanglement islands~\cite{Geng:2020qvw,Geng:2020fxl,Geng:2021hlu}, finding higher curvature corrections to holographic complexity~\cite{Hernandez:2020nem}, the formulation of quantum corrected BTZ black holes~\cite{Emparan:2020znc} and the proposal of quantum corrections to holographic complexity~\cite{Emparan:2021hyr,Chen:2023tpi}. Moreover, because of the additional contact terms that arise when including an intrinsic gravity term on the brane~\cite{Almheiri:2019hni,Chen:2020uac,Hernandez:2020nem}, the behaviour of entanglement and complexity becomes sensitive to the parameters of the brane action. 

To be concrete, we focus on the case of an AdS$_3$ black hole \emph{i.e.,} the Ba\~{n}ados-Teitelboim-Zanelli (BTZ) geometry \cite{Banados:1992wn}, with an embedded ETW brane that cuts off the second asymptotic region of the maximally extended spacetime. In the simplest possible scenario, the ETW brane is endowed with a constant tension, which fixes its location within the bulk spacetime. As discussed in detail in \cite{Lee:2022efh}, and summarized in section \ref{sec:basic_setup} of the present work, the brane can have three distinct trajectories in the bulk spacetime, depending upon the brane tension. For tension less than unity, the brane completely cuts off the second asymptotic region of the maximally extended planar BTZ geometry - see fig.~\ref{fig:subcritical}. This is referred to as the \emph{subcritical} case. On the other hand, for tension greater than unity, only part of the second asymptotic region is cut off by the brane - see fig.~\ref{fig:supercritical}. This \emph{supercritical} geometry is holographically tantamount to including degrees of freedom from a second copy of the CFT \cite{Maldacena:2001kr}. The \emph{critical} case, fig.~\ref{fig:critical}, which corresponds to the brane tension being exactly equal to unity, amounts to the brane reaching the boundary of the second asymptotic region in the infinite past/future. The physical picture of the ETW brane cutting off the entire second asymptotic region of the extended spacetime geometry is thus unambiguously met by the subcritical case only, on which we focus our attention in this paper.

Additionally, in the spirit of constructing an effective bottom-up model, we also allow for the possibility of the ETW brane to carry intrinsic gravitational dynamics. Given that Einstein gravity is purely topological in two dimensions, the simplest possibility is to consider Jackiw-Teitelboim (JT) gravity \cite{JACKIW1985343, TEITELBOIM198341} to be localized on the ETW brane that cuts off the BTZ spacetime. This leads to interesting changes in the properties of the dual CFT state. For instance, it was found in \cite{Lee:2022efh} that the time dependent behaviour of entanglement entropy for the dual CFT state gets modified in the presence of JT gravity on the brane. In particular, it was found that only a finite subspace of the bulk parameter space for the brane location and the suitably defined JT coupling lead to entanglement dynamics compatible with known bounds on entanglement growth in two-dimensional CFTs \cite{Hartman:2015apr}. In other words, constraints on entanglement dynamics for the CFT state translate into constraints on the effective bulk description, which a priori might appear unconstrained.

Our focus in the present work is on understanding the time dependent behaviour of holographic complexity for these states via the CV proposal. Once again, as is the case for entanglement, the presence of JT gravity on the brane affects the behaviour of complexity as a function of time for the dual CFT state. It turns out that to extract the full time dependence of holographic complexity, it is best suited to follow a numerical approach. However, the asymptotically early/late time dynamics of complexity, corresponding to time scales $t_{\rm bdy} \to \pm \infty$, can be obtained analytically, as we discuss below. We find that only a finite subspace of the bulk parameter space for the brane location and 
the JT coupling allows for the Lloyd bound on the rate of change of complexity to hold true at all times. For bulk parameter values outside this restricted subspace, we find that although the rate of change of complexity still reaches the value dictated by the Lloyd bound at asymptotically early/late times, the bound is violated during intermediate stages of time evolution of the state. More specifically, we find that at asymptotically late times the Lloyd bound value is reached from above, whereas for asymptotically early times it is approached from below, thus violating the bound. If one demands the Lloyd bound to be satisfied for the states of our interest, one gets a reduction in the bulk parameter space of the brane location and the JT coupling. When combined with constraints from entanglement dynamics obtained in \cite{Lee:2022efh}, one ends up getting a significantly reduced parameter space. This highlights how the holographic duality can constrain the effective bulk description following information theoretic constraints on the dual CFT state.

This paper is organized as follows. In section \ref{sec:basic_setup}, we detail the setup of the problem, describe the BTZ geometry with a subcritical ETW brane, and summarize the CV proposal and the Lloyd bound on complexity growth. This section also helps set up the notation for the rest of the paper. Subsequently, in section~\ref{sec:complexity_growth}, we perform a detailed analytic investigation of the asymptotic behaviour of the rate of change of holographic complexity. Our approach is based on treating the search for the maximal volume surface as finding the trajectory of a classical particle scattering off an effective potential. The violation of the Lloyd bound occurs when the energy of the particle is higher than the potential barrier. We derive an analytic expression for the critical curve within the space of bulk parameters, namely the brane location and the JT coupling, that separates the Lloyd bound respecting region from the Lloyd bound violating region. We additionally delve into several illustrative examples, deriving analytic expressions for the complexity growth rate. Section~\ref{sec:numerical_approach} then provides a detailed analysis of the full time dependence of holographic complexity extracted using a numerical approach, confirming the results of section~\ref{sec:complexity_growth}. Note that sections \ref{sec:complexity_growth} and \ref{sec:numerical_approach} can be read independently. In section~\ref{sec:Engelhardt-Folkestad}, we connect the violation of the Lloyd bound found in the previous sections for part of the bulk parameter space to the violation of the WEC in the bulk by studying an alternate description of the system. More specifically, we consider a wormhole spacetime with two asymptotically AdS$_3$ external regions and with a matter source. The matter is fine-tuned in such a way that the extremal volume slice is identical to that of two copies of the original system consisting of the ETW brane carrying intrinsic JT gravity. Next, in section \ref{sec:compare_bounds}, we summarize known constraints on the JT coupling and the brane location for the subcritical geometry, which were obtained in \cite{Lee:2022efh} using the boundedness of entanglement velocity. We discuss the possibility of further constraining the parameter space of the bulk effective description by combining the entanglement velocity constraints with the constraint from the Lloyd bound, assuming it to hold true at all times. Section \ref{sec:discussion} concludes the paper with a discussion and an outlook towards various future possibilities. Appendices \ref{sec:simple}-\ref{app:ext_curve} contain several technical details utilized in performing the calculations. 


\section{Basic setup}
\label{sec:basic_setup}
We begin by reviewing the black hole solution for three-dimensional Einstein gravity with a negative cosmological constant, given by the Ba\~{n}ados-Teitelboim-Zanelli (BTZ) geometry \cite{Banados:1992wn}. The metric for the planar BTZ black hole in Schwarzschild coordinates $(t,r,x)$ is given by
\be
\label{eq:planarBTZ}
    d s^2 = -f(r)\,dt^2 + \frac 1 {f(r)} \, d r^2 + \frac{r^2}{L^2} \, d x^2, \quad \text{with} \quad f(r) = \frac{r^2 - r^2_0} {L^2}, 
\ee
where the coordinates $t, x \in (-\infty, \infty)$ and $r \in [r_0, \infty)$. Here $r_0$ denotes the location of the horizon, and the constant $L$ is the AdS length scale. The black hole has the Hawking temperature $T_H = f'(r_0)/4\pi =r_0/2\pi L^2$. The entropy density associated with the horizon is $s = r_0/2G_N L$, and the energy density is $\varepsilon = r_0^2/8\pi G_N L^3$. 

\begin{figure}[t]
    \centering
    \includegraphics[width=\textwidth]{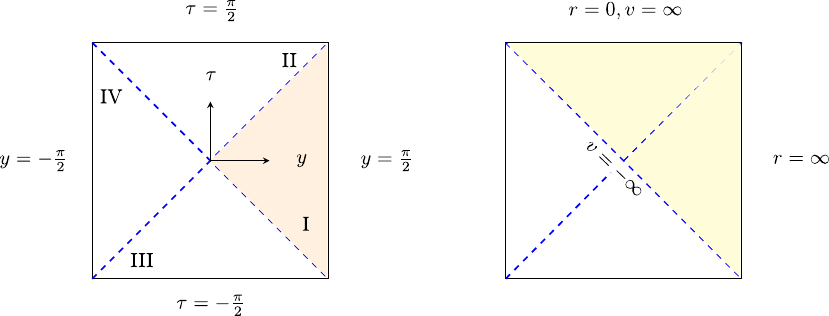}
    \caption{The maximally extended planar BTZ spacetime. On the left panel, in terms of the global coordinates $(\tau, y)$, the horizons are at $\tau = \pm y$ (dashed blue), and the asymptotic AdS boundaries at $y = \pm \pi/2$. The colored region denotes the patch covered by the Schwarzschild coordinates 
    $(t, r)$. On the right panel, the colored region depicts the portion of the geometry covered by the ingoing Eddington-Finkelstein coordinates $(v, r)$. The definition of the Schwarzschild time, as outlined in equation~\eqref{eq:Sch to global}, requires a specific choice of sign within each coordinate patch: positive sign for patches I and III, and negative sign for patches II and IV.}
    \label{fig:BTZspacetime}
\end{figure}

The Schwarzschild coordinates $(t, r, x)$ cover only the region of spacetime outside the horizon; see fig.~\ref{fig:BTZspacetime} left panel. In order to also cover the black hole interior, one can introduce the ingoing Eddington-Finkelstein (EF) coordinates, which cover the shaded region in the right panel of fig.~\ref{fig:BTZspacetime}. This coordinate system is defined by the radial coordinate $r$, which now extends all the way to the singularity, $r \in (0, \infty)$, while the ingoing EF time coordinate is defined via
\begin{equation}\label{eq:v and t}
    v=t-\int_{r}^{\infty} \frac{d r'}{f(r')}~ \in~ (-\infty, \infty)\,, \quad {\rm for}  \quad  r > r_0\,.
\end{equation}
The metric in the EF coordinates is given by
\begin{equation}\label{eq:metric.EF}
d s^2 = -f(r)d v^2+2dv dr+\frac{r^2}{L^2}\,dx^2\,.
\end{equation}
This metric is smooth across the future horizon, and the geometry extends past the horizon. We can define an interior Schwarzschild time via
\begin{equation}
    v=t+\int_{0}^{r} \frac{d r'}{f(r')}~ \in~ (-\infty, \infty)\, \quad {\rm for}  \quad r<r_0\,.
\end{equation}
Additionally, for the numerical approach used in section \ref{sec:numerical_approach}, we will find it useful to work with a global coordinate system that covers the entire maximally extended spacetime geometry. Following \cite{Cooper:2018cmb}, we first consider a coordinate transformation to Kruskal-like coordinates $(U, V)$, given by\footnote{This is the coordinate transformation for the right Schwarzschild patch only and needs to be written separately for the other three patches.}
\be
\frac{r}{r_0} = \frac{1-UV}{1+UV}\, , \qquad t = \frac{L^2}{2r_0} \log\left(-\frac{U}{V}\right).
\ee
In terms of the Kruskal-like coordinates, the metric takes the form
\be
d s^2 = - \frac{4L^2}{(1+UV)^2} \, d U d V + \frac{r_0^2}{L^2} \left(\frac{1-UV}{1+UV}\right)^{\!2} d x^2.
\ee
The $(U, V, x)$ coordinates run from $-\infty$ to $\infty$, and cover the entire maximally extended black hole spacetime. Note that the horizons in these coordinates are at $UV = 0$, the two asymptotic boundaries at $UV = -1$, and the future/past singularities at $UV = 1$. 
Next, to simplify things further, we introduce $U = \tan \alpha, V = \tan \beta$, with the coordinate range $-{\pi}/{2} \le \alpha, \beta \le {\pi}/{2}$. Defining $\tau = \alpha+\beta, \, y = \beta - \alpha$, we finally get the metric in the form
\be
d s^2 = \frac 1 {\cos^2 y} \left(- L^2 d\tau^2 + L^2 d y^2 + \frac{r_0^2}{L^2} \cos^2(\tau)\, d x^2 \right). \label{MetricGlobal}
\ee
Here $-\pi/2 \le \tau,\, y \le \pi/2$, with the horizons at $\tau = \pm y$, the two asymptotic AdS boundaries at $y = \pm \pi/2$, and the future/past singularities at $\tau = \pm \pi/2$. fig.~\ref{fig:BTZspacetime} depicts the maximally extended spacetime in the $(\tau, y, x)$ global coordinates. More explicitly, the relation between the $(t,r)$ and $(\tau,y)$ coordinates is 
\be\label{eq:Sch to global}
\begin{aligned}
    & \frac{r}{r_0}=\frac{\cos\tau}{\cos y}, \\
    & t= \frac {L^2}{2 r_0} \log\left(\pm\frac{\sin y+\sin\tau}{\sin y-\sin \tau}\right).
    \end{aligned}
\ee
The signs are chosen based on which of the four patches in the maximally extended BTZ spacetime one wants to cover with the Schwarzschild coordinates, see fig.~\ref{fig:BTZspacetime}.

The maximally extended geometry of fig.~\ref{fig:BTZspacetime} is dual to the thermofield double state, which is a state obtained by entangling the two copies of the CFT living on the two asymptotic boundaries \cite{Maldacena:2001kr}. We are, however, interested in states of a single copy of the CFT, whose dual description includes part of the region beyond the horizon, and would therefore like to cut off the left asymptotic region of the maximally extended BTZ geometry. In a bottom-up approach, as mentioned in the introduction section~\ref{sec:intro}, this can be done by introducing a constant tension ETW brane. In the presence of the brane, the total action also includes a brane term and is given by\footnote{The action needs to be supplemented by appropriate counter terms to make it finite on-shell \cite{Balasubramanian:1999re}.}
\be
\label{eq:totalaction}
\begin{aligned}
    I = \frac{1}{16\pi G_N} \Bigg[\int_{\text{bulk}} d^3x \sqrt{-g} \left(R +\frac{2}{L^2}\right) + 2 \int_{\partial AdS} d^2x \sqrt{-h} \, K& \\+ 2 \int_{\rm brane} d^2x \sqrt{-h} \left(K-\frac{T_0}{L}\right)&\Bigg].
\end{aligned}
\ee
Here $K$ is the trace of the extrinsic curvature, and $T_0$ denotes the brane tension. When extremizing the action to obtain eq.~\eqref{eq:planarBTZ} as the solution, we impose the usual Dirichlet boundary conditions on the metric variation at the AdS boundary, \emph{i.e.,} the metric variation vanishes on the AdS boundary. However, at the location of the brane, rather than imposing Dirichlet boundary conditions, we impose Neumann boundary conditions, which allows the brane to localize at a position determined by its tension, via the Israel junction conditions\cite{Israel:1966rt} \be
\label{eq:brane_eom}
K_{ij} = \frac{T_0}{L} h_{ij}~.
\ee
In other words, the brane is located at a position such that the trace of its extrinsic curvature is constant. 

Depending upon the tension, the brane can have three distinct trajectories, as discussed in \cite{Lee:2022efh}. For $0 \le T_0 < 1$, the ETW brane cuts off the entire left asymptotic region and is located at a fixed value of the $y$-coordinate, given by 
\begin{equation}
\sin y_{\rm brane} = - T_0~.    
\label{eq:brane_loc}
\end{equation}
In Schwarzschild coordinates, this equation takes the form 
\begin{equation}
\begin{aligned}
\frac{r_{\rm brane}^2}{r_0^2} &= \frac{1-T_0^2 \,{\rm tanh}^2\left(\frac{r_0 t_{\rm brane}}{L^2}\right)}{1-T_0^2} \quad \quad \quad  \text{outside the horizon}\,,
\\
\frac{r_{\rm brane}^2}{r_0^2} &= \frac{1-T_0^2 \,{\rm coth}^2\left(\frac{r_0 t_{\rm brane}}{L^2}\right)}{1-T_0^2} \quad \quad \quad \text{inside the horizon}\,.
\end{aligned}\label{eq: brane location Schw}
\end{equation}

\begin{figure}[t]
    \centering
    \subcaptionbox{Subcritical: $0\le T_0 < 1$\label{fig:subcritical}}[0.3\textwidth]{
    \includegraphics[width=0.25\textwidth]{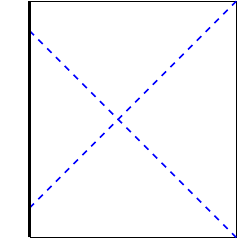}}\hspace{0.35cm}
    \subcaptionbox{Critical: $T_0 = 1$\label{fig:critical}}[0.3\textwidth]{
    \includegraphics[width=0.25\textwidth]{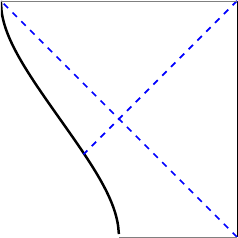}
}\hspace{0.35cm}
    \subcaptionbox{Supercritical: $T_0 > 1$\label{fig:supercritical}}[0.3\textwidth]{
    \includegraphics[width=0.25\textwidth]{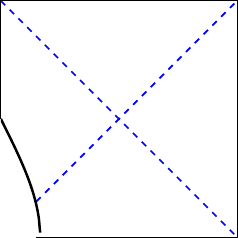}}
    \caption{The subcritical, critical, and supercritical ETW brane trajectories in BTZ spacetime. Figure adapted from \cite{Lee:2022efh}.
    }
    \label{fig:trajectories}
\end{figure}

For $T_0 = 1$, the ETW brane emanates out of the past singularity at $\tau = -\pi/2$ and reaches the left asymptotic boundary at $\tau = \pi/2$.\footnote{The trajectory obtained by performing a reflection about the $\tau = 0$ axis, wherein the brane emanates from the left asymptotic boundary at $\tau = -\pi/2$ and ends into the future singularity at $\tau = \pi/2$, is also a valid solution.} Finally, when $T_0 > 1$, the ETW brane emanates from the past singularity at $\tau = -\pi/2$, and reaches the left asymptotic boundary in the far future, without cutting off the second asymptotic region entirely. Following the nomenclature of \cite{Lee:2022efh}, we label the three possibilities as subcritical $(0 \le T_0 < 1)$, critical $(T_0 = 1)$ and supercritical $(T_0 > 1)$. The three cases are illustrated in fig.\ \ref{fig:trajectories}. 

It is interesting to note that the cosmological constant on the subcritical, critical, and supercritical brane is negative, zero, and positive, respectively. Additionally, the induced metric on the brane for the three cases takes the following form \cite{Lee:2022efh}
\begin{subequations}
\begin{align}
    &\text{Subcritical:}\,\, d s_{\text{ind}}^2 = - L^2 d \lambda^2 + \frac{r_0^2}{L^2} \frac{\cos^2 \left(\lambda \sqrt{1 - T^2_0}\right)}{1 - T^2_0}\,d x^2, \quad \lambda \in \left[\frac{-\pi}{2\sqrt{1-T_0^2}} , \frac{\pi}{2\sqrt{1-T_0^2}} \right]\,, \label{MetricBraneSubcr}\\
    &\text{Critical:} \quad\,\,\,\, d s_{\text{ind}}^2 = -\,L^2 d \lambda^2 + \frac{r_0^2}{L^2} \lambda^2 d x^2, \quad \lambda \in [0,\infty)\,, \label{MetricBraneCr}\\
    &\text{Supercritical:} \,\,  d s_{\text{ind}}^2 = - L^2 d \lambda^2 + \frac{r_0^2}{L^2} \frac{\sinh^2(\lambda \sqrt{T^2_0 - 1})}{T^2_0 - 1} \, d x^2,\quad \lambda \in [0,\infty)\,,\label{MetricBraneSupercr}
\end{align}
\end{subequations}
where $\lambda$ denotes the proper time on the brane.

The induced metric on the subcritical brane is that of a big bang-big crunch cosmology \cite{Cooper:2018cmb}. For the critical and supercritical cases, one gets an expanding spacetime, with a late time exponentially expanding de Sitter phase for the supercritical case.\footnote{The trajectories obtained after a reflection about the $\tau = 0$ axis for the critical and supercritical cases are also valid solutions, which now represent contracting spacetimes, with $\lambda \in (-\infty, 0]$. The supercritical case now admits an early time de Sitter phase.}

As mentioned earlier, our interest in the present work is in high-energy pure states for a single copy of a CFT. However, as is evident from fig.\ \ref{fig:supercritical}, the supercritical case includes part of the second asymptotic boundary as well and is thus holographically tantamount to including the degrees of freedom associated with a second copy of the CFT. Because of this, we will not be considering the supercritical case in our subsequent discussion. Also, the critical case fig.\ \ref{fig:critical}, though interesting, lacks a moment of time reflection symmetry. It is therefore difficult to imagine how one can prepare such a state using an Euclidean path integral, even though the Lorentzian geometry exists. On the other hand, the subcritical geometries of fig.\ \ref{fig:subcritical} can be constructed starting from boundary states with limited entanglement in a CFT and evolving them in Euclidean time \cite{Kourkoulou:2017zaj, Almheiri:2018ijj}, in what is also known as the AdS/BCFT correspondence \cite{Karch:2000gx, Takayanagi:2011zk}. Henceforth, we will work solely with the subcritical case, since it is only the subcritical geometry that most unambiguously satisfies our requirement for the bulk to only include part of the second asymptotic region.

\subsection{Subcritical ETW brane with JT gravity}
\label{sec:JTbrane}
Let us now consider the possibility for intrinsic gravitational dynamics to exist on the ETW brane. Since Einstein gravity in two dimensions is purely topological, the simplest possibility to consider is the presence of Jackiw-Teitelboim (JT) gravity \cite{JACKIW1985343, TEITELBOIM198341} on the brane. The action for JT gravity is given by
\be
\label{eq:JTaction}
    I_\text{JT} = \frac{1}{16\pi G_N^{\rm brane}} \int d^2x \sqrt{-h} \left[\Phi_0 R^{\rm brane} + \varphi \left( R^{\rm brane} - 2\Lambda^{\rm brane} \right)\right].
\ee
Here, $G_N^{\rm brane}$ is Newton's gravitational constant on the brane, and $\Lambda^{\rm brane}$ is the (negative) cosmological constant on the brane. $R^{\rm brane}$ is the Ricci scalar on the brane, and the scalar $\varphi$ is the dilaton, with the constant piece $\Phi_0$ associated with ground state entropy of an extremal black hole, if one considers JT gravity to be arising in the near-horizon limit of a near-extremal black hole \cite{Maldacena:2016upp, Nayak:2018qej, Moitra:2018jqs, Moitra:2019bub}. Since the term proportional to $\Phi_0$ is purely topological, we will not be considering it in the subsequent discussion.

Note that the dilaton equation of motion fixes the background geometry on the ETW brane to be AdS$_2$, with $R^{\rm brane} = 2\Lambda^{\rm brane} = -2/L_2^2$, with $L_2$ being the AdS$_2$ length scale. Interestingly, once the action eq.\ \eqref{eq:totalaction} is augmented with the JT action eq.\ \eqref{eq:JTaction}, the location of the brane $y_{\rm brane}$ can be adjusted by tuning either $\Lambda^{\rm brane}$ or the tension $T_0$. Furthermore, for the purpose of computing complexity in this model, it is irrelevant which of the two parameters is used to adjust $y_{\rm brane}$, and one can simply set $T_0 = 0$. The brane is still located at a fixed value of the $y$-coordinate, now given by
\be
\label{eq:lam_y_rel}
- \frac{1}{L^2} \cos^2 y_{\rm brane} = \Lambda^{\rm brane}\,,
\ee
or in Schwarzschild coordinates 
\begin{equation}
\begin{aligned}
\frac{r_{\rm brane}^2}{r_0^2} &= -\frac{1-(L^2 \Lambda^{\rm brane} + 1) \,{\rm tanh}^2\left(\frac{r_0 t_{\rm brane}}{L^2}\right)}{L^2 \Lambda^{\rm brane}} \quad \quad \quad  \text{outside the horizon}\,,
\\
\frac{r_{\rm brane}^2}{r_0^2} &= -\frac{1-(L^2 \Lambda^{\rm brane} + 1) \,{\rm coth}^2\left(\frac{r_0 t_{\rm brane}}{L^2}\right)}{L^2 \Lambda^{\rm brane}} \quad \quad \quad \text{inside the horizon}\,.
\end{aligned}\label{eq: brane location Schw with JT}
\end{equation}
Additionally, the variation of the metric gives an equation for the dilaton,
\be
\label{eq:eom_dilaton}
\nabla_i \nabla_j \varphi + \Lambda^{\rm brane} \varphi h_{ij} = \frac{G_N^{\rm brane}}{G_N} K_{ij},
\ee
with the covariant derivative $\nabla_i$ taken with respect to the induced metric $h_{ij}$ on the brane. As discussed in \cite{Lee:2022efh}, this admits the solution
\be
\label{eq:dilaton_sol}
\varphi(\tau_{\rm brane}) = \varphi_0 + \varphi_1 \sin \tau_{\rm brane},
\ee
where $\varphi_0 = {G_N^{\rm brane} K}/\left({2 G_N \Lambda^{\rm brane}}\right)$ and $\varphi_1$ are constants. 
In the subsequent discussion, we will be referring to the dimensionless parameter
\be
\alpha \equiv \frac{G_N \, \varphi_1}{G_N^{\rm brane} L}
\label{def_alpha}
\ee
as the ``JT coupling." 

\subsection{The CV proposal}
\label{sec:review_CV}
As mentioned in the introduction section~\ref{sec:intro}, several proposals holographically capture the key aspects of complexity for the CFT state, such as a late time linear growth and the switchback effect \cite{Chapman:2021jbh}. One of the prominent proposals is the complexity equals volume (CV) proposal \cite{Stanford:2014jda}, which is at the core of our interests in the present paper. The CV proposal states that the complexity of the CFT state on a given time slice $\Sigma_{\rm CFT}$ on the boundary is captured by the volume of the maximal volume codimension-one surface $\mathcal{B}$ in the bulk such that $\partial \mathcal{B} = \Sigma_{\rm CFT}$ \emph{i.e.,} the bulk surface is anchored on the same boundary time slice on which the CFT state is defined. The precise definition is given by\footnote{Note that there is an ambiguity in the choice of the length scale that appears in the denominator of eq.~\eqref{def:CV1} to make $\mathcal{C}_V$ dimensionless. In most of the literature, it is chosen to be the AdS length scale $L$, a choice which we adhere to as well.}
\be
\mathcal{C}_V = \frac{{\rm Vol}(\mathcal{B})}{G_N L}~.
\label{def:CV1}
\ee
The presence of an ETW brane in the bulk with intrinsic gravitational dynamics brings in a new element to the CV proposal. As argued in \cite{Hernandez:2020nem} from a doubly-holographic perspective, the complexity for the CFT state now includes a contact term at the location of the brane, proportional to the volume of the region of intersection of the bulk extremal surface and the brane,
\be
\mathcal{C}_V = \frac{{\rm Vol}(\mathcal{B})}{G_N L} + \frac{{\rm Vol}(\mathcal{B} \cap {\rm brane})}{G_N^{\rm brane} L^{\rm brane}}~.
\label{def:CV2}
\ee
Here $L^{\rm brane}$ is an appropriate length scale on the brane that renders the contact term dimensionless. For our setup, with the ETW brane endowed with JT gravity, we follow the convention of picking the length scale associated with the relevant geometry, and so we pick the length scale $L^{\rm brane}$ to correspond to the AdS$_2$ length scale $L_2$. The effect of keeping an arbitrary $L^{\rm brane}$ on the main results is discussed in footnote~\ref{foot:Lbrane dependence}.
The bulk codimension-one surface $\mathcal{B}$ anchored at $\Sigma_{\rm CFT}$ now maximizes the RHS of eq.~\eqref{def:CV2}. The presence of the contact term does not alter the bulk equations of motion, but it does affect the boundary conditions, and therefore value of the complexity $\mathcal{C}_V$ itself. 

\subsection{The Lloyd bound on complexity growth}
\label{sec:Lloyd bound}
We now briefly comment upon the Lloyd bound on the rate of change of complexity, which is the final ingredient that we will need for our study. The Lloyd bound was originally proposed as a fundamental bound on the maximum possible rate of computation based on the average energy of the quantum system \cite{lloyd2000ultimate}. It was subsequently generalized to quantum computational complexity in \cite{Brown:2015lvg}, where it was interpreted as providing a bound on the maximum possible rate of change of complexity for a quantum system. For holographic theories with a dual gravitational description provided by an AdS black hole, the energy of the system could be replaced by the mass $M$ of the black hole, leading to the following statement for the Lloyd bound on the rate of change of complexity \cite{Brown:2015lvg}, 
\be
\label{Lloyd_Bound_1}
\left| \dv{\mathcal{C}}{t}\right| \leq \gamma M.
\ee
Here $\gamma$ is a numerical factor that depends upon the specifics of the case. The bound above can be made tighter if the system carries additional conserved charges, such as an electric charge or angular momentum \cite{Engelhardt:2021mju}. 

Our interest in the present work is in planar AdS$_3$ black hole geometries with ETW branes. The dual two-dimensional CFT state is also translationally invariant, and thus it makes sense to recast the bound in eq.~\eqref{Lloyd_Bound_1} in terms of complexity per unit length on the boundary time slice $\Sigma_{\rm CFT}$, which we denote by $c_{{}_V}$. Using the fact that the energy density for the state is $\varepsilon = {r_0^2}/{8\pi G_N L^3}$ in eq.~\eqref{Lloyd_Bound_1}, the statement of the Lloyd bound for our setup becomes
\be
\label{Lloyd_Bound_2}
\left| \dv{c_{{}_V}}{t}\right| \leq \frac{r_0^2}{2 G_N L^3}\, ,
\ee
where we have chosen $\gamma = 4\pi$. This choice corresponds to the value associated with the saturation of the Lloyd bound for eternal black holes \cite{Brown:2015bva}. For our setup, the rate of complexity growth also saturates the bound in eq.~\eqref{Lloyd_Bound_2} when there is no intrinsic gravitational dynamics on the brane, as can be seen in fig.~\ref{fig:No_JT_plots_2}. This makes the choice $\gamma = 4\pi$ natural.

As mentioned in the Introduction section~\ref{sec:intro}, with intrinsic gravitational dynamics present on the brane, we observe a violation of the Lloyd bound eq.~\eqref{Lloyd_Bound_2} for a certain range of values for the bulk parameters comprising the brane location $y_{\rm brane}$ and the JT coupling $\alpha$. As discussed in detail in sections \ref{sec:complexity_growth} and \ref{sec:numerical_approach}, we find that although the asymptotically early/late time rate of change of complexity does approach the value set by the Lloyd bound, for part of the bulk parameter space a violation occurs during the intermediate stages of time evolution of the state. In particular, for positive (negative) values of $\alpha$ violating the bound, the rate of change of complexity reaches the Lloyd bound value from above (below) at asymptotically late (early) times, signaling a violation of the Lloyd bound during intermediate stages of time evolution - see for instance the figs. \ref{fig:dCVdt_With_JT_1} and \ref{fig:With_JT_plots}.\footnote{This pattern of violation of the Lloyd bound is similar to other instances observed in \cite{Carmi:2017jqz, Yang:2017czx, An:2018xhv, Alishahiha:2018tep, Auzzi:2018pbc, Bernamonti:2021jyu, Auzzi:2023qbm}. It is worth pointing out that aside from~\cite{Auzzi:2023qbm}, all of the aforementioned violations of the Lloyd bound were in the context of the CA proposal, and not for the CV proposal, which is seemingly more robust in meeting the bound. See also the discussion in \cite{Auzzi:2018zdu}, where CV respects the Lloyd bound in a warped AdS$_3$ setup as well, but with the warping parameter playing a nontrivial role in eq.~\eqref{Lloyd_Bound_1}.} In section~\ref{sec:Engelhardt-Folkestad}, we argue that this violation of the Lloyd bound is tied to the violation of the WEC by the associated bulk parameters. More precisely, we introduce an equivalent description of our setup with an ETW brane in terms of a two-sided asymptotically AdS$_3$ geometry, with the brane replaced by a suitably chosen thin shell of matter, such that the rate of change of complexity is identical for the two cases. The violation of the WEC in the two-sided setup can then be seen to translate into a violation of the Lloyd bound for the single-sided setup with a brane. Further, demanding the WEC be met for the two-sided setup is equivalent to requiring the Lloyd bound to be met by complexity growth for the single-sided setup with a brane, leading to an effective reduction in the allowed bulk parameter space $(y_{\rm brane}, \alpha)$.


\section{Asymptotic behaviour of complexity growth}
\label{sec:complexity_growth}
We now proceed to perform a detailed analysis of the asymptotic behaviour of the rate of change of the volume complexity for our setup, employing some of the techniques laid out in \cite{Belin:2021bga, Belin:2022xmt}. After introducing some relevant definitions, we discuss the necessary boundary conditions that need to be imposed for computing extremal surfaces in section~\ref{sec:bdy_conds}. Subsequently, in section~\ref{sec:mech}, we introduce an effective mechanical picture describing the problem of interest in terms of a particle scattering off a potential barrier. This helps build an intuitive understanding of the situation. This is followed by the details of the asymptotic analysis in section~\ref{sec:asymptotics}, where we establish a clear distinction between parts of the bulk parameter space, comprising the brane location $y_{\rm brane}$ and the JT coupling $\alpha$, which respect or violate the Lloyd bound on complexity growth, eq.~\eqref{Lloyd_Bound_2}.

Employing the Eddington-Finkelstein coordinates~\eqref{eq:metric.EF}, and assuming the codimension-one maximal volume surface is parametrized in terms of a parameter $\sigma$, \emph{i.e.}
\begin{equation}
\label{eq:bulkcodim1surf}
x^\mu_{\rm extremal}=(v(\sigma),r(\sigma),x),
\end{equation}
the induced metric on the extremal surface is given by
\begin{equation}
    d s^2_{\text{ind}}=\left(-f(r)\dot{v}^2+2\dot{v}\dot{r}\right) d \sigma^2 +\frac{r^2}{L^2}dx^2\,, 
\end{equation}
where an overhead dot denotes a derivative with respect to the parameter $\sigma$. 

We study the complexity volume functional of the form (\ref{def:CV2}). Specifically for the case of JT gravity on the brane, the contact term in eq.~\eqref{def:CV2} reads
\be
\frac{{\rm Vol}(\mathcal{B} \cap {\rm brane})}{G_N^{\rm brane} L^{\rm brane}} = \frac{\ell \, \varphi(r_{\rm brane}, v_{\text{brane}})}{G_N^{\rm brane} L_2} \,. \label{eq:contact}
\ee
Here $L_2$ denotes the AdS$_2$ length scale on the brane, and $\varphi$ is the dilaton solution rewritten in the $(r, v)$-coordinates. The volume complexity~\eqref{def:CV2} per unit length for the boundary CFT state is then given by
\begin{equation} \label{eq:cv_EF}
    c_{{}_V }= \frac{\varphi(r_{\rm brane},\,v_{\rm brane})}{G_N^{\rm brane} L_2}+\frac{1}{G_NL} \int d \sigma~ \mathcal{L}~,
\end{equation}
where $r_{\rm brane}$ and $v_{\rm brane}$ are the location of the intersection of the maximal volume surface and the brane, and the ``effective Lagrangian'' $\mathcal{L}$ is given by
\be
\mL = \frac{r}{L}\sqrt{-f(r)\dot{v}^2+2 \dot{v} \dot{r}}\,. \label{eq:L_v}
\ee

The functional~\eqref{eq:cv_EF} is invariant with respect to the reparametrizations $\sigma \to h(\sigma)$. This gauge freedom can be fixed by imposing the condition 
\be\label{eq:gauge-choice}
{\cal L}  \equiv g \,, 
\ee
where $g$ is some function of the coordinates $(v,r)$ and velocities $(\dot{v},\dot{r})$. 

Notably, the Lagrangian in~\eqref{eq:L_v} is independent of $v$, which implies that the corresponding conjugate momentum 
\be\label{eq:Pv definition}
P_v\equiv\frac{\pd \mL}{\pd \dot{v}}
\ee
is conserved along the trajectory, \ie it is $\sigma$ independent. In terms of $P_v$, the equations of motion that follow by extremizing eq.~\eqref{eq:cv_EF} can be written as
\begin{subequations}\label{eq:velocities1}
\begin{align}
    \dot{v} &= \left(-P_v \pm \sqrt{\frac{f r^2}{L^2}+P_v^2 }\right) \frac{L^2}{r^2} \frac{g}{f}~,\label{eq:vdot}\\
    \dot{r}^2 &=\left( \frac{f r^2}{L^2}+P_v^2 \right) \frac{L^4}{r^4} g^2 \,.\label{eq:rdot}
\end{align}
\end{subequations}
Solutions to the above equations correspond to the desired extremal surfaces.

\subsection{Boundary conditions}
\label{sec:bdy_conds}
The equations of motion given in eq.~(\ref{eq:velocities1}) are to be supplemented with appropriate boundary conditions at the locations where the extremal surface intersects the brane as well as the asymptotic AdS$_3$ boundary. We have two Dirichlet boundary conditions on the asymptotic AdS$_3$ boundary and mixed (one Dirichlet and one Neumann) boundary conditions on the brane. 

\paragraph{Asymptotic boundary conditions in Schwarzschild coordinates.} 
The Dirichlet boundary conditions on the asymptotic boundary are 
\bea
r(0) &=& \frac{L}{\epsilon}=r_{\rm max}\,, \quad \epsilon \to 0 \label{eq: Dirichlet-r}\\ 
t(0) &=& v(0) - \int_{0}^{r(0)} \frac{d x}{f(x)} = t_{\text{bdy}}\,. \label{eq: Dirichlet-t}
\eea
Here $\epsilon$ acts as a regulator for otherwise divergent volumes of surfaces due to the hyperbolic nature of the asymptotically AdS geometry. We have defined $t_{\rm bdy}$, which is the boundary time at which the maximal volume surface anchors at the asymptotic boundary. Importantly, for each $t_{\rm bdy}$, the corresponding maximal volume surface will have specific conserved momentum $P_v$. In other words, while the conserved momentum is independent of the path parameter $\sigma$, it does depend on the boundary time $t_{\rm bdy}$ at which the surface is anchored.

\paragraph{Brane boundary conditions in global coordinates.}
To discuss the brane boundary conditions, it is useful for a moment to go to global coordinates first, and then rewrite them in Schwarzschild coordinates. Recall that in global coordinates the dynamical variables are $y(\sigma)$ and $\tau(\sigma)$, and the effective Lagrangian is
\begin{equation}
    {\cal L} = \frac{r_0 \cos(\tau)}{\cos^2(y)} \sqrt{\dot{y}^2-\dot{\tau}^2}\,. \label{eq:L_tau}
\end{equation}
The Dirichlet boundary condition on the brane imposes that the extremal surface is anchored on the brane at the location $y = y_{\rm brane}$. The position of the brane $y_{\rm brane}$ is related to the tension $T_0$ by
\begin{equation}\label{eq:replacement Tension}
    \sin^2(y_{\rm brane}) =T_0^2 \,,
\end{equation} 
or in the presence of JT gravity on the brane, to the brane cosmological constant $\Lambda^{\rm brane}$ by
\begin{equation}\label{eq:replacement Lambda}
    \cos^2(y_{\rm brane}) = - L^2\Lambda^{\rm brane}~.
\end{equation}

The Neumann boundary condition on the brane is obtained by varying the functional~(\ref{eq:cv_EF}) and setting to zero the generalized momentum conjugate to the normal coordinate of the brane. In global coordinates, the normal coordinate is simply $\tau$, and the boundary condition reads
\begin{equation}
    \frac{G_N L}{G^{\rm brane}_N L_2} \eval{\varphi'(\tau)}_{\rm brane} + \eval{P_\tau}_{\rm brane}=0\,,
\end{equation}
where $\varphi' \equiv \frac{d\varphi}{d\tau}$, which we can determine from eq.~\eqref{eq:dilaton_sol}, and $P_\tau \equiv \frac{\partial {\cal L}}{\partial \dot{\tau}}$, which we can determine from eq.~\eqref{eq:L_tau}. After putting all these ingredients together, we find that the Neumann boundary condition in global coordinates is given by 
\begin{equation}
\label{eq:Neumann boundary global coord}
\frac{ L^2 \alpha }{r_0 L_2}-  \frac{\cos\left(\tau_{\rm brane}\right)}{\cos^4 \left(y_{\rm brane}\right)} \eval{\frac{\dot{\tau}}{g}}_{\rm brane} = 0\, ,
\end{equation}
where we have used the definitions of $\alpha$~\eqref{def_alpha} and of $g$~\eqref{eq:gauge-choice}.

\paragraph{Brane boundary conditions in Schwarzschild coordinates.}
The Dirichlet boundary condition on the brane is determined by the brane profile, given in Schwarzschild coordinates by eq.~(\ref{eq: brane location Schw}) when there is no intrinsic gravity on the brane and eq.~(\ref{eq: brane location Schw with JT}) for the case of JT gravity on the brane. 

To rewrite the Neumann boundary condition~(\ref{eq:Neumann boundary global coord}) in Schwarzschild coordinates, we use the coordinate transformation given by eq.~(\ref{eq:Sch to global}), and find that it becomes
\begin{equation}
\begin{split}
\label{eq:Neumann boundary EF0}
&\frac{L^2 \alpha\, r_{\rm brane}}{r^2_0 L_2}\cos(y_{\rm brane}) = \\
&\frac{r_{\rm brane}^2}{r_0\cos\left(y_{\rm brane}\right)} \frac{\sqrt{r_0^2-r_{\rm brane}^2 \cos^2\left(y_{\rm brane}\right)}}{r_{\rm brane}^2-r_0^2} \eval{\frac{\dot{r}}{g}}_{\rm brane} + \frac{r_{\rm brane}}{r_0} \frac{L^2\tan\left(y_{\rm brane}\right)}{r_{\rm brane}^2-r_0^2} \, P_v\,.
\end{split}
\end{equation}

\subsection{Effective mechanical picture}\label{sec:mech}
\begin{figure}
    \centering
    \includegraphics[width=0.55\textwidth]
    {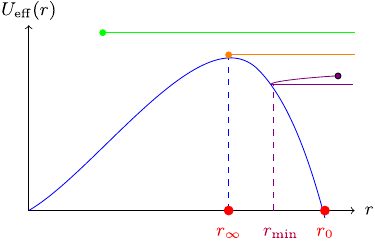}\hspace{5mm}
    \includegraphics[width=0.35\textwidth]
    {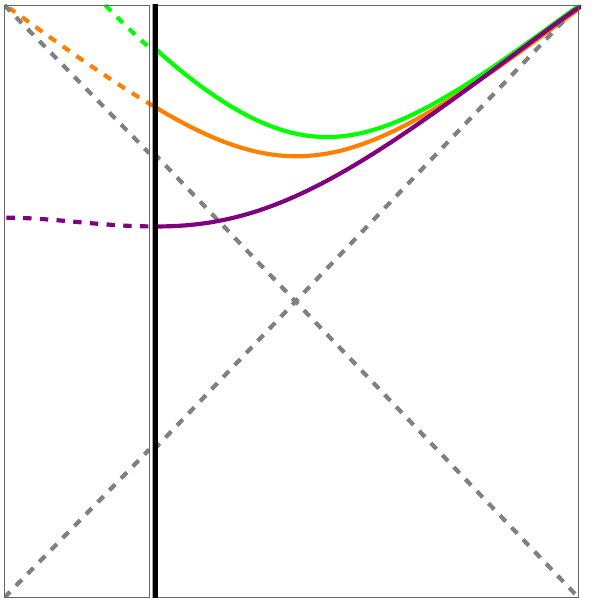}\\
    \hspace{1cm}(a) \hspace{6cm} (b)
    \caption{(a) Characteristic effective potential~(\ref{eq:U_0}) for the CV proposal with the gauge choice $g=r^2/L^2$. Here $r_0$ is the location of the horizon, and $r_{\infty}$ is the location of the maximum of the potential. Three possible particle trajectories are shown in purple, orange, and green. The purple trajectory depicts a particle with energy $P_v^2<U_{\rm eff}(r_\infty)$, which scatters at the potential. After scattering on the potential, the particle reaches the brane at $r_{\rm brane}>r_\infty$, indicated by the purple dot. The orange trajectory corresponds to a particle with energy $P_v^2=U_{\rm eff}(r_\infty)$, which asymptotically reaches the maximum of the potential at $r_\infty$. In this case, the location of the brane is indicated by the yellow dot and asymptotes to $r_\infty$ at late times. Finally, the green trajectory corresponds to a particle with energy $P_v^2>U_{\rm eff}(r_\infty)$, which passes over the potential barrier. The particle moves towards values of $r$ that are smaller than $r_\infty$ until it eventually terminates at the brane, visualized by a green dot. The extremal surfaces determined by these trajectories are sketched in panel (b).}
    \label{fig:Ueff}
\end{figure}

Our goal is to find the bulk codimension-one surface that maximizes the functional~\eqref{eq:cv_EF} under the specified boundary conditions discussed above. Let us look at this problem in Eddington-Finkelstein coordinates and with the reparametrization freedom of $\sigma$ fixed such that $g = \frac{r^2}{L^2}$. With this parametrization, the profile of the maximal volume surface can be considered as the trajectory of a classical particle moving under the influence of an effective potential. This analogy arises from interpreting the projection of the bulk codimension-one surfaces parametrized by $x^\mu=(v(\sigma), r(\sigma), x)$ to the $(v,r)$ plane as the trajectory followed by a classical particle on a line. The effective Lagrangian $\mathcal{L}$~\eqref{eq:L_v} describes the particle's dynamics and its motion is governed by the equations of motion~\eqref{eq:velocities1}. More specifically, eq.~\eqref{eq:rdot} with the gauge choice $g=\frac{r^2}{L^2}$ becomes\footnote{The gauge $g=\frac{r^2}{L^2}$ ensures that the $\dot{r}$ equation contains a constant $P_v^2$ which is then interpreted as the energy of the non-relativistic particle.}
\be
\dot{r}^2 + U_{\rm eff}(r) = P_v^2 . \label{eq: particle-EOM}
\ee 
The total energy of the particle is given by $P_v^2$. It is conserved along the trajectory of the particle, or, equivalently, along the given extremal surface. The effective potential in eq. (\ref{eq: particle-EOM}) is given by
\begin{align}
U_{\rm eff}(r) = -\frac{f(r)r^2}{L^2}\,, \label{eq:U_0}
\end{align}
and characterizes the ``force'' experienced by the particle as it moves in the radial direction. The profile of the potential is depicted in fig.~\ref{fig:Ueff}. There are three possibilities for the particle trajectory shown by purple, yellow, and green lines in Fig.~\ref{fig:Ueff}, which are realized depending on the value of the energy $P_v^2$ in relation to the maximum value of the potential at $r_\infty = \frac{r_0}{\sqrt{2}}$, which we denote as 
\begin{equation}\label{eq:U(rinfty)new}
	P_\infty^2 \equiv U_{\rm eff}(r_\infty) =\frac{r^4_0}{4 L^4}.
\end{equation}
These possibilities are the following.
\begin{itemize}
    \item[\textbf{a)}] When $P_v< P_\infty$, the particle scatters off the potential at a turning point $r_{\min}$, turns around, and continues to larger values of $r$ until terminating at the brane at $r_{\rm brane}>r_\infty$. The turning point $r_{\text{min}}$ of this trajectory is determined by $\dot{r} = 0$, so that eq: (\ref{eq: particle-EOM}) gives 
\begin{align}\label{eq:Pv def}
P_v^2 = U_{\rm eff}(r_{\rm min}).
\end{align}    
    The trajectory corresponding to this scattering case is shown by the purple line in Fig.~\ref{fig:Ueff}(a). 
    \item[\textbf{b)}] When $P_v=P_\infty$, the particle moves towards the top of the potential at $r_\infty$, where it eventually terminates. In this case $r_{\text{brane}}$ is asymptotically close to $r_\infty$. The trajectory corresponding to this case is shown by the orange line in Fig.~\ref{fig:Ueff}(a). 
    \item[\textbf{c)}] When $P_v>P_\infty$, the particle passes over the potential barrier and moves towards values of $r$ that are smaller than $r_\infty$ until it eventually terminates at the brane at $0 < r_{\text{brane}} < r_\infty$. This trajectory is shown by the green line in Fig.~\ref{fig:Ueff}(a).
\end{itemize}

To investigate the rate of complexity growth we evaluate the variation of the complexity functional~(\ref{eq:cv_EF}) with respect to boundary time,\footnote{That is, we vary eq.~\eqref{eq:cv_EF}, and find
\begin{equation}
    \delta c_{{}_V} = \frac{1}{G_N L} \int d\sigma \left(\text{Eq.~of motion} \right) \delta x^\mu + \frac{1}{G_N L} \eval{\left(P_v \delta v^\mu + P_r \delta r\right)}_{\rm brane}^{\rm bdy} + \frac{1}{G_N^{\rm brane} L_2} \eval{\partial_\mu \varphi \, \delta x^\mu}_{\rm brane}\,,
 \end{equation} together with the fact that $\eval{v}_{\rm bdy} = t_{\rm bdy}$ to find eq.~\eqref{eq:growth CV Pt}.} which gives
\begin{equation}\label{eq:growth CV Pt}
	\dv{c_{{}_V}}{t_{\rm bdy}}= \frac{P_v}{G_N L}\,.
\end{equation}
Eq.~\eqref{eq:growth CV Pt} then implies that to analyze the time dependence of complexity, we need to find the relation between $P_v$ and $t_{\rm bdy}$. We can obtain this by recasting eq.~\eqref{eq:vdot} in Schwarzschild coordinates, without fixing a gauge,  
\begin{equation}\label{eq:exptoint}
   \dot{t}(\sigma)=-\frac{P_v}{f(r)}\frac{L^2}{r^2}g \,.
\end{equation}
Employing the relation~(\ref{eq:rdot}), one can integrate eq.~(\ref{eq:exptoint}) to obtain
\begin{equation}\label{eq:tbdy integral1}
    t_{\rm bdy}-t_{\rm brane}=-\int^{r_{\rm bdy}}_{r_{\rm brane}} d r \frac{P_v}{f(r)\sqrt{\frac{f(r)r^2}{L^2}+P_v^2} }~,
\end{equation}
where $r_{\rm brane}$ and $t_{\rm brane}$ are the coordinates of the intersection of the maximal volume surface and the brane. Notice that the result above is independent of the particular choice of gauge $g$. At late times ($t_{\rm bdy}\to \infty$), the right-hand side of eq.~\eqref{eq:tbdy integral1} diverges, and so the integral in the right side should diverge too. This occurs when $P_\infty^2$ approaches the maximum of the effective potential $P_\infty^2 \equiv U_{\rm eff}(r_\infty)$. Therefore the late time value of $P_v$ is 
\begin{equation}\label{eq:limitlatepv}
	\lim_{t_{\rm bdy} \to \infty} P_v = P_\infty.    
\end{equation}

The possibilities \textbf{a)}-\textbf{c)} for the particle trajectory depending on the value of the energy $P_v^2$, as described above, reflect in the integral (\ref{eq:tbdy integral1}) as different singularity structures of the integrand. This makes it so each of the cases \textbf{a)}, \textbf{b)} and \textbf{c)} requires a different integration contour and a different asymptotic expansion for late times $t_{\text{bdy}}$. Thus, the different qualitative options for the effective particle trajectory correspond to different qualitative behaviours of holographic complexity at $t_{\text{bdy}} \to \infty$. As we will see below, the case \textbf{a)} describes the complexity evolution which satisfies the Lloyd bound, the case \textbf{b)} describes the complexity evolution in the edge case where the Lloyd bound is satisfied marginally, and the case \textbf{c)} describes the Lloyd bound violation. Let us now proceed to analyze these options in detail. 

\subsection{Asymptotic analysis}
\label{sec:asymptotics}

Our primary objective is to compute the late time expansion of ${\rm d} c_{{}_V}/{\rm d} {t_{\rm bdy}}$ with JT gravity on the brane with a general coupling $\alpha$, as well as a cosmological constant $\Lambda^{\rm brane}$ which plays the role of nonzero tension when $\Lambda^{\rm brane}<\Lambda^{\rm bulk}$. In doing so, we uncover three qualitatively different behaviours of ${\rm d} c_{{}_V}/{\rm d} {t_{\rm bdy}}$ depending on the parameters of the model. Namely, the Lloyd bound~\eqref{Lloyd_Bound_2} can be satisfied, marginally satisfied, or violated which corresponds to the cases \textbf{a)}, \textbf{b)} and \textbf{c)} in section ~\ref{sec:mech} respectively. These turn out to be correlated with the depth at which the maximal volume surface probes the geometry. Specifically, the Lloyd bound holds true when the maximal volume surface stays above $r_\infty$, while its violation occurs precisely when the surface delves deeper than $r_\infty$. The Lloyd bound is only marginally satisfied when the maximal volume surface is precisely at $r_\infty$.

Before going into the details of how to compute the late time expansion of ${\rm d} c_{{}_V}/{\rm d} {t_{\rm bdy}}$, let us explain briefly the effect of modifying $\Lambda^{\rm brane}$ and $\alpha$ on the behaviour of the brane and the maximal volume surface. Of particular interest will be the asymptotic value of the intersection of these two, which will determine if the maximal volume surface probes deeper than $r_\infty$. Therefore, we are interested in the late time limit of $r_{\rm brane}$, which we denote as $r_{\rm brane}^* = \lim_{t_{\rm bdy}\to \infty} r_{\rm brane}$. 
\begin{itemize}
    \item When $\Lambda^{\rm brane}<\Lambda^{\rm bulk}=-1/L^2$ (or equivalently $T_0>0$), the brane moves away from $y_{\rm brane} = 0$ to a position determined by eq.~\eqref{eq:replacement Lambda}. Consequently, decreasing $\Lambda^{\rm brane}$ (or increasing tension) results in an increase in $r_{\rm brane}$. In particular, in the absence of JT gravity (or if $\alpha = 0$), $r_{\rm brane}^*>r_\infty$.

    \item When $\alpha\neq0$, the angle at which the maximal volume surface intersects the brane changes as determined by the Neumann boundary condition on the brane~\eqref{eq:Neumann boundary global coord}. Consequently, for $t_{\rm bdy}>0$, increasing $\alpha$ decreases $r_{\rm brane}$. In particular, for $\Lambda^{\rm brane} = \Lambda^{\rm bulk}$ and $\alpha>0$, $r_{\rm brane}^*<r_\infty$. Note that although a nonzero JT coupling breaks time reversal symmetry in the model, there is a spurious symmetry $(t_{\rm bdy},\alpha)\to (-t_{\rm bdy},-\alpha)$, and therefore $\lim_{t_{\rm bdy}\to -\infty} r_{\rm brane}<r_\infty$ for $\alpha<0$. 
	
\end{itemize}
In the discussion below, we will study the three qualitatively different regimes separately. Concretely, the steps required to compute the late time expansion of  ${\rm d} c_{{}_V}/{\rm d} {t_{\rm bdy}}$ in each regime are:

\begin{enumerate}
    \item[\S 1.] Finding the position of the intersection of the maximal volume surface and the brane \emph{i.e.} $(r_{\rm brane}, t_{\rm brane})$ as a function of the conserved momentum  $P_v$. This can be done by combining eqs.~\eqref{eq:rdot}, \eqref{eq:Neumann boundary EF0} and \eqref{eq:quartic}.
    \item[\S 2.] Calculating the time difference between the boundary and the brane along the maximal volume surface (denoted by $t_{\rm bdy} - t_{\rm brane}$) in terms in of $P_v$, using eq.~\eqref{eq:tbdy integral1} in the late time regime.
    \item[\S 3.] Combining the results above to derive an expression for $P_v$ in the late time regime, and using it to determine ${\rm d} c_{{}_V}/{\rm d} {t_{\rm bdy}}$ via eq.~\eqref{eq:growth CV Pt}.
\end{enumerate}

Let us now proceed to systematically address each of these steps, and apply this procedure to several illustrative examples. For detailed calculations, we refer the reader to Appendices~\ref{sec:simple} and \ref{app: analytics}.

\paragraph{\S 1.~Intersection of maximal volume surface and brane:}
The first step is to find the location of the intersection of the brane and the extremal surface for CV as a function of $P_v$. First, we use the equation of motion for $\dot{r}$~\eqref{eq:rdot} and the Neumann boundary conditions on the brane~(\ref{eq:Neumann boundary EF0}) to obtain
\begin{equation}\label{eq:brane position}
    P_v^2-{ U}_{\rm eff}(r_{\rm brane}) = \frac{r_{\rm brane}^2}{r_0^2-r_{\rm brane}^2 \cos^2(y_{\rm brane})} \left[\frac{r_0^2-r_{\rm brane}^2}{L_2 r_0}\,\alpha \cos^2(y_{\rm brane})+ P_v \sin(y_{\rm brane}) \right]^2.
\end{equation}

Solving this expression determines the location $r_{\rm brane}$ as a function of the momentum $P_v$, for a given tension/cosmological constant (related to $y_{\rm brane}$), and JT coupling $\alpha$. It can be rewritten as
\begin{equation}\label{eq:quartic}
    (r_0^2-r_{\rm brane}^2)\left(a-b \frac{r_{\rm brane}^2}{r_0^2} + c \frac{r_{\rm brane}^4}{r_0^4}\right)=0\,,
\end{equation}
where
\begin{equation}
    \begin{aligned}
        a & = \frac{P_v^2}{\cos^2(y_{\rm brane})}\,,\\
        b & =  \frac{r_0^4}{L^4} \sec^2(y_{\rm brane})+ \frac{r_0}{L}  \alpha P_v \sin(2y_{\rm brane}) + \frac{r_0^2}{L^2} \alpha^2 \cos^4(y_{\rm brane})\,, \\
        c & = \frac{r_0^4}{L^4}+ \frac{r_0^2}{L^2} \alpha^2 \cos^4(y_{\rm brane})\,.
    \end{aligned}
\end{equation}
The solutions to the quartic equation within the second parentheses in eq.~\eqref{eq:quartic} determine the location of the intersection $r_{\rm brane}$. 

With the behaviour of $r_{\rm brane}$ as a function of $P_v$ under control, we can determine the $P_v$ dependence of $t_{\rm brane}$ using the brane embedding~\eqref{eq: brane location Schw} which in terms of $y_{\rm brane}$ looks like\footnote{Note that in the most relevant case, and in particular whenever we are close to the boundary in parameter space between Lloyd bound violating and respecting regions, determined by eq.~\eqref{eq:alpha condition}, the intersection of the extremal surface computing the volume complexity and the brane happens inside the horizon, but we keep both expressions here for completeness.}
\begin{equation}
\begin{aligned}
\label{eq:embedding-y}
        \frac{r_{\rm brane}^2}{r_0^2}& = \frac{1-\sin^2(y_{\rm brane}) \,{\rm tanh}^2\left(\frac{r_0 t_{\rm brane}}{L^2}\right)}{\cos^2(y_{\rm brane})} \quad {\rm outside\ the\ horizon}\,\\
         \frac{r_{\rm brane}^2}{r_0^2}& = \frac{1-\sin^2(y_{\rm brane}) \,{\rm coth}^2\left(\frac{r_0 t_{\rm brane}}{L^2}\right)}{\cos^2(y_{\rm brane})} \quad {\rm inside\ the\ horizon}\,.
\end{aligned}
\end{equation}

\paragraph{\S 2.~Time difference between the boundary and the brane:} We now need to evaluate the time difference between the boundary and the brane along the extremal surface computing the volume complexity, denoted by $t_{\text{bdy}} - t_{\rm brane}$, using the integral in~\eqref{eq:tbdy integral1}. We only need to evaluate this integral in the late time limit for computing the asymptotic behaviour of complexity growth. To do this, we expand the momentum around its late time value,\footnote{Recall that the momentum $P_v$ is conserved as a function of the path parameter $\sigma$ along its trajectory. But different trajectories have different values of $P_v$ which at late times $\left(t_{\rm bdy}\to \infty \right)$ approaches $P_\infty$, as explained around \eqref{eq:limitlatepv}}
\begin{equation}\label{eq:Pinfty}
P_{v}^2 = P_{\infty}^2 \pm \delta P^2\,,
\end{equation}
where $P_{\infty} = \frac{r_0^2}{2L^2} = \frac{r_{\infty}^2}{L^2}$, and work perturbatively in $\delta P$.

In terms of the effective mechanical picture of section~\ref{sec:mech}, when the particle has energy greater than the maximum of the potential, we use the plus sign in eq.~\eqref{eq:Pinfty}. This positive sign corresponds to a violation of the Lloyd bound, in the sense that ${\rm d}c_{{}_V}/{\rm d}t_{\rm bdy}$ approaches its asymptotic value from above, and is therefore not bounded by it. Conversely, when the particle scatters off the potential, we use the negative sign in eq.~\eqref{eq:Pinfty}, and the Lloyd bound is satisfied. The integral of interest~\eqref{eq:tbdy integral1} can then be expressed as
\begin{equation}\label{eq:main integral}
    I=-\int_{r_{\rm brane}}^{r_{\rm bdy}}\frac{P_\infty~d r}{L^{-2}(r^2-r_0^2)\sqrt{L^{-4}(r^2-r_\infty^2)^2\pm\delta P^2}}+\mathcal{O}(\delta P^2)
\end{equation}
where we have used $P_{\infty} = \frac{r_0^2}{2L^2} = \frac{r_{\infty}^2}{L^2}$ in the denominator. The contour of integration depends on the location of $r_{\rm brane}$ with respect to $r_\infty$, which as we will explain below is intricately tied to whether the Lloyd bound is violated or respected. The derivation of the late time behaviour of $t_{\text{bdy}} - t_{\rm brane}$ in different cases can be found in appendix~\ref{app:late time}. This results in an expression for $\left(t_{\text{bdy}} - t_{\rm brane}\right)$ as a function of $\left(r_{\rm brane},\delta P\right)$.

\paragraph{\S 3.~Late time expansion of $P_v$ and ${\rm d} c_{{}_V}/{\rm d} {t_{\rm bdy}}$:}
In this final step, we utilize the results obtained thus far to determine $P_v$. 
Inverting the result obtained in step two, we can find $\delta P$ as a function of $\left(r_{\rm brane},t_{\text{bdy}}-t_{\rm brane}\right)$. Plugging the asymptotic expansion of $r_{\rm brane}(\delta P,\alpha,y_{\rm brane})$ and $t_{\rm brane}(\delta P,\alpha,y_{\rm brane})$ from step one then gives us the late time expansion for $P_v$. Finally, we use the relation~\eqref{eq:growth CV Pt} in order to find the late time rate of growth of volume complexity.

Employing the procedure outlined in \S 1 - \S 3 above, we now summarize the late time rate of growth of volume complexity in various regions of the bulk parameter space $(y_{\rm brane}, \alpha)$. In each case, we begin with a simple example before working out the more general problem.

Before moving on to the analysis, let us note that whether the Lloyd bound is respected or violated for a certain choice of the bulk parameters can be easily seen via the Neumann boundary conditions on the brane~(\ref{eq:Neumann boundary EF0}). More specifically, the analogy with the mechanical picture in section~\ref{sec:mech} directly implies that the Lloyd bound is violated if and only if the particle has energy higher than the maximum of the potential barrier in fig.~\ref{fig:Ueff}, \emph{i.e.} $P_v^2 > P_{\infty}^2$. Conversely, if the Lloyd bound is to be respected, the particle must have lower energy than $P_\infty^2$ and will therefore reach a minimum $r_{\rm min} > r_\infty$ and bounce off from the potential barrier to a larger $r_{\rm brane}$. The marginal case corresponds to the extremal surface reaching the curve $r=r_\infty$ at late times. For this to be consistent with the Neumann boundary condition, we need $\dot{r} \big|_{\rm brane}=0$ at late times. Note that $P_v = P_\infty$ for $r= r_\infty$ which we will use in eq.~\eqref{eq:Neumann boundary EF0}. This gives\footnote{\label{foot:Lbrane dependence} The following is dependent on the choice $L^{\rm brane} = L_2$ made in the normalization of the contact term in eq.~\eqref{def:CV2}. For other choices, there should be an extra overall factor of $\frac{L^{\rm brane}\cos(y_{\rm brane})}{L}$.}
\begin{equation}\label{eq:alpha condition}
    |\alpha| = \alpha_c =  \frac{-\sin(y_{\rm brane})}{\cos^3(y_{\rm brane})} \frac{r_0}{L}= \frac{r_0}{L} \frac{\sqrt{-(L^{2}\Lambda^{\rm brane})^{-1}-1}}{(-L^2 \Lambda^{\rm brane})} \,,
\end{equation}
which is the boundary in the parameter space between the Lloyd bound respecting and the Lloyd bound violating region, as we will discuss further below.

\subsubsection{Boundary in the parameter space}
\label{sec:boundary}
We begin with the boundary between the Lloyd bound respecting vs violating regions in the bulk parameter space of $(y_{\rm brane}, \alpha)$, which is given by eq.~\eqref{eq:alpha condition}. In this case, the Lloyd bound is marginally satisfied.

\paragraph{No JT gravity, no tension.} We begin by summarizing the simplest example,
an ETW brane without intrinsic JT gravity or tension. Specifically, this is the $\alpha =0$ and $y_{\rm brane} = 0$ limit. Following the analysis of Appendix~\ref{app: analytics}, the late time dependence of the intersection of the brane and the extremal surface computing the volume complexity is
\begin{equation}\label{eq: rbrane simple}
     \frac{r_{\rm brane}^2}{r_0^2} = \frac{1}{2} + 2\left(3-2\sqrt{2}\right)^{\frac{1}{2}} e^{-\frac{\sqrt{2}r_0}{L^2}t_{\rm bdy}}.
\end{equation}
The intersection $r_{\rm brane}$ asymptotically reaches $r_\infty= r_0/\sqrt{2}$ from above. 

Integrating~\eqref{eq:main integral} as in Appendix~\ref{app:late time} and using the location of intersection~\eqref{eq: rbrane simple} and solving for the late time rate of complexity growth then gives
\begin{equation}
	\label{eq:dCVdt_no_dil-app}    \dv{c_{{}_V}}{t_{\rm bdy}} = \frac{r_0^2}{2G_N L^3}\left(1-8  \left(3-2\sqrt{2}\right)^{\sqrt{2}}\,e^{-\frac{2\sqrt{2}r_0}{L^2} t_{\rm bdy} }\right)\,+\mathcal{O}\left(e^{-4\sqrt{2}\frac{r_0}{L^2} t_{\rm bdy} }\right)\,.
\end{equation}
It is clear that for the simple example we considered here, the complexity growth reaches its asymptotic value from below, and therefore the Lloyd bound~\eqref{Lloyd_Bound_2} is satisfied. This case corresponds to exactly half of the double-sided black hole geometry~\cite{Belin:2021bga} because the ETW brane simply cuts the wormhole geometry in half at the $y=0$ slice.

\paragraph{General case.} More generally, the boundary in parameter space between the Lloyd bound violating and respecting regions is given by eq.~\eqref{eq:alpha condition}. For this more general case, the position of the intersection of the brane and the extremal surface is
\begin{equation}\label{eq:rbrane-boundary}
\begin{aligned}
    \frac{r_{\rm brane}^2}{r_{\infty}^2} &= 1 + \frac{4\left(3-2\sqrt{2} \right)^{\frac{1}{\sqrt{2}}}}{\sqrt{1+\sin^2(y_{\rm brane})}}\sqrt{D(y_{\rm brane})} \, e^{-\frac{\sqrt{2}r_0}{L^2}t_{\rm bdy}}\,,
\end{aligned}
\end{equation}
where
\begin{equation}
    D(y_{\rm brane})=\left[\frac{\sec(y_{\rm brane})}{\sqrt{6}}\left(\sqrt{4+2\cos^2(y_{\rm brane})}+ 2 \sin(y_{\rm brane})\right) \right]^{2\sqrt{2}}.
\end{equation}
At late times, $r_{\rm brane}$ approaches $r_\infty$ from above. Note that the tensionless limit ($y_{\rm brane} \to 0$) with no JT coupling discussed earlier agrees with the analysis above.

Proceeding as the example above we find
\begin{equation}
\label{eq:Delta t marginal2}
\dv{c_{{}_V}}{t_{\rm bdy}} = \frac{P_\infty}{G_N L}- \frac{4 r_0^2 }{L^3 G_N \left(1+\sin^2 (y_{\rm brane})\right)^2 }\left(3-2 \sqrt{2}\right)^{\sqrt{2}} D\left(y_{\rm brane}\right)  
 e^{-\frac{2\sqrt{2}r_0}{L^2}t_{\rm bdy}}\,.
\end{equation}
Thus, for the general case too, the Lloyd bound value is reached asymptotically from below, respecting the bound. Of course, in the tensionless limit ($y_{\rm brane} \to 0$), the coefficient $D(y_{\rm brane}) \to 1$, in agreement with the $y_{\rm brane}=0,\alpha=0$ case.

\subsubsection{Lloyd bound respecting region}
\label{sec:Lloyd}
When the magnitude of the JT coupling $\alpha$ is smaller than the critical value~\eqref{eq:alpha condition}, the volume complexity satisfies the Lloyd bound~\eqref{Lloyd_Bound_2}, in the sense that its time derivative is always bounded by the asymptotic value, which we elucidate now. As before, we first discuss the simpler case, namely when there is no JT coupling but the tension of the brane is nonzero, after which we generalize our findings to the case of nonzero JT coupling.

\paragraph{No JT gravity, nonzero tension.} For the case with JT coupling $\alpha=0$, and nonzero tension, $y_{\rm brane} \neq 0$, the position of the intersection of the brane and the extremal surface computing the volume complexity is
\begin{equation}\label{eq:simple tension}
\begin{aligned}
   \frac{r^2_{\rm brane}}{r^2_\infty} & =\frac{1}{1+\sin(y_{\rm brane})} - \frac{8(3-2\sqrt{2})^{\sqrt{2}}}{L^2 \sin(y_{\rm brane})} \, B(y_{\rm brane}) \, D(y_{\rm brane})\, e^{-\frac{\sqrt{2}r_0}{L^2}t_{\rm bdy}}\,,
\end{aligned}
\end{equation}
where
\begin{equation}
\begin{split}
     B(y_{\rm brane}) &= \left(2 \csc(y_{\rm brane}) \left(\sqrt{1+\sin(y_{\rm brane})}-1\right)-1\right)\times\\
     &\qquad\qquad\left(\frac{3+2\sin(y_{\rm brane}) + 2\sqrt{2 + 2 \sin(y_{\rm brane})}}{|1+2 \sin(y_{\rm brane})|}\right)^{\frac{1}{\sqrt{2}}},\\
    D(y_{\rm brane}) &= \exp\left(\sqrt{2} \coth^{-1} \left(-\csc(y_{\rm brane}) \sqrt{\frac{3-\sin(y_{\rm brane})}{2}} \right)\right).
\end{split}
\end{equation}
Keeping in mind that $-1<\sin(y_{\rm brane})<0$, we see that $r_{\rm brane} > r_\infty$, and it reaches its final value from above. To derive eq.~\eqref{eq:simple tension}, we are in the regime $e^{-\frac{\sqrt{2}r_0}{L^2}t_{\rm bdy}} \ll y_{\rm brane}$.

Integrating eq.~\eqref{eq:main integral} as in appendix~\ref{app:late time} and using eq.~\eqref{eq:simple tension}, we find
\begin{equation}\label{eq:qqq}
     \dv{c_{{}_V}}{t_{\rm bdy}} = \frac{P_\infty}{G_N L} - \frac{4 r_0^2}{G_NL^3} \left(3-2 \sqrt{2} \right)^{\sqrt{2}} B(y_{\rm brane}) \, D(y_{\rm brane}) \, e^{-\frac{\sqrt{2} r_0t_{\rm bdy}}{L^2}}\,.
\end{equation}
Note that $\lim_{y_{\rm brane}\to 0}B(y_{\rm brane}) \propto -y_{\rm brane}$, which vanishes in the tensionless limit. In this case, the subleading corrections of order $\mathcal{O}\left(\exp\left(- \frac{2\sqrt{2}r_0}{L^2}t_{\rm bdy}\right)\right)$ in eq.~\eqref{eq:qqq} become relevant, which is in agreement with the result with zero tension~\eqref{eq:dCVdt_no_dil-app}. 

\paragraph{Near the boundary in parameter space.} Next, we consider small deviations from the boundary into the Lloyd bound respecting region in parameter space. For concreteness, we focus on the late time expansion $t_{\rm bdy} \to \infty$ with $\alpha>0$.\footnote{ The present analysis in the late time regime constrains $\alpha<\alpha_c$, where $\alpha_c$ is positive, which is why we pick $\alpha>0$ for this analysis. A similar analysis for $t_{\rm bdy}\to -\infty$ provides a lower bound $\alpha>-\alpha_c$.}

First, we find the location of the brane for
\begin{equation}\label{eq:alpha epsilon}
    \alpha = \alpha_c - \delta \alpha \,,
\end{equation}
with a small $\delta \alpha>0$ but at times late enough such that $e^{-\frac{\sqrt{2}r_0}{L^2} t_{\rm bdy} } \ll \delta\alpha$. The position of the intersection of the brane and the extremal surface computing the volume complexity is
\begin{equation}\label{eq: rbrane Lloyd}
\begin{aligned}
    \frac{r_{\rm brane}^2}{r_\infty^2} & = 1 + \frac{L}{r_0} \cos^3(y_{\rm brane}) \left(\sqrt{1+\sin^2(y_{\rm brane})}-\sin(y_{\rm brane})\right) \delta \alpha \\
    & \quad +  \frac{8(3-2\sqrt{2})^{\sqrt{2}}}{L^2 \delta \alpha} \frac{1+\sin^2(y_{\rm brane})}{\cos^3(y_{\rm brane})} \, B(y_{\rm brane},\delta \alpha)\, D(y_{\rm brane},\delta\alpha) \, e^{-\frac{\sqrt{2}r_0}{L^2}t_{\rm bdy}}\,,
\end{aligned}
\end{equation}
where
\begin{align}
    &\log B(y_{\rm brane},\delta \alpha) = \nonumber\\
    &\quad \sqrt{2} \tanh^{-1}\left(\frac{1}{\sqrt{2}} + \frac{1}{\sqrt{2}} \frac{L}{r_0} \cos^3(y_{\rm brane})\left(\sqrt{1+\sin^2(y_{\rm brane})}-\sin(y_{\rm brane})\right)\delta\alpha\right)\nonumber\\
    & \quad -2 \coth^{-1}\left(1 + \frac{L}{r_0} \cos^3(y_{\rm brane})\left(\sqrt{1+\sin^2(y_{\rm brane})}-\sin(y_{\rm brane})\right)\delta\alpha\right),\\
    &\log D(y_{\rm brane},\delta\alpha) = \nonumber\\
    & \quad -\sqrt{2} \, {\rm coth}^{-1}\sqrt{\frac{2+ \cos^2(y_{\rm brane})}{2\sin^2(y_{\rm brane})} + \frac{L}{2r_0} \frac{\cos^5(y_{\rm brane})}{\sin^2(y_{\rm brane})}\left(\sqrt{1+\sin^2(y_{\rm brane})} -\sin(y_{\rm brane}) \right)\delta \alpha}\nonumber.
\end{align}
This result is consistent with eq.~\eqref{eq:simple tension} in the limit $\alpha \to 0$, which implies $y_{\rm brane}$ is ${\cal O}(\delta \alpha)$. The intersection of the extremal surface and the brane remains above $r_\infty$ and approaches its asymptotic value from above as a function of time. Notice that the late time value of $r_{\rm brane}$ approaches $r_\infty$ from above as $\delta \alpha$ becomes smaller. 

Integrating eq.~\eqref{eq:main integral} and putting the above results together, we find that
\begin{equation}\label{eq:qqqq}
    \dv{c_{{}_V}}{t_{\rm bdy}} = \frac{P_\infty}{G_N L} - \frac{4 r_0^2}{G_NL^3} \left(3-2 \sqrt{2} \right)^{\sqrt{2}} B(y_{\rm brane},\delta \alpha) \, D(y_{\rm brane},\delta \alpha)\, e^{-\frac{\sqrt{2} r_0t_{\rm bdy}}{L^2}}\,.
\end{equation}
Importantly, close to the boundary in parameter space, when $\delta \alpha \to 0$, the late time value of $r_{\rm brane}$ approaches $r_\infty$ and the coefficient in front of the exponential in eq.~\eqref{eq:qqqq} vanishes since $B(y_{\rm brane},\delta \alpha)  \sim \delta \alpha$. When this happens, the subleading corrections would become important, which is in agreement with the results of the boundary in parameter space in eq.~\eqref{eq:Delta t marginal2}. This is consistent with our findings for the simple example of ($\alpha=0,y_{\rm brane} \neq 0$) discussed above. There the asymptotic value of $r_{\rm brane}$ approached $r_\infty$ for $y_{\rm brane}\to 0$ which led to a vanishing leading exponential correction to $\delta P$ and to ${\rm d}c_{{}_V}/{\rm d} t_{\rm bdy}$.

\subsubsection{Lloyd bound violating region}
\label{sec:no Lloyd}
When the JT coupling is bigger than the critical value~\eqref{eq:alpha condition}, the volume complexity violates the Lloyd bound~\eqref{Lloyd_Bound_2}, and ${\rm d} c_{{}_V}/{\rm d} {t_{\rm bdy}}$ reaches its asymptotic value at late times from above.\footnote{For $\alpha<-\alpha_c$, the asymptotically early time $(t_{\rm bdy} \to -\infty)$ value is reached from below, also in violation of the Lloyd bound. An identical analysis applies to the $\alpha<0$ setting at early times.} The simplest example of this class is when there is a nonzero JT coupling and the brane is positioned at $y_{\rm brane}=0$. This occurs when the cosmological constant $\Lambda^{\rm brane}$ on the brane matches the cosmological constant $\Lambda^{\rm bulk}$ of the bulk AdS$_3$. 

\paragraph{Nonzero JT coupling, equal cosmological constants.} We begin with the case $\Lambda^{\rm brane}=\Lambda^{\rm bulk}$, so the location of the brane is set to $y_{\rm brane}=0$. In this limit, for any $\alpha\neq 0$, the intersection of the brane and the extremal surface computing the volume complexity is given by
\begin{equation}\label{eq:simple JT}
\begin{aligned}
    \frac{r_{\rm brane}^2}{r_\infty^2} = 1-\frac{L|\alpha|}{\sqrt{r_0^2+L^2\alpha^2}} + \frac{4r_0^2 }{L|\alpha| \sqrt{r_0^2+L^2\alpha^2}} \left(3-2\sqrt{2}\right)^{\sqrt{2}}B(\alpha)e^{-\frac{\sqrt{2}r_0}{L^2}t_{\rm brane}}\,,
\end{aligned}
\end{equation}
where
\begin{equation}
    B(\alpha) = \left(2\beta-2\sqrt{\beta^2-\beta}-1\right)\left(\frac{3\beta+2 \sqrt{2}\sqrt{\beta^2-\beta}-1}{1+\beta}\right)^{\frac{1}{\sqrt{2}}}\,\,
\end{equation}
with $\beta^2=1+\frac{r_0^2}{L^2\alpha^2}$.
Contrary to the Lloyd bound respecting cases, the point of intersection reaches deeper towards the singularity, $r_{\rm brane}< r_\infty$, due to the contact term associated with JT gravity pulling it inwards.

Putting these expressions together, we find that
\begin{equation}\label{eq: dcdt JT}
    \dv{c_{_V}}{t_{\rm bdy}} =  \frac{P_\infty}{G_N L} + \frac{4 r_0^2}{G_NL^3} \left(3-2 \sqrt{2} \right)^{\sqrt{2}} B(\alpha) \, e^{-\frac{\sqrt{2} r_0t_{\rm bdy}}{ L^2}}\,.
\end{equation}
As mentioned, this violates the Lloyd bound~\eqref{Lloyd_Bound_2} as the asymptotic value is reached from above. Notice that for $\alpha \to 0$, the coefficient $B(\alpha) \sim \alpha$ vanishes as expected because the asymptotic value of $r_{\rm brane}$ approaches $r_\infty$ for small $\alpha$.  This is in agreement with the results of the boundary in parameter space in eq.~\eqref{eq:dCVdt_no_dil-app}.

\paragraph{Near the boundary in parameter space.} We now generalize our findings and perform a similar analysis to the one presented in section~\ref{sec:Lloyd}, using a perturbative expansion for the JT coupling,
\begin{equation}\label{eq:alpha epsilon2}
    \alpha = \alpha_c + \delta \alpha \,,
\end{equation}
to locate the boundary in parameter space where there occurs a violation of the Lloyd bound via eq.~\eqref{eq:quartic}, assuming a small $\delta \alpha>0$ but at late enough times such that $e^{-\frac{\sqrt{2}r_0}{L^2} t_{\rm bdy} } \ll \delta\alpha$. The location of the intersection of the brane and the extremal surface computing the volume complexity is given by
\begin{align}\label{eq: intersec rbrane 4}
    \frac{r_{\rm brane}^2}{r_\infty^2} & = 1 - \frac{L}{r_0} \cos^3(y_{\rm brane}) \left(\sqrt{1+\sin^2(y_{\rm brane})}-\sin(y_{\rm brane})\right) \delta \alpha\\
    &\qquad + \frac{8 r_0}{L \delta \alpha} \frac{\sqrt{1+\sin^2(y_{\rm brane})}}{\cos^3(y_{\rm brane})} 
    \left(3-2\sqrt{2}\right)^{\sqrt{2}} B(y_{\rm brane},\delta \alpha) D(y_{\rm brane},\delta \alpha) e^{-\frac{\sqrt{2}r_0}{L^2} t_{\rm bdy}}\,,\nonumber
\end{align}
where
\begin{align}
    &\log B(y_{\rm brane},\delta \alpha) = \nonumber\\
    &\quad \sqrt{2} \tanh^{-1}\left(\frac{1}{\sqrt{2}} - \frac{1}{\sqrt{2}} \frac{L}{r_0} \cos^3(y_{\rm brane})\left(\sqrt{1+\sin^2(y_{\rm brane})}-\sin(y_{\rm brane})\right)\delta\alpha\right)\nonumber\\
    &\quad  -2 \tanh^{-1}\left(1- \frac{L}{r_0} \cos^3(y_{\rm brane})\left(\sqrt{1+\sin^2(y_{\rm brane})}-\sin(y_{\rm brane})\right)\delta\alpha\right),\\
    &\log D(y_{\rm brane},\delta\alpha) = \nonumber\\
    &\quad -\sqrt{2} \, {\rm coth}^{-1}\sqrt{\frac{2+ \cos^2(y_{\rm brane})}{2\sin^2(y_{\rm brane})} + \frac{L\cos^5(y_{\rm brane})\left(\sqrt{1+\sin^2(y_{\rm brane})} -\sin(y_{\rm brane}) \right)}{2r_0\sin^2(y_{\rm brane})}\,\delta \alpha}\nonumber.
\end{align}
This is consistent with eq.~\eqref{eq:simple JT} for $y_{\rm brane} = 0$ and small $\alpha \sim \delta \alpha$. The asymptotic value of $r_{\rm brane}$ is smaller than $r_\infty$ for nonzero $\delta \alpha$ and it is reached from above.

Integrating eq.~\eqref{eq:main integral} and utilizing the above results gives
\begin{equation}\label{eq:dcdt no Lloyd}
    \dv{c_{_V}}{t_{\rm bdy}} =  \frac{P_\infty}{G_N L} + \frac{4 r_0^2}{G_NL^3} \left(3-2 \sqrt{2} \right)^{\sqrt{2}} B(y_{\rm brane},\delta \alpha) \, D(y_{\rm brane},\delta\alpha) \, e^{-\frac{\sqrt{2} r_0t_{\rm bdy}}{ L^2}}\,,
\end{equation}
violating the Lloyd bound. 
As before, note that in the limit $\delta \alpha \to 0$, the coefficient $B(y_{\rm brane},\delta \alpha)\sim \delta \alpha$ vanishes because the value of $r_{\rm brane}$ asymptotes to $r_\infty$, and the subleading corrections would become important for the late time dependence of ${\rm d}c_{{}_V}/{\rm d}t_{\rm bdy}$.  This is in agreement with the results of the boundary in parameter space in eq.~\eqref{eq:Delta t marginal2}.

\subsection{Summary of the analytic results}
To summarize the analysis of the previous subsections, whether the Lloyd bound is satisfied or not depends ultimately on whether the Neumann boundary condition~\eqref{eq:Neumann boundary EF0} sets $\eval{\dot{r}}_{{\rm brane}}$ to a positive or negative value at very late times. This condition can be translated to a bound on the magnitude of the JT coupling $\alpha$ in terms of the position of the brane $y_{\rm brane}$ (or equivalently, the brane cosmological constant $\Lambda^{\rm brane}$), and the size of the black hole in units of AdS$_3$ length $r_0/L$, see eq.~\eqref{eq:alpha condition}. The bounds for positive (negative) $\alpha$ are found from the late time (early time) behaviour of the rate of change of complexity. The features for $\alpha < 0$ at $t_{\rm bdy}<0$ are symmetric to the ones for $\alpha>0$ at $t_{\rm bdy}>0$, so for simplicity we focus on the $\alpha>0$ starting from $t_{\rm bdy} \geq 0$.

Consider $\alpha<\alpha_c$ and the extremal surface anchored at some fixed $t_{\rm bdy}$ at the right asymptotic boundary at $r_{\rm bdy}$. As a function of $\sigma$, it initially falls to a smaller radius and crosses the event horizon. Inside the black hole, it reaches a turning point $r_{\rm min}> r_\infty$ and turns to a larger radius towards the excised left asymptotic boundary. Depending on the parameters of the model, and the boundary time $t_{\rm bdy}$, it may or may not leave the left horizon. In either case, it reaches the ETW brane at $r_{\rm brane} > r_\infty$ as in eq.~\eqref{eq: rbrane Lloyd}, at which point its trajectory stops. If we now change the anchoring point by increasing $t_{\rm bdy}$, the turning point $r_{\rm min}$ approaches $r_\infty$ from above. In these cases, the extremal surface never probes the region of the black hole closer to the singularity than $r_\infty$, and the Lloyd bound is respected at all times.

For $\alpha>\alpha_c$, the extremal surfaces anchored at small fixed $t_{\rm bdy}$ follow qualitatively similar trajectories to the ones for $\alpha<\alpha_c$. However, for late times (large $t_{\rm bdy}$), this behaviour changes. In particular, there is no turning point $r_{\rm min}$ and they simply fall from $r_{\rm bdy}$ to $r_{\rm brane}<r_\infty$, see~\eqref{eq: intersec rbrane 4}. This qualitative change in the trajectory of the extremal surfaces occurs precisely when the Lloyd bound is violated. Moreover, the trajectory of the maximal volume slice tends towards the singularity at $r=0$ but is stopped at the location of the brane. This follows from the analogy with the mechanical picture in section~\ref{sec:mech}, as shown in fig.~\ref{fig:Ueff}. In this setting, the maximal volume slices probe the geometry deeper than $r_\infty$, a region that is not accessible in the Lloyd bound respecting cases.

For $\alpha=\alpha_c$, we are at the boundary between the two cases. The features of the extremal volume surfaces are similar to the $\alpha<\alpha_c$ case for early times. The features of the maximal volume surface at late times is a limit between the two cases described above. For a fixed, large $t_{\rm bdy}$ the maximal volume surface falls from $r_{\rm bdy}$ to a turning point $r_{\rm min}$ and barely turns around before intersecting the brane at $r_{\rm brane}$ and ending their trajectory there. The location of $r_{\rm brane}$ is asymptotically close to its turning point, and both approach $r_\infty$ as the boundary time $t_{\rm bdy}$ goes to infinity. In this case, we find that the order of the correction to the rate of change of complexity is subleading compared to the cases in each region. Specifically, the coefficient in the exponential time falloff in eq.~\eqref{eq:Delta t marginal2} is twice the coefficient found in the other cases~\eqref{eq:qqqq} and \eqref{eq:dcdt no Lloyd}. In this sense, the Lloyd bound is marginally satisfied when $\alpha=\alpha_c$.


\section{Numerical results for full time dependence of volume complexity}
\label{sec:numerical_approach}
After analytically examining the late time behaviour of complexity growth in the previous section, we now proceed to numerically compute the full time dependence of the volume complexity for the planar BTZ geometry with a subcritical ETW brane. Considering first the simpler case of no intrinsic gravitational dynamics on the brane, we numerically extract the time dependence of the volume complexity. Subsequently, we endow the brane with JT gravity and again compute the full time dependence of complexity. When the magnitude of the JT coupling is larger than $\alpha_c$ in  eq.~\eqref{eq:alpha condition}, we observe a violation of the Lloyd bound~\eqref{Lloyd_Bound_2}, confirming the findings of section~\ref{sec:asymptotics}. The global coordinate system introduced in section \ref{sec:basic_setup}, which covers the entire maximally extended planar BTZ geometry, is well suited for the numerical approach we take in the present section.

\subsection{No intrinsic dynamics on the brane}
\label{sec:no_dyn}
Consider the CFT state defined on a time slice $\tau = \tau_{\rm bdy}$. To compute the associated volume complexity, we need to look for bulk codimension-one surfaces anchored at $\tau_{\rm bdy}$ which extremize eq.~\eqref{def:CV1}. We will parametrize this bulk surface by $\tau = \tau(y)$. We also introduce the cutoff near AdS boundary $y_\epsilon$ which in global coordinates can be related to the radial cutoff in  Schwarzschild coordinates imposed in (\ref{eq: Dirichlet-r}) at $r_{\rm max} = L/\epsilon$:
\be
\label{eq:ycutoff1}
y_\epsilon = \frac{\pi}{2} - \frac{\epsilon \, r_0}{L} \cos \tau_{\rm bdy}.
\ee
We gauge fix the extremal surface parameter introduced in section~\ref{sec:complexity_growth} as $\sigma = y_\epsilon - y$. From eq.\ \eqref{MetricGlobal}, the induced metric on this surface is \
\be\label{eq:induced metric}
d s^2_{\rm ind} = \frac{1}{\cos^2 y} \left(L^2\left[1-\dot{\tau}(y)^2\right]d y^2+\frac{r_0^2}{L^2} \cos^2 \tau(y) \,d x^2\right),
\ee
where an overhead dot denotes a derivative with respect to $y$. Thus, the volume of the surface is 
\be\label{eq:Volume no JT}
{\rm Vol} = \ell r_0 \int_{y_{\rm brane}}^{y_\epsilon} d y \, \frac{\cos\tau(y)}{\cos^2 y} \sqrt{1-\dot{\tau}(y)^2}~.
\ee
Here $\ell$ denotes the length of the interval on the boundary along the $x$ direction, which factors out due to translation invariance. The location of the brane is given by $y_{\rm brane}$ and $y_\epsilon$ will regulate the otherwise divergent volume.

\begin{figure}
    \centering
    \subcaptionbox{$y_{\rm brane} = 0$}{\includegraphics[height=0.3\textwidth]{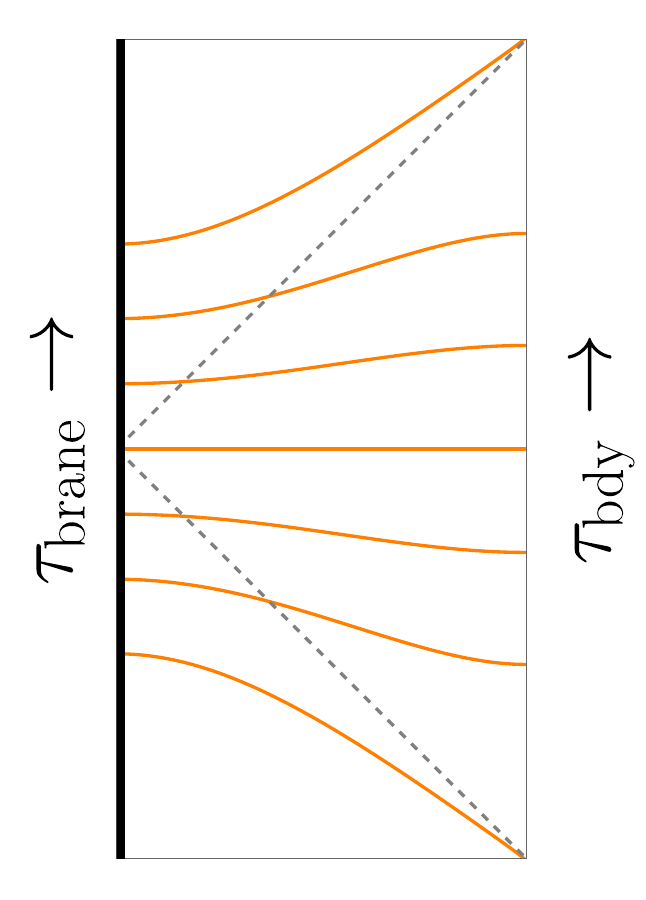}}\qquad
    \subcaptionbox{$y_{\rm brane} = -\frac{\pi}{4}$}{\includegraphics[height=0.3\textwidth]{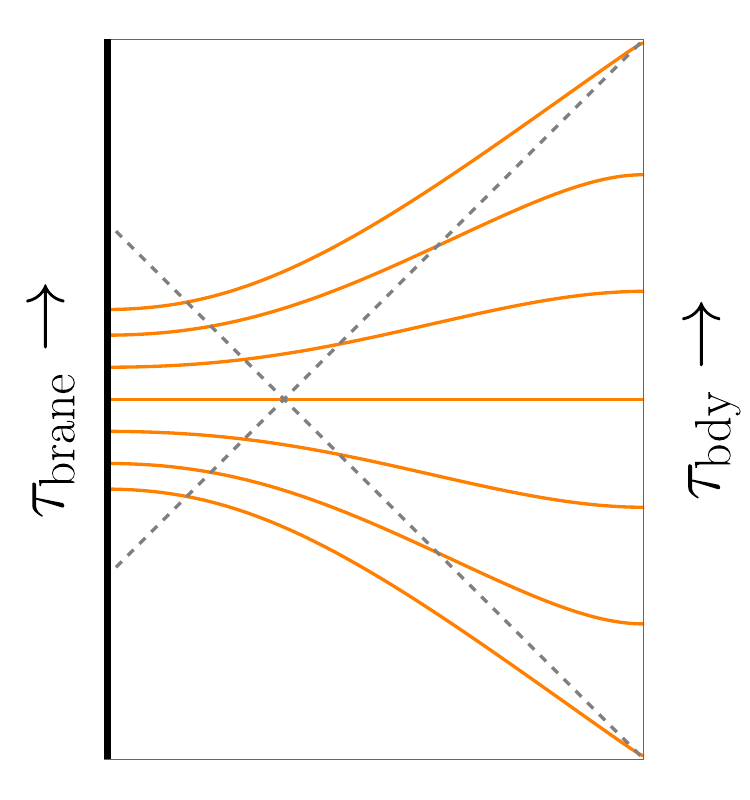}}\qquad
    \subcaptionbox{$y_{\rm brane} = - 1.5$}{\includegraphics[height=0.3\textwidth]{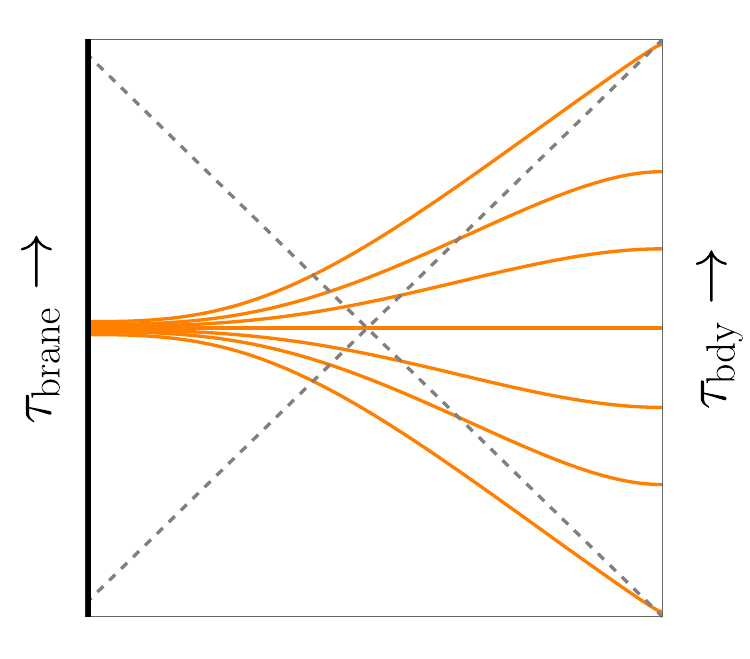}}
    \caption{Plots representing the maximal volume surfaces for different values of the brane location. We have suppressed the transverse direction, which respects translation invariance. The larger the brane tension (\ie more negative $y_{\rm brane}$), the smaller the region spanned by the extremal surfaces on the brane.}
    \label{fig:No_JT_trajectories}
\end{figure}

The dependence on the Schwarzschild time $t_{\text{bdy}}$ can be reintroduced by using the relation with $\tau_{\rm bdy}$ on the boundary,
\be
\label{eq:ttotau}
t_{\rm bdy} = \frac{L^2}{2r_0} \log \left(\frac{1+\sin \tau_{\rm bdy}}{1-\sin \tau_{\rm bdy}}\right).
\ee
The cutoff $y_\epsilon$ can then be expressed as
\be
\label{eq:ycutoff2}
y_\epsilon = \frac{\pi}{2} - \frac{\epsilon \, r_0}{L}\,\, {\rm sech} \left(\frac{r_0 t_{\rm bdy}}{L^2}\right).
\ee
Now, for $\tau(y)$ to be an extremal surface, the variation of the volume~\eqref{eq:Volume no JT} should vanish under small variations of the surface. Replacing $\tau(y) \rightarrow \tau(y)+\delta\tau(y)$ in eq.~\eqref{eq:Volume no JT} gives 
\be
\begin{split}
\delta {\rm Vol} = \,\, &\ell r_0 \int_{y_{\rm brane}}^{y_\epsilon} d y \left[\dv{y}\left(\frac{\cos\tau(y)}{\cos^2 y} \frac{\dot{\tau}(y)}{\sqrt{1-\dot{\tau}(y)^2}}\right) - \frac{\sin\tau(y)}{\cos^2 y} \sqrt{1-\dot{\tau}(y)^2}\right] \delta\tau(y) \\
&-\ell r_0 \,\frac{\cos\tau(y)}{\cos^2 y} \frac{\dot{\tau}(y)}{\sqrt{1-\dot{\tau}(y)^2}} \,\delta\tau(y) \bigg|^{y_{\epsilon}}_{y_{\rm brane}}.
\end{split}
\label{eq:varvol}
\ee
The second line above denotes the two boundary terms that are generated in computing the variation $\delta{\rm Vol}$. For the surface to be extremal, we must have the variation $\delta{\rm Vol} = 0$. This is achieved by setting the integrand in the first line of eq.\ \eqref{eq:varvol} to vanish, which can equivalently be expressed as
\be
\ddot{\tau}(y) = \left(1-\dot{\tau}(y)^2\right) \left[\tan\tau(y) - 2\tan y \, \dot{\tau}(y)\right],
\label{eq:bulk_traj}
\ee 
along with a Dirichlet boundary condition at the AdS boundary, as well as a Neumann boundary condition at the location where the surface intersects the ETW brane. These are given by
\begin{equation}
\tau(y_\epsilon) = \tau_{\rm bdy}~,\qquad\dot{\tau}(y)|_{y_{\rm brane}} = 0~,  
\label{eq:bdy_cond}
\end{equation}
respectively.
The Dirichlet boundary condition at the AdS boundary ensures that the extremal surface is anchored on the boundary at $\tau_{\rm bdy}$, while the Neumann boundary condition at the location of the brane implies that the extremal surface and the brane intersect at a right angle. Eq.\ \eqref{eq:bulk_traj} determines the trajectory of the extremal surface in the bulk. Though a full analytic solution seems difficult to achieve, the equations of motion can be solved numerically in a straightforward manner. See fig.\ \ref{fig:No_JT_trajectories} for trajectories of the extremal surfaces as they appear in the bulk, corresponding to different values of the brane location. As it turns out, the smaller the region of spacetime beyond the horizon cut off by the brane, or equivalently the larger the brane tension, the smaller the region spanned by the extremal surfaces on the brane. Furthermore, the intersection of the maximal volume surface and the brane occurs at a larger distance from the singularity for larger tension, in agreement with eq.~\eqref{eq:simple tension}. 

\begin{figure}
    \centering
    \subcaptionbox{\label{fig:No_JT_plots_1}}{\includegraphics[height=0.31\textwidth]{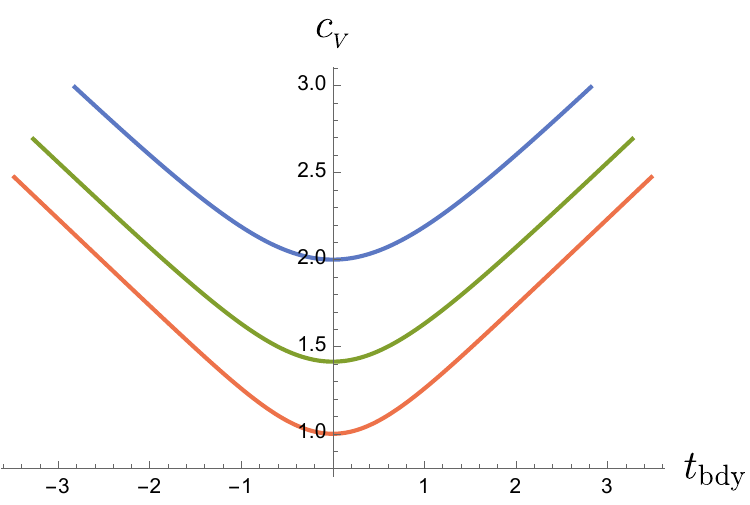}}\quad
    \subcaptionbox{\label{fig:No_JT_plots_2}}{\includegraphics[height=0.31\textwidth]{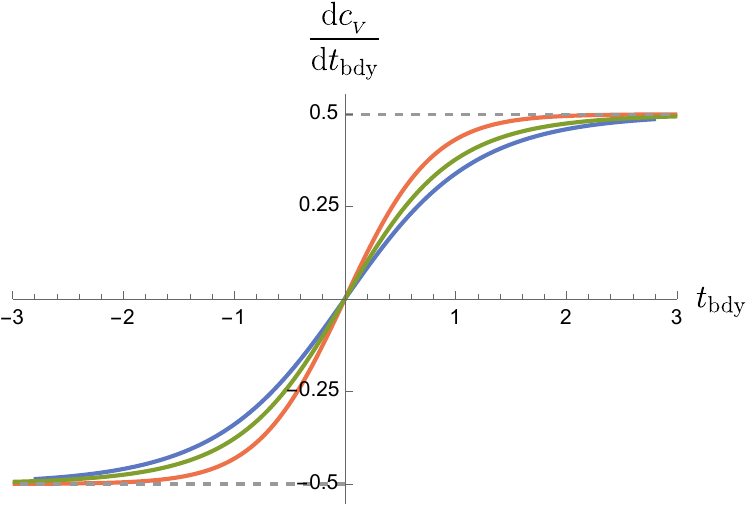}}
    \caption{The plots above depict the (renormalized) volume complexity and its rate of change for the translationally invariant boundary CFT state as a function of the boundary time. The red, green, and blue curves correspond to the brane locations $y_{\rm brane} = 0, - \frac{\pi}{8}, - \frac{\pi}{4}$ respectively. In (b), the asymptotically early/late time value of ${\rm d} c_{{}_V}/{\rm d} t_{\rm bdy}$ corresponds to the Lloyd bound. We have chosen to work with $L = r_0 = G_N = 1$.}
    \label{fig:No_JT_plots}
\end{figure}

We can also compute the volumes of the extremal surfaces anchored at different instants of time on the boundary, eq.~\eqref{eq:Volume no JT}, by first numerically determining the trajectory of the surface using eq.~\eqref{eq:bulk_traj}, with the boundary conditions~\eqref{eq:bdy_cond}, and then evaluating the volume~\eqref{eq:Volume no JT}. 
This can, in turn, be used to extract the time dependent behaviour of volume complexity~\eqref{def:CV1}, which is plotted in fig.\ \ref{fig:No_JT_plots_1} as a function of the boundary time.\footnote{To obtain an $\epsilon$-independent result, we first define a ``renormalized" volume, which is obtained by subtracting off the volume contribution of an extremal surface in the global AdS$_3$ geometry,
\begin{equation*} 
{\rm Vol}_{\rm renorm.} = {\rm Vol} - {\rm Vol}_{{\rm AdS}_3},\label{eq:RenVol}
\end{equation*} 
where ${\rm Vol}_{{\rm AdS}_3}$ is given by
\begin{equation*} 
{\rm Vol}_{{\rm AdS}_3} = \ell L \left(\frac{1}{\epsilon} - 1\right).
\end{equation*}
With this prescription, the renormalized volume complexity, also known as the \emph{complexity of formation}, per unit length is then simply given by 
${{\rm Vol}_{\rm renorm.}}/{G_N L \, \ell}$.} 
Also, in fig. \ref{fig:No_JT_plots_2}, we plot the rate of change of the volume complexity with respect to the boundary time for different choices of the brane location. As is evident from the plots, the magnitude of ${\rm d} c_{{}V}/{\rm d} t_{\rm bdy}$ asymptotes to the same value for both early and late times ($t_{\rm bdy}\to \pm \infty$), which is independent of the brane's location. This asymptotic value corresponds to the Lloyd bound on complexity growth~\eqref{Lloyd_Bound_2}.  Furthermore, the magnitude of ${\rm d} c_{{}_V}/{\rm d} t_{\rm bdy}$ reaches its asymptotic values from below. Thus, we see that in the absence of any intrinsic dynamics on the brane, the Lloyd bound is always respected by the volume complexity, as expected from the analysis of section~\ref{sec:Lloyd}. In the tensionless case, this result agrees with the asymptotic analysis performed in appendix \ref{sec:simple}, see eq.~\ref{eq:cv-simple}. 

\subsection{JT gravity on the brane}
\label{sec:with_JT}
Let us now explore in detail the consequences of including intrinsic gravitational dynamics localized on the brane, in particular, the Jackiw-Teitelboim (JT) model of two-dimensional gravity, introduced in section \ref{sec:JTbrane}. As discussed in section \ref{sec:review_CV}, the presence of intrinsic dynamics on the brane modifies the volume complexity proposal with an additional contact term at the location of the brane~\eqref{def:CV2}. We write the contact term in eq.~\eqref{def:CV2} in the form
\be
\frac{{\rm Vol}(\mathcal{B} \cap {\rm brane})}{G_N^{\rm brane} L^{\rm brane}} = \frac{\ell \, \varphi(\tau_{\rm brane})}{G_N^{\rm brane} L_2} \,, \label{eq:contact2}
\ee
where $\varphi(\tau_{\rm brane})$ is given in (\ref{eq:dilaton_sol}). If $\tau = \tau(y)$ is the surface that extremizes eq.~\eqref{def:CV2}, then the (renormalized) volume complexity per unit length for the CFT state on the boundary is now given by
\be
c_{{}_V} = \frac{\varphi(\tau_{\rm brane})}{G_N^{\rm brane} L_2} + \frac{r_0}{G_N L} \int_{y_{\rm brane}}^{y_\epsilon} d y \, \frac{\cos\tau(y)}{\cos^2 y} \sqrt{1-\dot{\tau}(y)^2} - \frac{1}{G_N} \left(\frac{1}{\epsilon} - 1\right).
\label{eq:CV_JT}
\ee
The extremal slice $\tau(y)$ still satisfies eq.\ \eqref{eq:bulk_traj}, with the Dirichlet boundary condition $\tau(y_\epsilon) = \tau_{\rm bdy}$ at the AdS boundary. However, the Neumann boundary condition at the intersection of the extremal slice with the brane now becomes
\be
\frac{\varphi'(\tau_{\rm brane})}{G_N^{\rm brane} L_2} + \frac{r_0}{G_N L} \frac{\cos\tau_{\rm brane}}{\cos^2 y_{\rm brane}} \frac{\dot{\tau}(y_{\rm brane})}{\sqrt{1-\dot{\tau}(y_{\rm brane})^2}} = 0\, ,
\label{eq:new_bc_1}
\ee
where the prime denotes a derivative with respect to $\tau$. Putting in the dilaton profile~\eqref{eq:dilaton_sol}, this leads to the boundary condition
\be
\dot{\tau}(y_{\rm brane}) = - \frac{\alpha L^2 \cos^2 y_{\rm brane}}{\sqrt{\alpha^2 L^4 \cos^4 y_{\rm brane} + r_0^2 L_2^2}}\, , 
\label{new_Neumann}
\ee
where the JT coupling $\alpha$ is defined in eq.~\eqref{def_alpha}. The presence of JT gravity thus leads to a nontrivial change in the boundary condition that the extremal surface is required to meet, apart from shifting the value of the complexity itself via the contact term contribution. 

\begin{figure}[t]
    \centering
    \subcaptionbox{$\alpha = 0$}{\includegraphics[height=0.3\textwidth]{Plots_for_paper/No_JT_Traj_1.pdf}}\quad
    \subcaptionbox{$\alpha = 1$}{\includegraphics[height=0.3\textwidth]{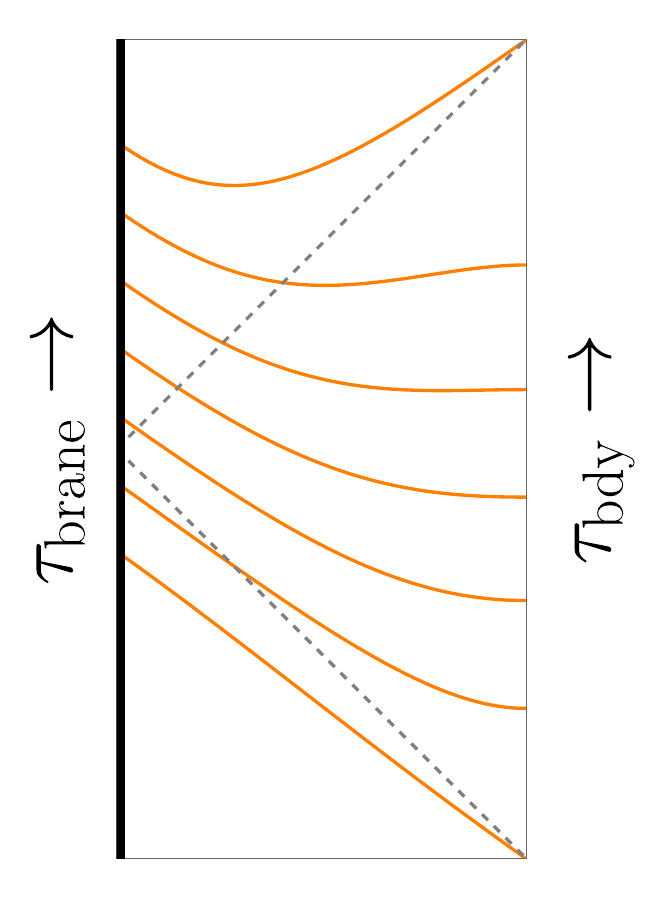}}\quad
    \subcaptionbox{$\alpha = 5$}{\includegraphics[height=0.3\textwidth]{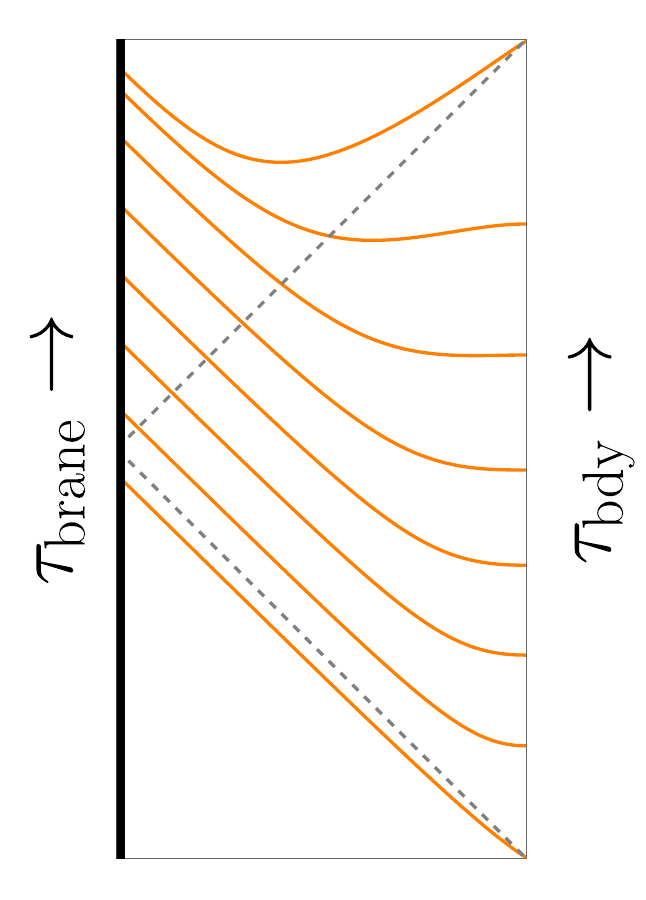}}\quad
    \subcaptionbox{$\alpha = 10$}{\includegraphics[height=0.3\textwidth]{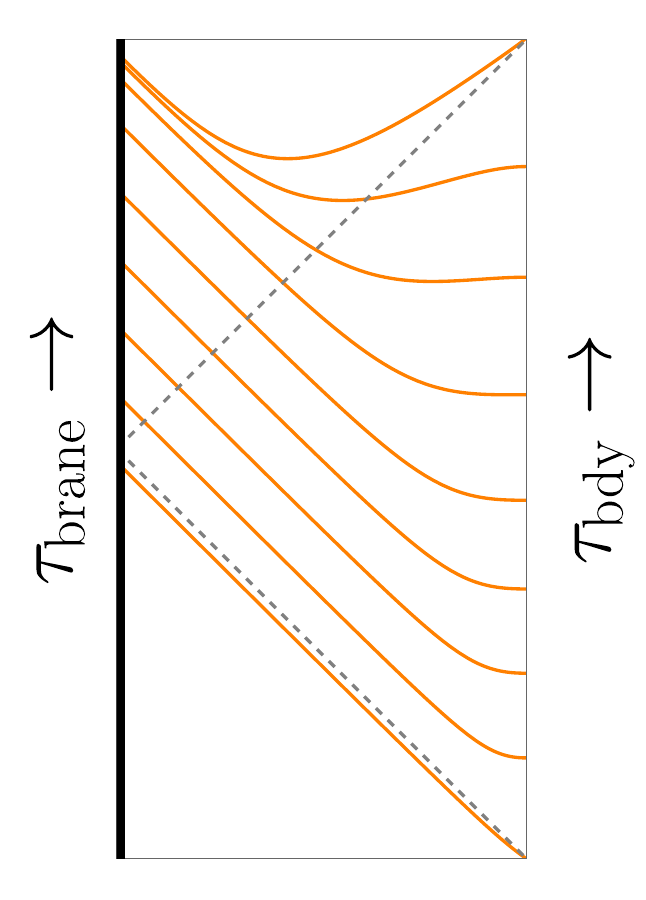}}
    \caption{Plots representing the maximal volume slices for different values of the JT coupling $\alpha$, with the brane located at $y_{\rm brane} = 0$. The plots for the corresponding negative values of $\alpha$ can be obtained by reflecting the ones above about the $\tau = 0$ horizontal axis. We have set $L = L_2 = r_0 = G_N = 1$.}
    \label{fig:With_JT_trajectories}
\end{figure}

\begin{figure}[t]
\centering
\includegraphics[height=0.5\textwidth]{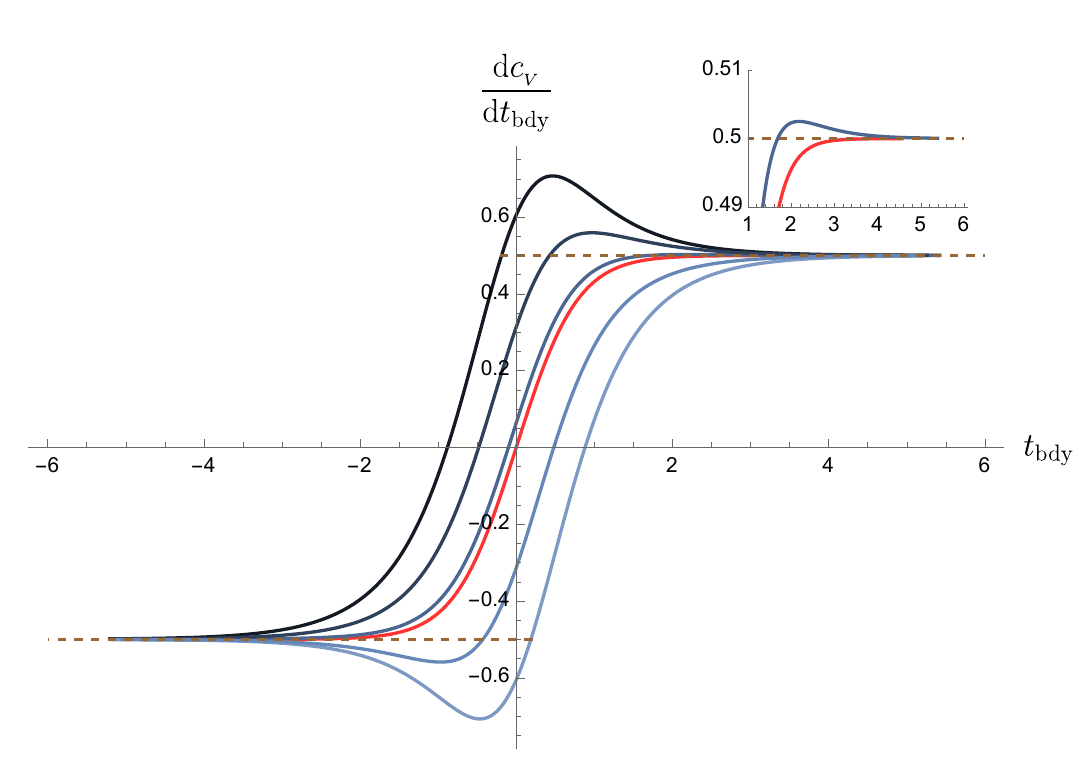}
\caption{Rate of change of volume complexity (per unit boundary interval length) as a function of the boundary time for different values of the JT coupling $\alpha$, with $y_{\rm brane} = 0$. From top to bottom, we have $\alpha = 1, 0.5, 0.1, 0 \, (\text{\emph{i.e.}\ no JT, shown in red}), - 0.5, - 1$ respectively. The inset shows the zoomed-in asymptotic behaviour for $\alpha = 0.1$, indicating that it too approaches the Lloyd bound from above, similar to the cases $\alpha = 0.5$ and $1$. For negative $\alpha$, the Lloyd bound is violated in the past. We have chosen $L = L_2 = r_0 = G_N = 1$.}
\label{fig:dCVdt_With_JT_1}
\end{figure}

We can now numerically explore the effects of the presence of JT gravity on the trajectories of the extremal surfaces as well as for the rate of change of the volume complexity itself. For instance, for the brane located at $y_{\rm brane} = 0$, fig.~\ref{fig:With_JT_trajectories} shows the plots of the extremal surfaces for different values of the JT coupling $\alpha$. The behaviour of the extremal surface trajectories is markedly different from the scenario with no JT gravity on the brane, see fig.~\ref{fig:No_JT_trajectories}. In particular, for $\alpha>0$ the intersection between the maximal volume surface and the brane goes closer to the future singularity for late time surfaces, and farther from the past singularity for early time surfaces, as was explained in section~\ref{sec:asymptotics}. fig.~\ref{fig:dCVdt_With_JT_1} depicts the behaviour of the rate of change of volume complexity per unit interval length as a function of the boundary time for different choices of $\alpha$, with the brane located at $y_{\rm brane} = 0$. Note that for the illustrative choice of the bulk parameters we make, $L = L_2 = r_0 = G_N = 1$, the Lloyd bound~\eqref{Lloyd_Bound_2} demands that $|{\rm d} c_{{}_V}/{\rm d} {t_{\rm bdy}}| \le 0.5$. The bound is not respected by the growth of complexity for any nonzero choice of the JT coupling $\alpha$, as expected from the analysis in section~\ref{sec:no Lloyd}, see eq.~\eqref{eq: dcdt JT}. 

\begin{figure}
\begin{minipage}{6in}
  \centering
$\vcenter{\hbox{\includegraphics[height=2.68in]{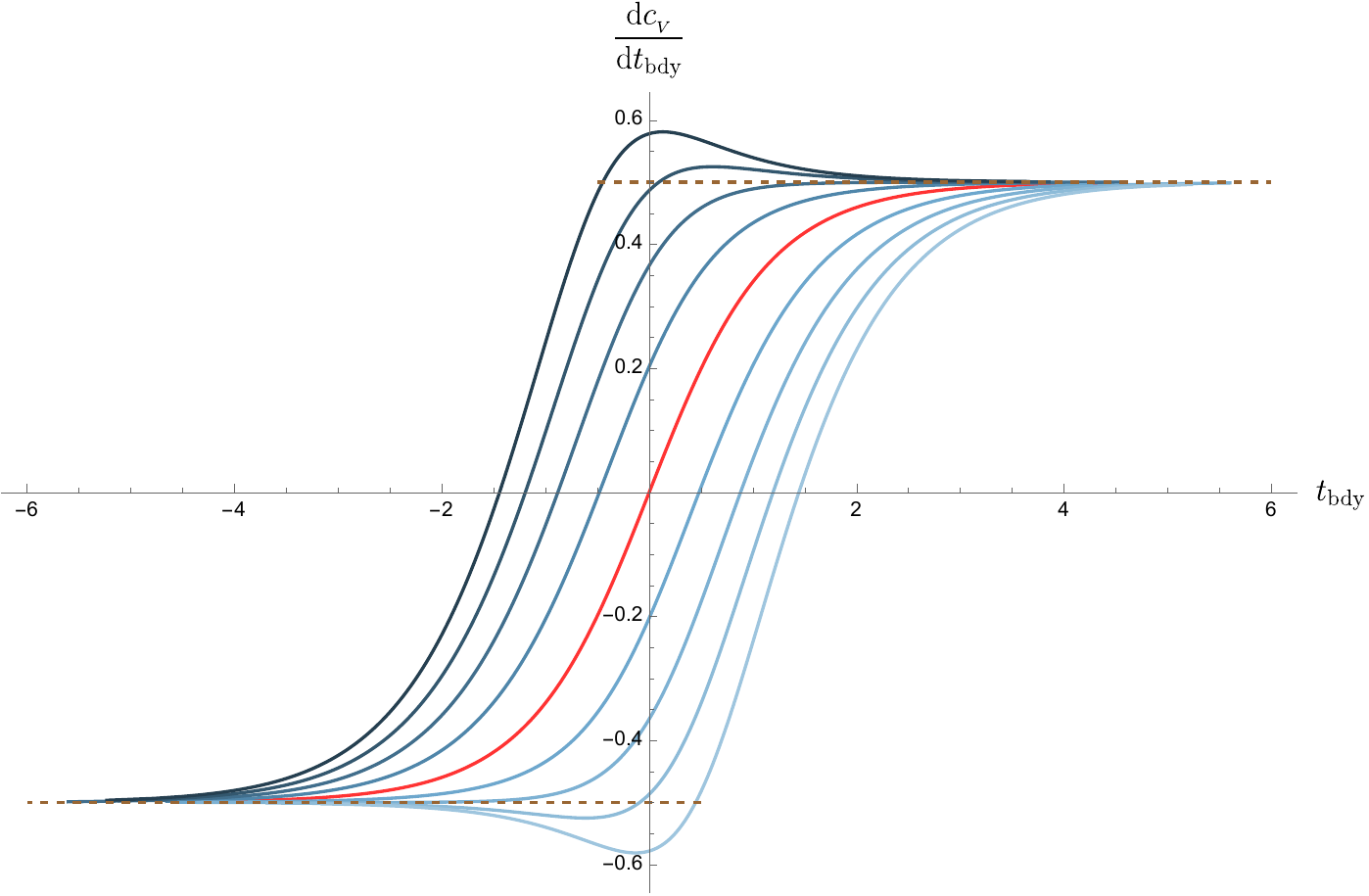}}}$
\,\,
$\vcenter{\hbox{\includegraphics[height=1.23in]{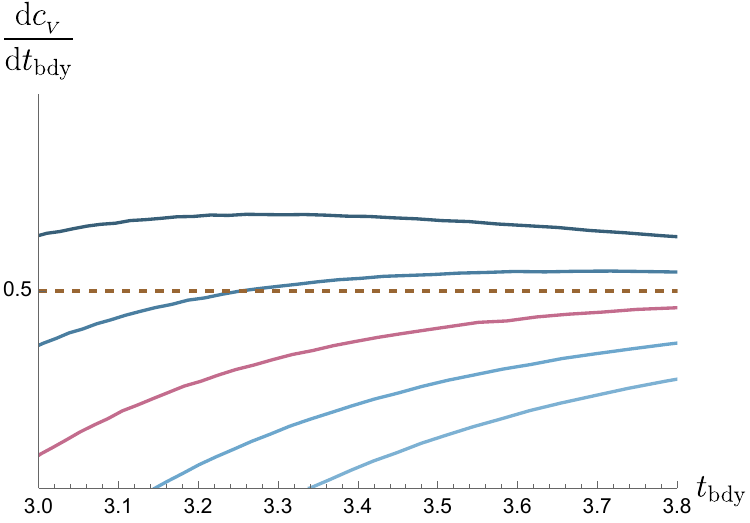}}}$
\end{minipage}
\caption{The left panel shows the rate of change of volume complexity as a function of the boundary time for different values of the JT coupling $\alpha$, with $y_{\rm brane} = - \pi/4$. From top to bottom, we have $\alpha = 4, 3, 2, 1, 0 \, (\text{\emph{i.e.}\ no JT, shown in red}), -1, -2, -3, -4$ respectively. In the right panel, from top to bottom, we have $\alpha = 2.02, 2.01, 2, 1.99, 1.98$ respectively. Clearly, for $|\alpha| \gtrsim 2$, the Lloyd bound gets violated. This is in agreement with the analytic results presented in section \ref{sec:complexity_growth}. We have chosen $L = L_2/\sqrt{2} = r_0 = G_N = 1$.}
\label{fig:With_JT_plots}
\end{figure}

Interestingly, unlike the case for $y_{\rm brane} = 0$, for $-\pi/2 < y_{\rm brane} < 0$, a finite range of values for the JT coupling $\alpha$ exists which leads to a rate of change of complexity in agreement with the Lloyd bound. For instance, fig.~\ref{fig:With_JT_plots} depicts the rate of change of complexity as a function of boundary time when $y_{\rm brane} = -\pi/4$. We numerically observe that the Lloyd bound is now respected for $|\alpha| \lesssim 2$. For $\alpha \gtrsim 2$ the Lloyd bound is violated in the future, and for $\alpha \lesssim -2$ it is violated in the past. This behaviour can further be checked numerically for other values of the bulk parameters $(y_{\rm brane}, \alpha)$ as well.  Thus, depending upon the location of the brane in the bulk spacetime, one gets a finite range of values for the JT coupling $\alpha$ which leads to complexity growth in agreement with the Lloyd bound. This agrees with the findings of section~\ref{sec:asymptotics}, and provides a numerical confirmation for the analytic results obtained in the previous section, in particular the bound given by eq.~\eqref{eq:alpha condition},
\be
|\alpha| \le \frac{r_0}{L} \, \frac{|\tan y_{\rm brane}|}{\cos^2 y_{\rm brane}}\,,
\label{eq:Lloyd_Bound_Again}
\ee
which defines the region of the parameter space $(y_{\rm brane}, \alpha)$ that respects the Lloyd bound.

\begin{figure}[t]
    \centering
    \subcaptionbox{$\alpha = 0$}{\includegraphics[height=0.24\textwidth]{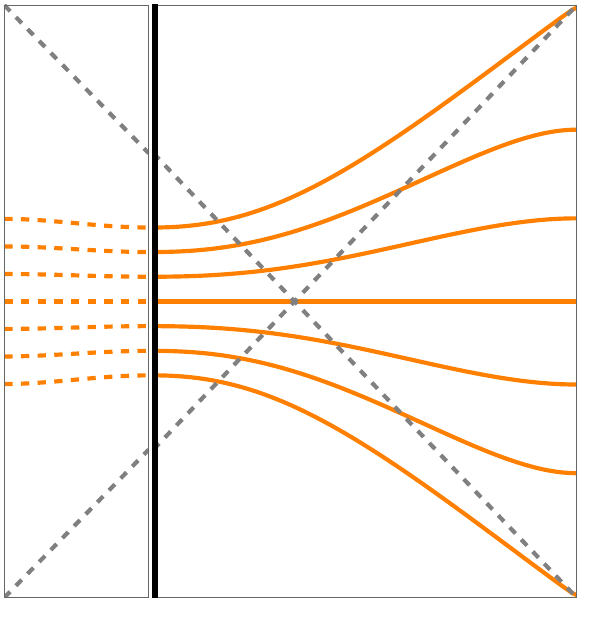}}\,
    \subcaptionbox{$\alpha = 1$}{\includegraphics[height=0.24\textwidth]{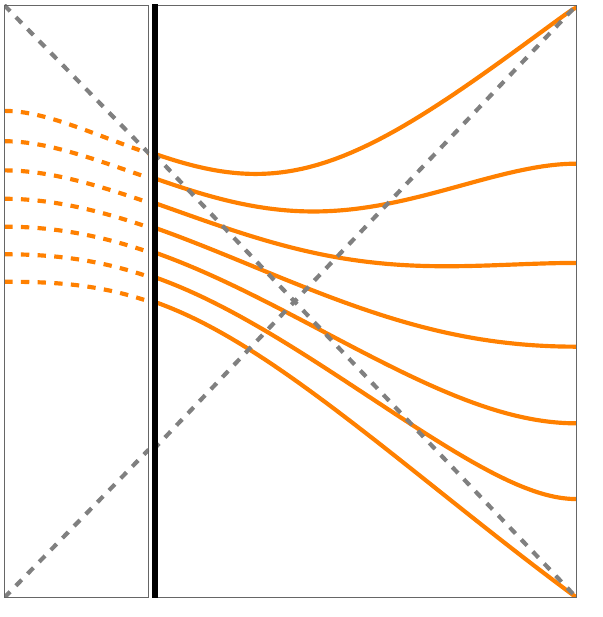}}\,
    \subcaptionbox{$\alpha = 2$}{\includegraphics[height=0.24\textwidth]{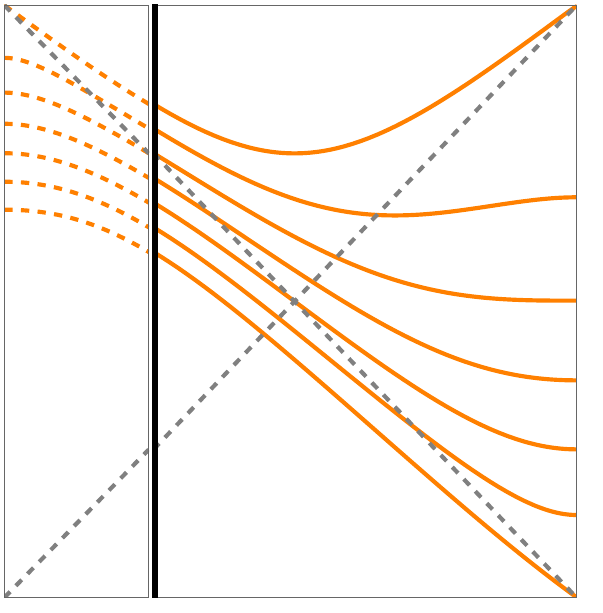}}\,
    \subcaptionbox{$\alpha = 5$}{\includegraphics[height=0.24\textwidth]{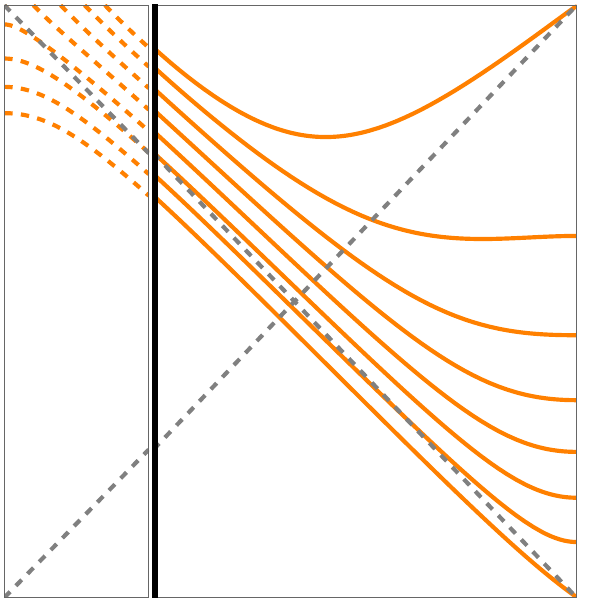}}
    \caption{Extension of the extremal volume surfaces beyond the ETW brane. In the plots above, we have chosen $y_{\rm brane} = - \pi/4$, and the parameters $L = L_2/\sqrt{2} = r_0 = G_N = 1$. For $0\le\alpha<2$, which belongs to the Lloyd bound respecting region of the bulk parameter space for the above parameter choices, the extremal surfaces reach the left asymptotic boundary when extended past the brane. At exactly $\alpha = 2$, which belongs to the boundary between the Lloyd bound respecting and violating regions, the latest extremal surface barely makes it to the left boundary. For $\alpha>2$, belonging to the Lloyd bound violating region, several extremal surfaces cannot make it to the left boundary and start falling into the singularity. The plots for $\alpha < 0$ can be obtained by reflecting the above about the $\tau = 0$ line.}
    \label{fig:surf_extension}
\end{figure}

Another aspect of the analytic understanding gained in the previous section that can also be observed via numerically studying the full time dependence of the volume complexity pertains to the effective mechanical picture presented in section \ref{sec:mech}. We found that depending upon the energy of the particle compared to the height of the potential barrier, see fig.~\ref{fig:Ueff}, the Lloyd bound was either satisfied, marginally satisfied, or violated. For these three cases, respectively, the particle either scattered off from the barrier, could marginally reach the top of the barrier, or could go past the barrier towards the singularity at $r=0$ until it terminates on the brane. The same behaviour can be observed in terms of the actual trajectories of the extremal volume surfaces computed numerically, as depicted in fig.~\ref{fig:surf_extension}. For illustrative purpose we choose the brane location to be $y_{\rm brane} = -\pi/4$, with $L = L_2/\sqrt{2}= r_0 = G_N = 1$. For this choice of parameters and following eq.~\eqref{eq:Lloyd_Bound_Again}, the Lloyd bound is met when $|\alpha| \le 2$. In fig.~\ref{fig:surf_extension}, we observe that if we extend the extremal volume surfaces past the ETW brane, they make it to the left asymptotic boundary when the Lloyd bound is respected, $0<\alpha<2$. When the Lloyd bound is marginally satisfied, $\alpha = 2$, the latest extremal surface barely makes it to the left asymptotic boundary, whilst for the Lloyd bound violating regime $\alpha>2$, the late time surfaces start falling into the singularity without being able to reach the left asymptotic boundary. This behaviour of the extremal surfaces either falling into or escaping the singularity is in exact analogy with the particle in the effective mechanical picture being able to go beyond the potential barrier to reach $r=0$ or being scattered off. 


\section{Wormhole perspective on the Lloyd bound and bulk energy conditions}
\label{sec:Engelhardt-Folkestad}
In the previous sections, we have mapped out the respective regions in the space of parameters $(\alpha, y_{\rm brane})$ which either obey or violate the Lloyd bound on complexity growth. However, so far we have not provided any physical argument for why the Lloyd bound gets violated in the first place. The goal of this section is to discuss the physical underpinnings of the Lloyd bound violation from the point of view of bulk gravitational dynamics. 

\subsection{The wormhole perspective}
\label{sec: wormhole}

Our approach here is motivated by results from the work of Engelhardt and Folkestad \cite{Engelhardt:2021mju}. They argue that the Lloyd bound for holographic volume complexity is related to the weak curvature condition (WCC) for a certain class of asymptotically AdS spacetimes:\footnote{In particular, for the Einstein-Maxwell-scalar theory with spherical symmetry, as well as for any general gravitational theory with matter that has compact support with spherical or planar symmetry in spacetime dimensions $d+1 \geq 4$, ref.~\cite{Engelhardt:2021mju} rigorously proves that the WCC is sufficient for the Lloyd bound to hold in asymptotically AdS$_{d+1}$ spacetimes.}
\be
\xi^\mu \xi^\nu \left(R_{\mu\nu} - \frac12 g_{\mu\nu} R - \frac{d (d-1)}{2 L^2} g_{\mu\nu}\right) \geq 0\,, \qquad \forall\ \text{timelike} \ \xi^{\mu}\,. \label{eq: WCC}
\ee

In Einstein gravity, the WCC becomes the weak energy condition (WEC). The BTZ spacetime with an ETW brane, which we have been studying so far, is quite different from the class of spacetimes for which the rigorous results of \cite{Engelhardt:2021mju} apply. However, as we argue below, one can replace the original BTZ black hole with an ETW brane for a spacetime that is closer to the regime of applicability of the Engelhardt-Folkestad statements, while displaying the same behaviour for volume complexity as our original model.

\begin{figure}
\centering
\subcaptionbox{\label{fig:double_up}}{\includegraphics[height=4.5cm]{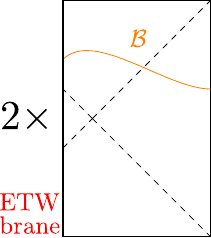}}
\hspace{20mm}
\subcaptionbox{\label{fig:effective wormhole}}{\includegraphics[height=4.5cm]{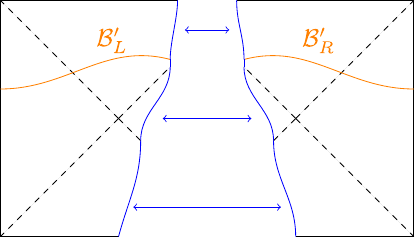}}
\caption{Two systems with the same volume complexity. (a) Two identical and disjoint copies of a BTZ black hole in the presence of ETW branes (thick black line) with an intrinsic JT gravity theory, each with an identical maximal volume slice $\mathcal{B}$. (b) A long wormhole formed by gluing two BTZ geometries along the worldvolume of a shell of matter (blue). The maximal volume slice is ${\cal B}'={\cal B}'_L\cup {\cal B}'_R$. The shell profile shown in (b) is fine-tuned to ensure that the volume of ${\cal B}'_L$ (or ${\cal B}'_R$) corresponds to
the volume of ${\cal B}$ plus the contact term evaluated at the intersection of ${\cal B}$ and the brane in (a).}
\label{fig:new_geometry}
\end{figure}

To apply the arguments of \cite{Engelhardt:2021mju}, we require all the boundaries of the spacetime to be asymptotically AdS. We can achieve this by doubling the original planar BTZ + ETW brane geometry. The construction of the new effective spacetime is shown in fig.~\ref{fig:new_geometry}. To emulate the presence of the brane in the two-sided geometry, we replace the brane with a thin shell of matter, which moves along some profile inside the event horizon when viewed from either boundary. The two halves of the geometry are glued along the worldvolume of the shell using the Israel junction conditions. Thus, the new spacetime can be thought of as a long asymptotically AdS wormhole, supported by a shell of matter with a given energy-momentum tensor. From the gravitational point of view, the profile of the shell determines the energy-momentum of the shell.

From the point of view of the volume complexity functional, the presence of the shell modifies the volume complexity of the two-sided BTZ geometry. In particular, we can fine-tune the profile of the shell in such a way that the resulting contribution to the complexity functional matches twice the contact term contribution from JT gravity present on the ETW brane in the single-sided setup, determined by the dilaton contribution to the volume~(\ref{eq:contact}) in the original geometry. In that case, the volume of the extremal surface $\mathcal{B}' = \mathcal{B}'_L \cup \mathcal{B}'_R $ is the same by construction as twice that for $\mathcal{B}$ plus two times the contact term evaluated at ${\cal B}\cap {\rm brane}$ at any asymptotic time. In what follows, we will refer to the newly constructed two-sided model as the \textit{wormhole picture}, and to the original ETW brane model as the \textit{brane picture}. 

We express the shell profile by\footnote{Note that the two sides of the wormhole are covered by their own sets of global $(\tau, y)$ coordinates. The $\tau$-coordinate is glued in such a way that equal$-\tau$ slices are continuous across the gluing surface.}
\begin{equation}
    y_{\rm shell}=y_0+y_1(\tau)
    \label{eq:shell_profile}
\end{equation}
where $y_0$ is the location of the original ETW brane, determined by $\Lambda^{\rm brane}$. The volume of any candidate extremal surface parametrized by $(\tau(\sigma),y(\sigma),x)$ in the wormhole picture is given by the functional~\eqref{eq:Volume no JT}, which in a general parametrization is
\be\label{eq: volume general}
{\rm Vol} = \ell r_0 \int d \sigma \, \frac{\cos\tau}{\cos^2 y} \sqrt{\dot{y}^2-\dot{\tau}^2}~,
\ee
where an overhead dot denotes a derivative with respect to the parameter $\sigma$.

Because of the $\mathbb{Z}_2$ symmetry in the setup, we can simply evaluate the volume of one of the components of ${\cal B}'$, say ${\cal B}'_R$, and multiply the final result by two. From this point of view, the extremal surface ${\cal B}'_R$ is that of half of the long wormhole geometry, with one boundary along the shell of matter and the other being the right asymptotic AdS$_3$ boundary. The on-shell variation of the volume functional~\eqref{eq: volume general} for $\mathcal{B}'_R$ gives
\begin{equation}
\label{eq:On shell variations wormhole}
    \delta {\rm Vol}_{\rm on-shell}=\int_{\rm shell}^{\partial {\rm AdS}_3} d\sigma \qty(P_\tau\delta\tau+P_y\delta y).
\end{equation}
This variation should be set to zero for extremality of ${\cal B}'_R$, which imposes boundary conditions at the asymptotic boundary and the shell worldvolume.  The boundary conditions at the asymptotic AdS$_3$ boundary are the same Dirichlet boundary conditions as in eq.~\eqref{eq:ttotau} and eq.~(\ref{eq:ycutoff2}),
\begin{equation}\label{eq:worm_dirichlet}
\begin{aligned}
    \tau_{\rm bdy} &= \arcsin \left(\tanh\left(\frac{r_0 t_{\rm bdy}}{L^2}\right)\right)\,,\\
    y_{\rm bdy} &= y_\epsilon= \frac{\pi}{2} - \frac{\epsilon \, r_0}{L}\,\, {\rm sech} \left(\frac{r_0 t_{\rm bdy}}{L^2}\right).
\end{aligned}
\end{equation}
With the shell profile~\eqref{eq:shell_profile}, the Dirichlet boundary condition along the worldvolume of the shell is
\begin{equation}
\label{eq:Neumann wormhole picture}
 \eval{\delta{y}-y_1'(\tau)\delta{\tau}}_{\text{shell}}=0.
\end{equation}
Next, we substitute the Dirichlet boundary condition from eq.~\eqref{eq:Neumann wormhole picture} into eq.~\eqref{eq:On shell variations wormhole}, and demand the total variation to vanish for extremal surfaces. This results in the following Neumann boundary condition along the shell
\begin{equation}
\label{eq:var_redef}
\eval{P_\tau\delta\tau+P_y\delta y}_{\rm shell}=\eval{(P_\tau+P_yy_1'(\tau))\delta\tau}_{\rm shell}=0.
\end{equation}
The canonical momenta can be found using the explicit Lagrangian~\eqref{eq:Volume no JT},
\begin{align}
    P_\tau&=\ell r_0\frac{\cos\tau}{\cos^2y}\frac{-\dot{\tau}}{\sqrt{\dot{y}^2-\dot{\tau}^2}} \, ,\label{eq:Ptau_worm}\\
    P_y&=\ell r_0\frac{\cos\tau}{\cos^2y}\frac{\dot{y}}{\sqrt{\dot{y}^2-\dot{\tau}^2}}~.\label{eq:Py_worm}
\end{align}
Using the above expressions, the Neumann boundary condition~\eqref{eq:var_redef} simplifies to
\begin{equation}
\label{eq:worm_condition}
    \eval{\dot{\tau}}_{\rm shell}=\eval{y'_1\dot{y}}_{\rm shell}\,,
\end{equation}
which simply states that the extremal surface ${\cal B}'_R$ intersects the worldvolume of the shell at a right angle.

We want to fine-tune the shell profile $y_1(\tau)$ so that the change in the volume of the extremal surface ${\cal B}'_R$ compared to $\cal B$ reproduces the $\tau$-dependent part of the contact term in the complexity associated to the surface $\cal B$ in the brane picture. To this end, we compute the volume contribution~\eqref{eq:Volume no JT} to complexity from the portion of the extremal surface between the shell and $y_0$, and match this to the $\tau$-dependent part of the contact term in eq.~\eqref{eq:CV_JT}. Using the dilaton profile~(\ref{eq:dilaton_sol}), we can express the relevant contribution to the contact term as
\begin{equation}\label{eq:contact JT}
   \ell r_0\varepsilon\sin\tau_{\rm brane}~,
\end{equation}
where $\varepsilon\equiv \frac{L^2}{r_0L_2}\alpha$.
We want to equate this contribution to the difference in the volumes of the extremal surface ${\cal B}'_R$ and $\cal B$. Though we expect the procedure outlined above for computing the shell profile to work in general, finding the exact form of the function $y_1(\tau)$ analytically is in practice quite involved, and we therefore resort to a perturbative approach in the following. We will assume that $y_1'(\tau) \ll 1$ in the region probed by the extremal surfaces, which, as we will see soon, is true for a small JT coupling $\alpha$. From eq.~\eqref{eq:worm_condition}, this assumption implies that $\tau$ is approximately constant between $y_0$ and $y_{\rm shell}$, so that the difference in volumes is
\begin{equation}\label{eq:new VOL}
\begin{aligned}
   \Delta {\rm Vol}&=\ell r_0\int_{y_{\rm shell}}^{y_0}d \tilde{y}\frac{\cos \tau}{\cos^2 \tilde{y}}\sqrt{1-\tau'(\tilde{y})^2}\\
    &\approx\ell r_0\cos\tau_{\rm brane}\qty(\tan y_0-\tan y_{\rm shell})~.
\end{aligned}
\end{equation}
Equating this difference in volume between $\cal B'_R$ and $\cal B$ with the $\tau$-dependent part of the contact term~\eqref{eq:contact JT} gives
\begin{equation}\label{eq: yyshell}
    y_{\rm shell}=\arctan(\tan y_0-\varepsilon\tan \tau_{\rm brane}).
\end{equation}
For $\abs{\varepsilon}\tan\tau_{\rm brane}\ll1$, we can approximate the shell profile as\footnote{On the other hand, for $\abs{\varepsilon}\tan\tau_{\rm brane}\gg1$, the shell profile becomes
\begin{equation}
    y_{\rm shell}\approx-\frac{\pi}{2}+\qty(\frac{\pi}{2}\tan y_0-1)\frac{\tau_{\rm brane}-\frac{\pi}{2}}{\varepsilon}.
\end{equation} However, the maximal volume surfaces ${\cal B}'_R$ will not be sensitive to this region of the shell profile.}
\be\label{eq:y new}
y_{\rm shell}\approx y_0-\varepsilon\cos^2y_0\tan\tau_{\rm brane},
\ee
where we recall that $\varepsilon\equiv \frac{L^2}{r_0L_2}\alpha$.

\begin{figure}[t]
\centering
\includegraphics[width=0.6\textwidth]{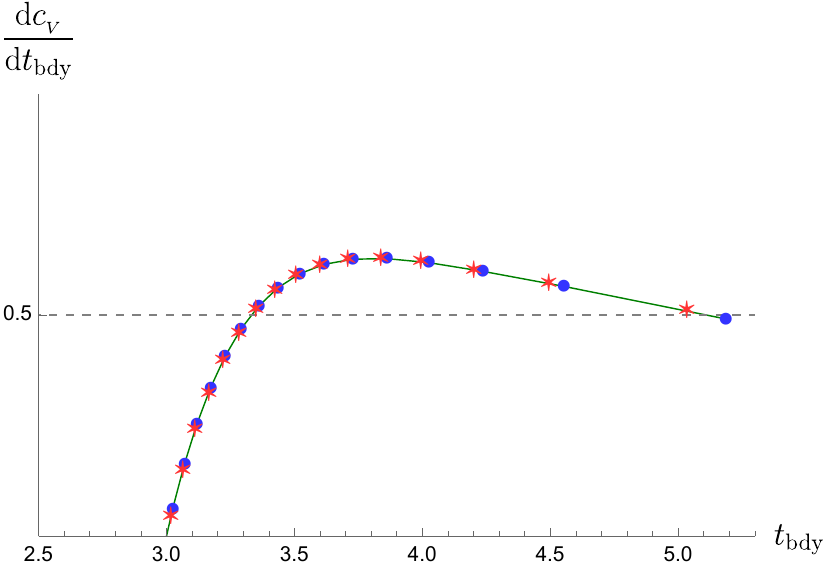}
\caption{Comparing the rate of change of complexity associated with the extremal surfaces $\cal{B}'_R$ and $\cal{B}$ when $\alpha \ll 1$ at very late times between the brane perspective (blue circles) and the wormhole perspective (red stars). For illustrative purposes we have chosen $\alpha = 0.01, y_{\rm brane} = 0$ here, with $L = L_2 = r_0 = G_N = 1$. Clearly, the wormhole description captures the rate of change of complexity quite well, including the Lloyd bound violating behaviour.} 
\label{fig:Matching complexity growth}
\end{figure}

Taking the derivative of eq.~\eqref{eq: yyshell} leads to 
\begin{equation}\label{eq:new new Neumann}
    y'_1(\tau)=\frac{-\varepsilon\sec^2\tau}{1+(\varepsilon\tan\tau-\tan y_0)^2}.
\end{equation}
Notice that $y_1'(\tau)\ll 1$ implies that $\varepsilon$ and thus the JT coupling $\alpha$ is small.

Following the method in section~\ref{sec:numerical_approach}, we can use the boundary conditions~\eqref{eq:worm_condition} and~\eqref{eq: yyshell} to solve for the trajectory of ${\cal B}'_R$ numerically. The volume of ${\cal B}'_R$ can then be evaluated via~\eqref{eq: volume general}, and its time dependence read off from the location at which ${\cal B}'_R$ reaches the asymptotic boundary via~\eqref{eq:worm_dirichlet}. The comparison between the rate of change of complexity in the wormhole picture and the brane picture is displayed in fig.~\ref{fig:Matching complexity growth}. We notice there is perfect agreement between the two approaches, which confirms that the wormhole picture is an alternative description of the original system, as far as computing holographic complexity is concerned.

\subsection{Weak energy condition in the wormhole picture}
\label{sec: WEC}

Having established the equivalence of the wormhole and the brane pictures for the CV computation, we now proceed to study the WCC (\ref{eq: WCC}) in the wormhole picture in connection with the violation of the Lloyd bound. The expression in parenthesis in eq.~(\ref{eq: WCC}) is the Einstein tensor, so in Einstein gravity WCC becomes the WEC: 
\begin{equation}\label{eq:WEC}
    T_{\mu\nu}\xi^\mu \xi^\nu\geq0~,\qquad \xi^2 = \xi^\mu \xi^\nu g_{\mu\nu}<0\,,
\end{equation}
where $\xi^\mu$ is an arbitrary timelike vector. In \cite{Engelhardt:2021mju}, the WEC is stated to be a sufficient condition for meeting the Lloyd bound in Einstein gravity
\begin{equation}\label{eq:WEC-Lloyd}
    {\rm WEC} \Rightarrow \text{Lloyd bound}\,.
\end{equation}

In contrast to the brane picture, the wormhole picture introduced in the previous subsection is asymptotically AdS$_3$, and one can apply the findings of \cite{Engelhardt:2021mju}.\interfootnotelinepenalty=10000\footnote{Strictly speaking, the results of \cite{Engelhardt:2021mju} apply to spacetime dimension $d+1\geq 4$, which does not include our case of $d=2$. Nonetheless, we can ask whether there is a relation between the WEC and the Lloyd bound in $d+1=3$.} We start by arguing that a nonzero JT coupling in the brane picture always leads to a violation of the WEC in the wormhole picture, see eq. (\ref{eq:Tmn xim xin}). Given that the violations of the Lloyd bound we observe occur at nonzero JT coupling, our results are therefore compatible with (\ref{eq:WEC-Lloyd}). However, in the above sections~\ref{sec:asymptotics} and~\ref{sec:numerical_approach}, and in particular eq.~\eqref{eq:alpha condition}, we have shown that a non-zero JT coupling does not always imply a violation of the Lloyd bound, provided that the brane cosmological constant is small enough to support it. This suggests that we can explore a weaker energy condition than WEC which nonetheless is sufficient to ensure that the Lloyd bound is satisfied.

To proceed in detail, let us assume that the worldvolume of the shell is embedded into the wormhole spacetime by $w^\mu=(\tau(\eta),~y(\eta),~x)$. For definiteness, we focus on the right half of the wormhole, so that $y_0 < 0$. The functions $\tau(\eta)$ and $y(\eta)$ should follow the profile of eq.~(\ref{eq: yyshell}):
\begin{equation}\label{eq:map}
    \tau(\eta)=\eta~,\quad y(\eta)=\arctan\left(\tan y_0-\varepsilon \tan \eta\right)~.
\end{equation}
The coordinates on the shell worldvolume are $z^a = (\eta,~x)$. The normal vector to the shell worldvolume $n^\mu$, pointing outwards, has the form
\begin{equation}
    n^\mu=-\frac{\cos y}{L\sqrt{\dot{\tau}^2-\dot{y}^2}}(\dot{y},~\dot{\tau},~0)~,
\end{equation}
where the dot denotes the derivative with respect to $\eta$. The matter stress tensor is related to the extrinsic curvature of the matter shell via the Israel junction conditions \cite{Israel:1966rt}:
\begin{equation}\label{eq:stress tensor-bulk}
    T_{\mu\nu} =\frac{1}{8\pi G_3}\int dz^2 \qty(\Delta K_{\mu\nu}-\Delta K h_{\mu\nu}) \delta(x^\mu -w^\mu(z^a))\,,  
\end{equation}
where the transverse projector $h_{\mu\nu}$ and the extrinsic curvature are defined in a standard way:
\begin{equation}
h_{\mu\nu}=g_{\mu\nu}-n_\mu n_\nu\,, \qquad  K_{\mu\nu} = \nabla_\mu n_\nu
\end{equation}
The stress tensor $T_{\mu\nu}$ is nontrivial only on the worldvolume of the shell, the rest of the bulk spacetime is a solution of the vacuum Einstein equations. Furthermore, eq.~\eqref{eq:stress tensor-bulk} explicitly shows that the stress tensor associated with the shell is tangential to the worldvolume of the shell, \ie $T_{\mu\nu}  n^\nu=0$. Therefore, in our wormhole spacetime the WEC condition (\ref{eq:WEC}) in the bulk is equivalent to the WEC on the shell worldvolume: 
\be
T_{ab} \xi^a \xi^b \geq 0\,, \label{eq: WEC-shell}
\ee
where $T_{ab} \equiv \frac{\pd {w^\mu}}{\pd z^a}\frac{\pd {w^\nu}}{\pd z^b}T_{\mu\nu} (w^\mu)$ is the pullback of the stress-energy tensor $T_{\mu\nu}$ on the shell worldvolume, and $\xi^a$ is any timelike vector on the shell worldvolume. Therefore, from this point onwards, we will focus on studying the condition (\ref{eq: WEC-shell}). 

To compute $T_{ab}$, we perform the pullback of the bulk Israel junction condition (\ref{eq:stress tensor-bulk}) to the worldvolume of the shell. We get
\begin{equation}\label{eq:stress tensor}
    T_{ab} =\frac{1}{8\pi G_3}\qty(\Delta K_{ab}-\Delta K h_{ab})\,,
\end{equation}
where the induced metric on the worldvolume of the shell $h_{ab}$ is given by
\begin{equation}\label{eq:inducedmetricshellh}
        ds^2_{\rm ind}=h_{ab} dz^a dz^b=\frac{1}{\cos^2y(\eta)}\qty((\dot{y}(\eta)^2-\dot{\tau}(\eta)^2)L^2d\eta^2+\frac{r_0^2}{L^2}\cos^2\tau(\eta) d x^2)~,
\end{equation} 
and $\Delta K_{ab}=K_{ab}^L-K_{ab}^R$ is the difference of the extrinsic curvatures on the two sides of the shell worldvolume, considering the direction of the normal vector on the shell to be outwards from the bulk, see fig. \ref{fig:effective wormhole}. Using the $\mathbb{Z}_2$-symmetry of the setup, this quantity can be expressed as $\Delta K_{ab}=2 K_{ab}$ where $K_{ab} = K_{ab}^L=- K_{ab}^R$. The perturbative evaluation of the extrinsic curvature $K_{ab}$ gives
\begin{equation}\label{eq:Kij explicitly}
    K_{ab}=-\frac{\sin y_0}{L} h_{ab}+\varepsilon \frac{\cos(y_0) \tan(\tau_{\rm brane})}{L} \left(\cos^2( y_0) h_{ab}-\frac{r_0^2}{L^2}\delta^x_a\delta^x_b\right)+\mathcal{O}(\varepsilon^2)~.
\end{equation}
Taking the trace of this quantity gives
\begin{equation}\label{eq:traceK}
    K= -\frac{2}{L}\sin y_0 + \varepsilon\frac{\cos^3 y_0}{L} \tan\tau_{\rm brane}\qty(1-\tan^2\left(\tau_{\rm brane}\right))+ {\cal O}(\varepsilon^2)\,.
\end{equation}
The detailed derivation of the expressions for $K_{ab}$ and $K$ is given in Appendix \ref{app:ext_curve}.

Now, we can compute the matter stress tensor in eq.~(\ref{eq:stress tensor}) using eq.~\eqref{eq:inducedmetricshellh}, eq.~(\ref{eq:Kij explicitly}) and eq.~\eqref{eq:traceK} as well as the $\mathbb{Z}_2$-symmetry. This yields
\begin{equation}\label{eq:cond Tij}
\begin{aligned}
    T_{ab}&= \frac{1}{4\pi G_3}\qty( K_{ab}- K h_{ab})\,,\\
    &= \frac{\sin y_0 }{4\pi L G_3}h_{ab}+\frac{\varepsilon \tan(\tau_{\rm brane})\cos y_0}{4\pi L G_3}\qty(\cos^2 y_0 \tan^2\left(\tau_{\rm brane}\right) h_{ab} -  \frac{r_0^2}{L^2}\delta^x_a\delta^x_b )+ {\cal O}(\varepsilon^2)~.
    \end{aligned}
\end{equation}
We are interested in evaluating:
\begin{equation}\label{eq:ki def}
    \xi^a\xi^b T_{ab}~,\quad\text{where}\quad \xi^2=\xi^a\xi^b h_{ab}< 0~.
\end{equation}
Using eq.~(\ref{eq:cond Tij}), we get
\begin{equation}\label{eq:Tmn xim xin}
    T_{ab}\xi^a\xi^b= \frac{\sin y_0 }{4\pi L G_3}\xi^2+\frac{\varepsilon \tan(\tau_{\rm brane})\cos y_0}{4\pi L G_3}\qty(\cos^2 y_0 \tan^2\left(\tau_{\rm brane}\right) \xi^2 - \frac{r_0^2(\xi^x)^2}{L^2} )+ {\cal O}(\varepsilon^2)\,.
\end{equation}
Consider a vector $\xi^a$ which is arbitrarily close to a null vector, meaning that $\xi^2$ can be arbitrarily close to zero (or at least $\abs{(\xi^a)^2}\lesssim\mathcal{O}(\epsilon^2)$) while $(\xi^x)^2>0$ and finite. For vectors in this limit, we find
\begin{equation}
   \lim_{\xi^2 \to 0} T_{ab}\xi^a\xi^b = -\frac{\varepsilon r_0^2 \tan(\tau_{\rm brane})\cos(y_0)}{4\pi L^3 G_3}(\xi^x)^2+ {\cal O}(\varepsilon^2)\,.
   \label{eq: WEC-eps}
\end{equation}
This quantity is negative for positive $\tau_{\rm brane}$, and we have therefore found a timelike vector for which $T_{ab}\xi^a\xi^b<0$, and therefore $T_{\mu\nu} \xi^\mu \xi^\nu < 0$ for the timelike bulk vector $\xi^\mu =\xi^a \frac{\pd w^\mu}{\pd z^a}$. Thus, it is evident that the WEC does not hold in the wormhole picture for any nonzero $\varepsilon$, which once again is related to the JT coupling in the brane picture $\varepsilon = \frac{L^2}{r_0L_2} \alpha$. This means that in the wormhole picture, the contrapositive of the implication~\eqref{eq:WEC-Lloyd} holds, allowing for violation of the Lloyd bound.

The results of previous sections show that indeed a Lloyd bound violation occurs. However, as shown in section~\ref{sec:complexity_growth}, when the brane is located at some $y_{\rm brane}<0$, there is a window of allowed values for the JT coupling, eq.~\eqref{eq:alpha condition}, such that the rate of complexity growth does not violate the Lloyd bound. We therefore propose a weaker positive energy condition (PEC) which we will argue to be sufficient to meet the Lloyd bound. Specifically, we require that 
\begin{equation}
    T_{\mu\nu} \zeta^\mu \zeta^\mu \geq 0\,,
\end{equation}
where $\zeta^\mu$ are timelike vectors normal to the maximal volume surface ${\cal B}'$
\begin{equation}
    \zeta^2 \equiv g_{\mu\nu}\zeta^\mu\zeta^\nu < 0\,, \qquad \gamma_{\mu\nu} \zeta^\nu = 0\,,
\end{equation}
with $\gamma_{\mu\nu} = g_{\mu\nu} - \frac{\zeta_\mu \zeta_\nu}{\zeta^2}$ the projector to the maximal volume surface.
The PEC requires that the energy density for coordinate systems defined by foliations of (part of) the geometry by the maximal volume slices is non-negative everywhere. In the wormhole picture, this translates to $T_{ab} \zeta^a \zeta^b \geq 0$ for the vector $\zeta^a (z)$, which is normal to the extremal volume slice, anchored on the shell worldvolume at any point $z$. In particular, this implies $\zeta^x=0$. Therefore in the wormhole picture, evaluating the PEC at the location of the shell using eq.~\eqref{eq:cond Tij}, we find
\begin{equation}
    T_{ab}\zeta^a\zeta^b= \frac{\zeta^2}{4\pi L G_3} \left(\sin(y_0) + \varepsilon \tan^3(\tau_{\rm brane}) \cos^3(y_0) \right) + {\cal O}(\varepsilon^2)\,.
\end{equation}

The magnitude of the second term becomes larger as $\abs{\tau_{\rm brane}}$ increases. Moreover, for small $\varepsilon$, the largest value of $|\tau_{\rm brane}|$ corresponds to $r_{\rm bane} = r_\infty = r_0/\sqrt{2}$. From eq.~\eqref{eq:Sch to global}, this implies that $\cos (\tau_{\rm brane}) \geq \cos(y_0)/\sqrt{2}$ and $|\tan(\tau_{\rm brane})| \leq \sqrt{2 \sec^2(y_0)-1}$. We therefore find
\begin{equation}
   T_{ab}\zeta^a\zeta^b \leq \frac{\zeta^2}{4\pi L G_3} \left(\sin(y_0)  + |\varepsilon| (1+\sin^2(y_0))^{3/2}\right) + {\cal O}(\varepsilon^2)\,.
\end{equation}
Hence, the PEC is violated when
\begin{equation}
    |\varepsilon| > \frac{-\sin(y_0)}{(1+\sin^2(y_0))^{3/2}}\,.
\end{equation}
In terms of the JT coupling $\alpha$, the PEC requires
\begin{equation}\label{eq:alpha PEC}
    |\alpha| \leq \frac{r_0}{L} \frac{-\tan(y_0)}{(1+\sin^2(y_0))^{3/2}}\,.
\end{equation}
This condition is stronger than the Lloyd bound constraint~\eqref{eq:alpha condition} with $y_{brane}=y_0$, giving support to our claim that a refined version of~\eqref{eq:WEC-Lloyd} would be
\begin{equation}\label{eq:WEC-PEC-Lloyd}
    {\rm WEC} \Rightarrow \text{PEC} \Rightarrow \text{Lloyd bound}\,.
\end{equation}

For small $|y_0|\sim |\varepsilon|$, the Lloyd bound constraint~\eqref{eq:alpha condition} and the PEC constraint~\eqref{eq:alpha PEC} match up to $\mathcal{O}\left(\varepsilon^3\right)$. 
On the other hand, recall that the equivalence between the brane and wormhole pictures only holds for small $|\alpha|$, as we have shown in section \ref{sec: wormhole}. Therefore, when $|y_0|$ is not small, both bounds are trivially satisfied in the regime where both pictures are equivalent. With these points in mind, we can conclude that the PEC provides a much tighter bound than the WEC on whether the Lloyd bound will be satisfied.


\section{Comparison with bounds from entanglement growth}
\label{sec:compare_bounds}

Despite there currently being no proof that the Lloyd bound holds for complexity = volume, it remains a conjecture that seems to hold in many physical models. Progress towards validating this conjecture has been made, such as in~\cite{Engelhardt:2021mju}, which established a connection between the Lloyd bound and the WEC within a specific class of holographic models. In the same spirit, we show that the Lloyd bound is violated when a weaker energy condition, the PEC, is not met in an equivalent formulation of our model in terms of a long wormhole with a shell of matter, see section~\ref{sec:Engelhardt-Folkestad}. 

So far, the violation of the Lloyd bound for volume complexity has only been found in bottom-up models which are not guaranteed to have a UV-complete microscopic description. It has been shown that bottom-up models may show pathologies, which, in the spirit of effective field theory, should be carved out of the parameter space of the effective model using constraints on physical quantities associated with the boundary CFT, see~\cite{Lee:2022efh, Geng:2023qwm, Geng:2023iqd, Miao:2023mui}. This translates to bounds on the allowed bulk parameters. 

The focus of the present work has been on the analysis of the growth of complexity in the boundary CFT with a particular emphasis on when the Lloyd bound is satisfied. We can turn this around and use the Lloyd bound to propose constraints on the parameter space of our model, keeping in mind that the Lloyd bound so far remains a conjecture. The constraints we propose are similar in spirit to the ones in~\cite{Lee:2022efh, Geng:2023qwm, Geng:2023iqd, Miao:2023mui}, but with the caveat that they rely on a conjectured bound rather than a theorem. 

Of particular relevance to our work are the results of  \cite{Lee:2022efh}. The authors considered a setup identical to the one presented here -- an ETW brane embedded in the BTZ geometry, cutting off the second asymptotic region, with JT gravity localized on the brane. The information-theoretic quantity they considered is the \emph{entanglement velocity} $v_E$, which is a measure of the rate of growth of entanglement in a translationally-invariant CFT state with a uniform energy density. It is defined via
\be
\label{vEdef}
v_E \equiv \frac{\partial_t S(A)}{s_{\rm eq} \rm{Vol}(\partial A)}\, ,
\ee
where $A$ denotes a spatial subregion of the CFT, whose entanglement with the rest of the system is given by the entanglement entropy $S(A)$. In addition, $\partial A$ denotes the boundary of $A$, and $s_{\rm eq}$ denotes the thermal entropy density for the CFT if it were in a thermal state with the same energy density as the pure state of interest. For two-dimensional CFTs, it was pointed out in \cite{Hartman:2015apr} that $v_E$ satisfies the instantaneous bound 
\be
\label{vEbound}
|v_E(t)| \le 1\, ,
\ee
a result that follows by carefully examining the monotonicity of relative entropy between the pure and thermal states of the CFT. By explicitly computing the entanglement entropy holographically using the Ryu-Takayanagi prescription \cite{Ryu:2006bv, Ryu:2006ef}, and demanding the constraint~\eqref{vEbound} to be met, the authors of \cite{Lee:2022efh} were able to compute bounds on the JT coupling $\alpha$, which we summarize below. 

For an arbitrary value of the JT coupling $\alpha$, even though the entanglement may grow smoothly as a function of time, it can still violate the bound in eq.~\eqref{vEbound}. As discussed in \cite{Lee:2022efh}, ruling this possibility out leads to the constraint
\be
\label{weak_bound}
|\alpha| \le \frac{1}{1+\sin y_{\rm brane}}\, ,
\ee
where $y_{\rm brane} \in (-\pi/2,0]$ denotes the location of the brane. Clearly, this bound on $\alpha$ is strongest when $y_{\rm brane} = 0$, implying $|\alpha| \le 1$, and becomes progressively weaker as $y \to -\pi/2$. It was consequently addressed as the \emph{weak bound} in \cite{Lee:2022efh}.

\begin{figure}[t]
\centering
\includegraphics[scale=0.7]{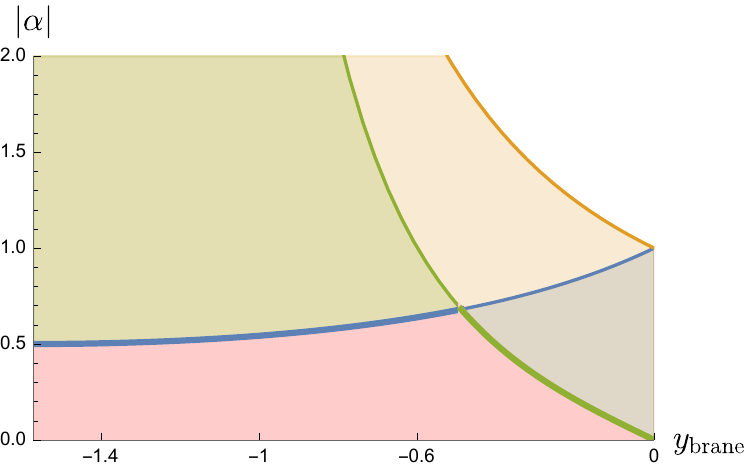}  
\caption{Combining various information-theoretic bounds to restrict the bulk parameter space for $r_0/L = 1$. The region bounded by the green curve respects the Lloyd bound, by the orange curve respects the weak entanglement bound, and by the blue curve respects the strong entanglement bound. Between the Lloyd bound and the weak bound from entanglement, the former is more constraining for $y_{\rm brane} \in [-1.083,0]$ (intersection of the green and orange curves not visible within the plotted range above). On the other hand, between the Lloyd bound and the strong bound from entanglement, the former is more restrictive only for $y_{\rm brane} \in [-0.488,0]$. The overall reduction in bulk parameter space is greater when the Lloyd bound is combined with the strong bound from entanglement - a pattern which is true for larger black holes $r_0/L > 1$ as well, see fig.~\ref{fig:comp_bounds_2}.}
\label{comp_bounds}
\end{figure}

A stronger bound on $\alpha$ was also obtained in \cite{Lee:2022efh} by disallowing discontinuous jumps in entanglement entropy as a function of time. It was observed that for certain values of the JT coupling $\alpha$, one would obtain multiple bulk surfaces extremizing the area functional as required by the Ryu-Takayanagi prescription for computing the holographic entanglement entropy, which on the one hand anchor at a particular instant of time on the boundary, while the other end either anchors on the ETW brane within $\tau_{\rm brane} \in [-\pi/2, \pi/2]$ \emph{i.e.}~the usual extremal surface, or it anchors on the brane extended into the previous/next universe (or coordinate patch) \emph{i.e.}~$\tau_{\rm brane} \in [-3\pi/2,-\pi/2]$ or $\tau_{\rm brane} \in [\pi/2,3\pi/2]$. In other words, the additional extremal surfaces pass through the past/future singularity.\footnote{Such surfaces would be automatically disallowed in a higher dimensional setup. However, in AdS$_3$, where there are no genuine curvature singularities, there is no a priori reason not to consider them.} Surprisingly enough, when such surfaces are present, they have the minimum area and thus are the candidate surfaces for computing the entanglement entropy. The entanglement entropy as a function of time then shows a discontinuous jump, or phase transition, between the phases where it is given by the usual extremal surface and the phase where it is given by the special surfaces ending in the previous/next universe. Since these discontinuous jumps in entanglement entropy are unphysical, by disallowing such behaviour one ends up with the constraint 
\be
\label{strong_bound}
|\alpha| \le \frac{1}{1-\sin y_{\rm brane}}\, .
\ee
This bound on $\alpha$ becomes progressively stronger as $y_{\rm brane} \to -\pi/2$, since it implies $|\alpha| \le 1$ when $y_{\rm brane}=0$ while $|\alpha| \le 1/2$ when $y_{\rm brane} = -\pi/2$. Consequently, it was addressed as the \emph{strong bound} in \cite{Lee:2022efh}.

\begin{figure}[t]
    \centering
    \subcaptionbox{\label{fig:comparison_2_1}}{\includegraphics[height=0.3\textwidth]{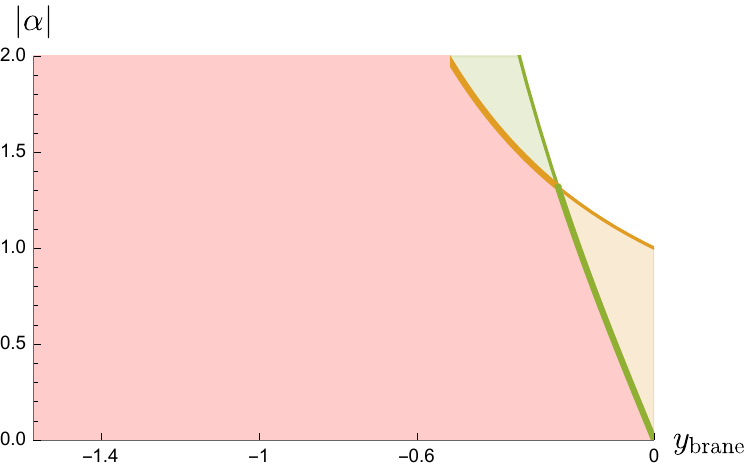}}\quad
    \subcaptionbox{\label{fig:comparison_2_2}}{\includegraphics[height=0.3\textwidth]{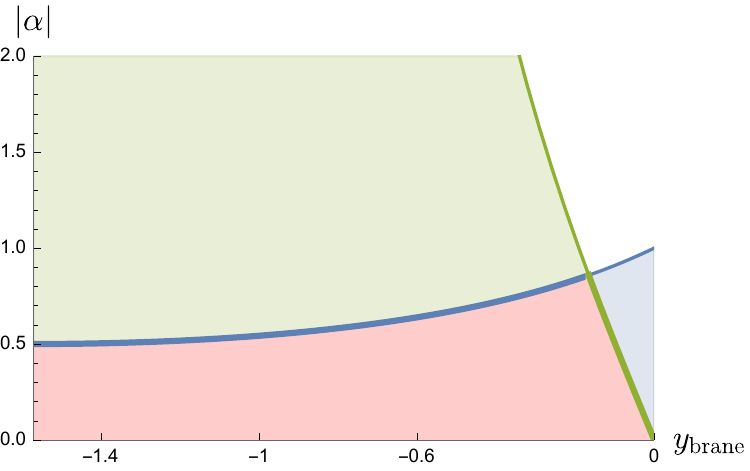}}
    \caption{Combining various information-theoretic bounds to restrict the bulk parameter space for $r_0/L = 5$. The pink region in (a) depicts the allowed bulk parameter space obtained by combining the Lloyd bound (green region) with the weak bound from entanglement (orange region), whilst the pink region in (b) depicts the allowed parameter space when the Lloyd bound is combined with the strong bound from entanglement (blue region).}
    \label{fig:comp_bounds_2}
\end{figure}

We can now combine the above constraints on $\alpha$ with the constraint we have computed in the present work by demanding that the boundary state respect the Lloyd bound, \emph{i.e.}
\be
\label{Lloyd_bound_alpha}
|\alpha| \le \frac{r_0}{L} \frac{|\tan y_{\rm brane}|}{\cos^2 y_{\rm brane}}  .
\ee
This is schematically represented in fig.~\ref{comp_bounds} for $r_0/L = 1$. In particular, the region bounded by the green curve corresponds to the bulk parameter space $(y_{\rm brane}, \alpha)$ such that the Lloyd bound~\eqref{Lloyd_bound_alpha} is met. On the other hand, the region bounded by the orange/blue curve denotes the bulk parameter space that respects the weak/strong bound from entanglement velocity, eqs.~\eqref{weak_bound} and \eqref{strong_bound}, respectively. Now, to restrict the bulk parameter space, one should combine the Lloyd bound constraint with either the weak bound or the strong bound from entanglement, depending upon whether one only wants to exclude the violation of eq.~\eqref{vEbound} or discontinuous jumps in entanglement entropy as well. Since the weak bound is less restrictive than the strong bound, the combination of the Lloyd bound and the weak bound provides weaker constraints on the parameters than the combination of the strong bound from entanglement and the Lloyd bound constraint. Further, the range of values for $y_{\rm brane}$ for which the strong bound from entanglement is more constraining than the Lloyd bound is also larger than the range of values for $y_{\rm brane}$ for which the weak bound from entanglement is more constraining than the Lloyd bound. In particular, while the strong bound from entanglement is more restrictive than the Lloyd bound for $y_{\rm brane} \in (-\pi/2, -0.488]$, the weak bound is more restrictive than the Lloyd bound only for $y_{\rm brane} \in (-\pi/2, -1.083]$.

The situation for larger black holes is shown in fig.~\ref{fig:comp_bounds_2} for $r_0/L = 5$. We can see that increasing the size of the black hole $r_0/L$ increases the importance of the entanglement bounds relative to the Lloyd bound when constraining the parameter space of the theory, \ie the ranges of $y_{\rm brane}$ for which the entanglement bounds are stronger than the Lloyd bound increases. Furthermore, the values of $y_{\rm brane}$, at which the two entanglement bounds become stronger than the Lloyd bound constraint, become closer. However, the strong bound from entanglement continues to be the more restrictive of the two when combined with the Lloyd bound constraint as expected.


\section{Discussion and outlook}
\label{sec:discussion}
In this paper, we have studied the time evolution of holographic complexity using the CV proposal for translationally invariant states of two-dimensional CFTs dual to the planar BTZ geometry with an ETW brane which is endowed with intrinsic gravitational dynamics, namely JT gravity. Let us highlight the key findings and conclusions of our analysis. 

\paragraph{Analytic results for the asymptotic rate of complexity growth.} We have performed a detailed analytic study of the late time behaviour of the rate of change of volume complexity. We have obtained asymptotic expressions for this rate in three distinct regions within the bulk parameter space, comprising the brane location and the JT coupling $(y_{\rm brane}, \alpha)$, where the Lloyd bound is either satisfied, marginally satisfied or violated. In particular, we have derived the critical curve in the space of bulk parameters which corresponds to the Lloyd bound being marginally satisfied, and separates the Lloyd bound respecting and violating regions, see eq.~(\ref{eq:alpha condition}). With reference to the effective mechanical picture introduced in section~\ref{sec:mech}, we have observed that the violation of the Lloyd bound is associated with the particle carrying energy greater than the height of the potential barrier, see fig.~\ref{fig:Ueff}, and going towards the singularity at $r=0$ before terminating on the brane. The Lloyd bound is satisfied when the particle scatters on the potential barrier before terminating on the brane. We have performed our calculations specifically for the planar BTZ geometry; however, we expect that the late time behaviour of the rate of complexity growth should be similar for higher-dimensional black holes with ETW branes that carry intrinsic Dvali-Gabadadze-Porrati (DGP) gravity \cite{Dvali:2000hr}, or similar models, as long as the Neumann boundary condition on the brane has a form that is qualitatively similar to eq.~(\ref{eq:Neumann boundary global coord}). 

\paragraph{Numerical study of the rate of complexity growth.} Using a numerical approach, we have elucidated the behaviour of volume complexity and its rate of change for our setup as a function of boundary time. Restricted to asymptotically early/late times, the numerical results are in agreement with the analytic observation that only part of the bulk parameter space respects the Lloyd bound on complexity growth at all times. In particular, for bulk parameters that violate the Lloyd bound, with $\alpha > 0$ the bound is reached from above at asymptotically late times, whilst for $\alpha<0$ it is reached from below at asymptotically early times - see figs.~\ref{fig:dCVdt_With_JT_1} and \ref{fig:With_JT_plots}. In line with the effective mechanical picture, numerical results show that the violation of the Lloyd bound occurs when the extension of the extremal volume surfaces beyond the ETW brane starts falling into the singularity instead of being able to reach the left asymptotic boundary, as depicted in fig.~\ref{fig:surf_extension}.

\paragraph{Connection between the Lloyd bound and bulk energy conditions.} In line with arguments of \cite{Engelhardt:2021mju}, we have explored the connection between the violation of the Lloyd bound with the violation of the weak energy condition in the bulk. We accomplished this by constructing a long wormhole geometry supported by a thin shell of matter hidden behind the horizons in such a way that the time dependent volume complexity matches with twice that of the original ETW brane setup, see fig.~\ref{fig:new_geometry}. This connects the WEC-based perspective of \cite{Engelhardt:2021mju} beyond purely asymptotically AdS spacetimes to include spacetimes with other types of boundaries, such as ETW branes. We have established that this long wormhole geometry violates the WEC for any choice of bulk parameters which also leads to a violation of the Lloyd bound. By examining the timelike directions that violate WEC when the Lloyd bound is satisfied, we have identified a weaker yet sufficient condition for the Lloyd bound to be respected -- the positive energy condition (PEC). Specifically, it is sufficient for the energy $T_{\mu\nu} \zeta^{\mu} \zeta^\nu$ to be non-negative in the directions $\zeta^\mu$, which are normal to the extremal volume surfaces. The directions $\zeta^\mu$ provide a time direction to the bulk manifold in a coordinate system which consists of a foliation by maximal volume surfaces, and the PEC then tells us that the energy density in these coordinates should be positive. The violation of PEC thus works as a refined necessary condition for the violation of the Lloyd bound, compared to the violation of WEC. The question of whether violation of the PEC is sufficient to guarantee there will be a violation of the Lloyd bound remains an open question. It might also be interesting to verify this observation in higher dimensional versions, or other extensions of our model.

\paragraph{Combining Lloyd bound constraint with entanglement growth.} We have illustrated the possibility of combining the Lloyd bound constraint with the entanglement bounds obtained in \cite{Lee:2022efh} to further constrain the bulk parameter space. We observe that the weak entanglement bound, which arises from bounding the rate of entanglement growth in terms of the entanglement velocity $v_E \le 1$, does constrain the bulk parameter space further when combined with the Lloyd bound, but not as much as the strong entanglement bound does, which is obtained by disallowing discontinuous jumps in entanglement growth. This illustrates the general idea that bounds on the rate of growth of information-theoretic measures for the boundary CFT state can lead to a significant reduction in the allowed bulk parameter space of couplings. An important research topic in quantum gravity is distinguishing low-energy effective field theories based on whether they admit a consistent UV completion or not. This goes by the name of the swampland program \cite{Vafa:2005ui}, wherein effective theories with no consistent UV completion belong to the swampland, whilst the ones that admit a UV completion into string theory belong to the landscape. Attempting to constrain low-energy effective bulk descriptions using information theoretic measures can broadly be thought of as providing another diagnostic for the swampland program, wherein models that otherwise seem innocuous but conflict with the said information theoretic measures should be thought of as belonging to the swampland \cite{Geng:2019bnn, Lee:2022efh, Geng:2023qwm, Geng:2023iqd}.

Let us now point out some interesting future directions. 
Among the most straightforward generalizations of our work are: first, to consider the possibility of more general dilaton gravity models to be localized on the ETW brane \cite{Grumiller:2021cwg}, and see how the associated couplings get constrained by demanding the Lloyd bound being met; second, to extend the analysis of the Lloyd bound for the \emph{complexity equals anything} family of observables proposed in \cite{Belin:2021bga,Belin:2022xmt}, and third, to consider higher dimensional generalizations of the setting we have considered here, where the simplest intrinsic braneworld theory would be DGP gravity \cite{Dvali:2000hr}. There have been recent developments on constraints for braneworld models \cite{Geng:2023qwm,Geng:2023iqd}, which would be interesting to compare and contrast with the higher dimensional version of our setup.

An interesting observation we have made is the connection between the violation of the Lloyd bound and the falling of the corresponding extremal surfaces into the future singularity of the black hole when extended beyond the ETW brane, see fig.~\ref{fig:surf_extension}. Note that in the black hole spacetime without the ETW brane, \emph{i.e.} the eternal BTZ black hole geometry, no mechanism would make the extremal volume surfaces curve towards and fall into the singularity, and so the Lloyd bound is always respected. The presence of the brane, or equivalently the shell in the long wormhole picture, adds matter that deforms the extremal surfaces through the corresponding boundary/gluing conditions. An interesting question to ask here is the microscopic interpretation of this behaviour. In terms of quantum circuits, one of the key assumptions justifying the Lloyd bound is the orthogonalizing property of the set of gates that are used to define complexity \cite{Cottrell:2017ayj}, and when the gates are not orthogonalizing, one in general does not expect the Lloyd bound to hold \cite{Jordan:2017vqh}. Having tunable matter in the bulk that can realize both Lloyd bound respecting and violating regimes would naively correspond to the presence of some high-energy excitation or source in the dual boundary state. Changing the parameters of this excitation, such as energy, would then allow the CFT to go from the Lloyd bound respecting to the violating regime. This can allow one to probe the orthogonalizing property of the given set of gates when acting on the holographic boundary state. Understanding the mechanism of this could potentially work as a pathway to establish a more precise connection between the orthogonalizing properties of quantum gates in a CFT, the Lloyd bound, and probing the singularity with extremal surfaces in the bulk description. Regardless of whether the orthogonalizing property of the gates holds or not, there is a possibility that a more fundamental bound on computational speed, and hence complexity growth, can be obtained by taking into account additional information theoretic measures and not just the energy scale \cite{Jordan:2017vqh}. It would then be interesting to see how such a universal complexity bound would manifest itself in the dual holographic description and what constraints the additional quantum information criteria would impose on the bulk gravity theory.

Another set of interesting questions arises from using the long wormhole construction discussed in section \ref{sec:Engelhardt-Folkestad} to relate the Lloyd bound to the WEC in the bulk. The long wormhole and the ETW brane geometries have matching volume complexity and rate of growth thereof (up to a factor of 2). The Lloyd bound violating regime in the brane picture thus matches the regime in the long wormhole picture where the WEC is violated. We expect that further investigation with a similar approach can be used to generalize the proof of the holographic Lloyd bound theorem of \cite{Engelhardt:2021mju} to Einstein gravity in $3$D. It would also be interesting to understand better how the bulk energy conditions work in the presence of branes with DGP couplings in higher dimensional setups, in which the results of~\cite{Engelhardt:2021mju} are more relevant. It is also worth noting that the long wormhole geometry we have discussed is similar to the geometry of black hole microstates discussed in \cite{Balasubramanian:2022gmo,Balasubramanian:2022lnw}. It would be interesting to further explore and build upon this connection. 

\acknowledgments
We would like to thank Irina Aref'eva, Shira Chapman, Dami\'{a}n Galante, Rob Myers, Andrew Svesko, and Nicolò Zenoni for helpful discussions. The work of SEAG is partially supported by the Research Foundation - Flanders (FWO) Research Project G0H9318N and the inter-university project iBOF/21/084. Work at VUB has been supported by FWO-Vlaanderen project G012222N and by the VUB Research Council through the Strategic Research Program High-Energy Physics. JH is supported by FWO-Vlaanderen through a Junior Postdoctoral Fellowship. The work of AS is supported by the European Research Council (ERC) under the European Union’s Horizon 2020 research and innovation program (grant agreement no.~758759). SEAG thanks the IFT-UAM/CSIC, the University of Amsterdam, the Delta Institute for Theoretical Physics, and the International Centre for Theoretical Physics for their hospitality and financial support during several phases of the project, and the FWO for also providing mobility support. AS would like to thank the organizers of the \emph{All Lambdas Holography} workshop at the Czech Academy of Sciences, Prague, and the \emph{Gravity Sagas} workshop at the University of Iceland, Reykjavik, where parts of this work were presented. AS would also like to acknowledge the warm hospitality of the Korea Institute for Advanced Study, Seoul, the Nordic Institute for Theoretical Physics (Nordita), Stockholm, the Tata Institute of Fundamental Research, Mumbai, and the University of Bristol, UK, during various stages of this project. 


\appendix


\section{Tensionless ETW brane without intrinsic gravity}\label{sec:simple}
In this appendix, we perform the asymptotic analysis for the simplest case of an ETW brane without intrinsic JT gravity nor tension ($T_0=0$). We proceed analogously to the analysis of \cite{Belin:2021bga}. The goal is to evaluate the leading contributions to the integral~\eqref{eq:tbdy integral1} in the late time limit. This is done by realizing that at late times the turning point $r_{\text{min}}$ approaches a local maximum of the effective potential $r_\infty$, defined in fig. \ref{fig:Ueff}. The corresponding maximal value of the potential is given by
\begin{equation}\label{eq:U(rinfty)}
	U_{\rm eff}(r_\infty) = P_\infty^2 =\frac{r^4_0}{4 L^4},
\end{equation}
where
\begin{equation}
	P_\infty \equiv \lim_{t\to \infty} P_v.    
\end{equation}
We will integrate the leading contribution to the integral~\eqref{eq:tbdy integral1} in the late time limit by expanding around the asymptotic values of $P_v$ and $U_{\rm eff}(r)$. To this end, we use eq.~(\ref{eq:U(rinfty)}) and 
\begin{align}\label{eq:PV rmin}
	U_{\rm eff}(r)&=P_v ^2+\frac{1}{2}U^{''}(r_{\rm min})(r-r_{\min})^2+\mathcal{O}\qty((r-r_{\min})^3)~,\\
	P_v^2&=-\frac{f(r_{\min})r_{\min}^2}{L^2}\\
&=P_\infty^2-\frac{2 r_0^2}{L^4} (r_{\min}-r_\infty)^2+\mathcal{O}\qty((r-r_{\min})^3)~.
\end{align}
In this simple case, it turns out that it is useful to consider the integral~(\ref{eq:tbdy integral1}) between the asymptotic boundary and the turning point instead
\begin{equation}\label{eq:tbdy integral2}
    t_{\rm bdy}-t_{\rm min}=-\int^{r_{\rm bdy}}_{r_{\rm min}} d r \frac{P_v}{f(r)\sqrt{\frac{f(r)r^2}{L^2}+P_v^2} }~.
\end{equation}
Let us introduce the expansion parameter
\begin{equation}\label{eq:expand delta r}
	\delta r \equiv r_{\rm min}-r_\infty~.
\end{equation}
With this parameter at hand, we find 
\begin{equation}
	\begin{aligned}
		t_{\rm bdy} - t_{\rm min} &= -\int_{r_{\rm min}}^{r_{\rm bdy}} dr \frac{P_\infty}{f(r)\sqrt{\frac{f(r)r^2}{L^2}+P_\infty^2} } + {\cal O}\left( \frac{\delta r^2}{r_0^2}\right)\\
		&= \frac{r_0^2 L^2}{2}\int_{r_{\rm min}}^{r_{\rm bdy}} dr \frac{1}{r^2-r_0^2} \frac{1}{r^2-\frac{r_0^2}{2}}+ {\cal O}\left( \frac{\delta r^2}{r_0^2}\right)\,.
	\end{aligned}
\end{equation}
The above integral should be evaluated separately in the interior region $r<r_0$ and the exterior region $r>r_0$. The integrand diverges at $r = r_0$, so we introduce a regulator $\epsilon$ for the two contributions to the integral. Explicitly,

\begin{equation}
	\frac{t_{\rm bdy} -t_{\rm min}}{L^2} = \frac{(t_{\rm bdy}-\eval{t}_{r_0+\epsilon}) -(t_{\rm min}-\eval{t}_{r_0-\epsilon})}{L^2}- \left(\eval{t}_{r_0+\epsilon} - \eval{t}_{r_0-\epsilon}\right)\,,
\end{equation}

\begin{equation}
	\frac{t_{\rm bdy} -t_{\rm min}}{L^2} = \frac{1}{r_0}\lim_{\varepsilon \to 0^+} \left( \Psi_{\rm in}\Big|_{r_{\rm min}}^{r_0-\varepsilon} + \Psi_{\rm out} \Big|_{r_0+\varepsilon}^{r_{\rm bdy}} \right) - \left(\eval{t}_{r_0+\epsilon} - \eval{t}_{r_0-\epsilon}\right)+ {\cal O}\left( \frac{\delta r^2}{r_0^2}\right)\,,
\end{equation}
where
\begin{equation}
	\begin{aligned}
		\Psi_{\rm in} = \sqrt{2} \,{\rm coth}^{-1}\left( \frac{\sqrt{2}r}{r_0}\right) - {\rm tanh}^{-1}\left(\frac{r}{r_0}\right)\,,\\
		\Psi_{\rm out} =  \sqrt{2} \,{\rm coth}^{-1}\left( \frac{\sqrt{2}r}{r_0}\right) - {\rm coth}^{-1}\left(\frac{r}{r_0}\right)\,.
	\end{aligned}
\end{equation}
Taking the $\epsilon\to 0$ limit, the time differences at $r_0\pm \epsilon$ vanishes,\footnote{This can be shown by approximating the trajectory $v(r)$ via a Taylor expansion up to second order across the horizon, using the velocities~\eqref{eq:velocities1} to obtain $\frac{dv}{dr}$ and differentiating the resulting expression with respect to r.} and the contributions from the regulators exactly cancel each other. The boundary contribution also vanishes, so we are left with
\begin{equation}
\label{eq:tbdy expanded}
\frac{t_{\rm bdy} -t_{\rm min}}{L^2} = 
\frac{1}{\sqrt{2}r_0} \log \left(\frac{\sqrt{2}r_0}{\delta r}\right) - \frac{1}{r_0}{\rm coth}^{-1}\left(\sqrt{2}\right) + {\cal O}\left( \frac{\delta r^2}{r_0^2}\right).
\end{equation}
Inverting the relation above in terms of the expansion parameter given in eq.~(\ref{eq:expand delta r}), we find the location of the turning point approaches $r_\infty$ to leading order as an exponential decay
\begin{equation}
	r_{\rm min} - r_\infty = \sqrt{2}r_0\,e^{-\frac{\sqrt{2}r_0}{L^2}\left(t_{\rm bdy}-t_{\rm min}+\frac{L^2}{r_0}{\rm coth}^{-1}\left(\sqrt{2}\right)\right)}\,+\mathcal{O}\left(e^{-2\sqrt{2}\frac{r_0}{L^2} (t_{\rm bdy} - t_{\rm min})}\right),
\end{equation}
and the late time limit of $P_v$ can be determined from the late time expansion of $P_v^2=U_{\rm eff}(r_{\rm min})$ in eq.~\eqref{eq:PV rmin} which leads to,
\begin{equation}\label{eq:Pv-simple-app}
	\begin{aligned}
		P_v &=\frac{r_0^2}{2 L^2}-4 \frac{r_0^2}{L^2} e^{-2\sqrt{2}\, {\rm coth}^{-1}\left(\sqrt{2}\right)}\,e^{-2\sqrt{2}\frac{r_0}{L^2} (t_{\rm bdy} - t_{\rm min})}\,+\mathcal{O}\left(e^{-4\sqrt{2}\frac{r_0}{L^2} (t_{\rm bdy} - t_{\rm min})}\right).
	\end{aligned}
\end{equation}
When $\alpha = y_{\rm brane}=0$, the Neumann boundary condition at the brane~\eqref{eq:Neumann boundary EF0} implies that $\eval{\dot{r}}_{\rm brane} = 0$ which means that the turning point is at $t_{\rm min}=t_{\rm brane}$. Together with eq.~\eqref{eq:Sch to global} which gives $t_{\rm brane}=0$, we can conclude that $t_{\rm min}=t_{\rm brane}=0$ in the simple case considered here.

Finally, from eq.~(\ref{eq:growth CV Pt}) the late time growth of complexity is 

\begin{equation}    
 \dv{c_{{}_V}}{t_{\rm bdy}} = \frac{r_0^2}{2G_N L^3}\left(1-8  \left(3-2\sqrt{2}\right)^{\sqrt{2}}\,e^{-\frac{2\sqrt{2}r_0}{L^2} t_{\rm bdy} }\right)\,+\mathcal{O}\left(e^{-4\sqrt{2}\frac{r_0}{L^2} t_{\rm bdy} }\right)\,. \label{eq:cv-simple}
\end{equation}
It is clear that for the simple example we considered here, the complexity growth reaches its asymptotic value from below, and therefore the Lloyd bound is satisfied. This case corresponds to exactly half of the double-sided black hole~\cite{Belin:2021bga} because the ETW brane simply cuts the wormhole geometry in half at the $y=0$ slice.


\section{Asymptotic analysis of the rate of complexity growth}
\label{app: analytics}
In this appendix, we provide details on the computation of the late time value of the growth rate of volume complexity in our setup for several illustrative examples as outlined in section~\ref{sec:asymptotics}.

\subsection{Examples}

\subsubsection{Boundary in parameter space}\label{sec:boundary-app}

We begin with the boundary in parameter space which is given by eq.~\eqref{eq:alpha condition}. In this case, the Lloyd bound is marginally satisfied, in the sense that ${\rm d} c_{{}_V}/{\rm d} {t_{\rm bdy}}$ reaches its asymptotic value from below exponentially twice as fast as in the other cases as we will see below. We start by exploring the simple example after which we generalize our findings.

\paragraph{No JT gravity, no tension.} To illustrate the approach to the asymptotic analysis, we first consider the simplest case, an ETW brane without intrinsic JT gravity nor tension. Specifically, this is the $\alpha =0$ and $y_{\rm brane} = 0$ limit. Let us now consider this limit following the three-step procedure outlined in section~\ref{sec:asymptotics}. In this case, the quadratic equation~\eqref{eq:quartic} reduces to
\begin{equation}
    P_v^2 - \frac{r_0^4}{L^4} \frac{r_{\rm brane}^2}{r_0^2} +  \frac{r_0^4}{L^4} \frac{r_{\rm brane}^4}{r_0^4} = 0\,.
\end{equation}
One finds
\begin{equation}
    \frac{r_{\rm brane}^2}{r_{\infty^2}} = 1 + \frac{L^2}{r_\infty^2}\delta P + {\cal O} \left( \delta P^2\right)\,.
\end{equation}

To determine the value of $t_{\rm brane}$ at late times, we need to use the brane embedding~\eqref{eq: brane location Schw} which in terms of $y_{\rm brane}$ looks like\footnote{We restore the $y_{\rm brane}$ dependence in the following and take the $y_{\rm brane} \to 0$ limit later to find $t_{\rm brane}$. We could have simply used the fact that the brane is located at $t_{\rm brane}=0$ for $y_{\rm brane}=0$ in the first place, but we keep  $y_{\rm brane}\neq 0 $ here to illustrate how to derive $t_{\rm brane}$ more generally in the other cases.}
\begin{equation}
    \frac{r_{\rm brane}^2}{r_0^2} = \frac{1-\sin^2(y_{\rm brane}) \,{\rm coth}^2\left(\frac{r_0 t_{\rm brane}}{L^2}\right)}{\cos^2(y_{\rm brane})}\,,
\end{equation}
as well as the location of the intersection of the brane and the maximal volume surface given~\eqref{eq: rbrane simple}. Solving these two equations asymptotically gives
\begin{equation}
    \coth^2 \left( \frac{r_0 t_{\rm brane}}{L^2}\right) = \frac{2- \cos^2(y_{\rm brane})}{2\sin^2(y_{\rm brane})} + {\cal O}\left( \delta P\right)\,,
\end{equation}
which gives $t_{\rm brane}=0$ for $y_{\rm min}\to 0$. 

The next step is to evaluate the time difference $t_{\rm bdy}-t_{\rm brane}$ in the late time regime. In this case, the contour integral~\eqref{eq:main integral} is a simple integral from $r_{\rm brane}$ to $r_{\rm bdy}$, where $r_{\rm brane}$ is asymptotically close to $r_{\infty}$. The details of the integration are in appendix~\ref{app:late time}. The integral is evaluated in eq.~\eqref{eq:Delta t marginal}, and gives
\begin{equation}\label{eq:Delta t simple}
   \frac{ t_{\rm bdy}-t_{\rm brane}}{L^2}=\frac{\log \left(\frac{2 r_0^2}{L^2 \delta P}\right)}{\sqrt{2} r_0} - \frac{{\rm arcoth}(\sqrt{2})}{r_0}\,.
\end{equation}
Inverting this equation and relating $\delta P$ to $P_v$ we find the late time behaviour of $P_v$ as
\begin{equation}\label{eq:Pv-simple}
	\begin{aligned}
		P_v &=\frac{r_0^2}{2 L^2}-4 \frac{r_0^2}{L^2} e^{-2\sqrt{2}\, {\rm coth}^{-1}\left(\sqrt{2}\right)}\,e^{-2\sqrt{2}\frac{r_0}{L^2} t_{\rm bdy}}\,+\mathcal{O}\left(e^{-4\sqrt{2}\frac{r_0}{L^2} t_{\rm bdy} }\right)\,.
	\end{aligned}
\end{equation}
The time dependence of complexity is then given by
\begin{equation}
 \dv{c_{{}_V}}{t_{\rm bdy}} = \frac{r_0^2}{2G_N L^3}\left(1-8  \left(3-2\sqrt{2}\right)^{\sqrt{2}}\,e^{-\frac{2\sqrt{2}r_0}{L^2} t_{\rm bdy} }\right)\,+\mathcal{O}\left(e^{-4\sqrt{2}\frac{r_0}{L^2} t_{\rm bdy} }\right)\,.
\end{equation}

It is clear that for the simple example we considered here, the complexity growth reaches its asymptotic value from below, and therefore the Lloyd bound is satisfied. This case corresponds to exactly half of the double-sided black hole~\cite{Belin:2021bga} because the ETW brane simply cuts the wormhole geometry in half at the $y=0$ slice.

\paragraph{General case.} More generally, the boundary in parameter space between the Lloyd bound violating and Lloyd bound respecting regions is given by eq.~\eqref{eq:alpha condition}. This time, we solve the condition~\eqref{eq:quartic} in the late time limit by substituting $P_v^2 = P_\infty^2-\delta P^2$ and perform an expansion in small $\delta P$,
\begin{equation}
    \frac{r_0^2}{L^2} \left(\frac{r_0^2-2r^2_{\rm brane}}{2L}\right)^2 - \left(1+2\frac{r_{\rm brane}^2}{r_0^2}\sin^2(y_{\rm brane})\right) \delta P^2 = 0\,.
\end{equation}
The solution is
\begin{equation}\label{eq:rbrane-boundaryapp}
    \frac{r_{\rm brane}^2}{r_{\infty^2}} = 1 + \frac{L^2}{r_\infty^2}\sqrt{1+\sin^2(y_{\rm brane})} \delta P + {\cal O} \left( \delta P^2\right)\,.
\end{equation}
The late time value is larger than $r_\infty$ and is reached from above linearly in $\delta P$. Note that the tensionless limit ($y_{\rm brane} \to 0$) agrees with the analysis above, the case without JT gravity and tension. The value of $t_{\rm brane}$ is given by
\begin{equation}
    \coth^2 \left( \frac{r_0 t_{\rm brane}}{L^2}\right) = \frac{2+ \cos^2(y_{\rm brane})}{2\sin^2(y_{\rm brane})} + {\cal O}\left( \delta P\right)\,.
\end{equation}

The contour integral to evaluate the time difference $t_{\rm bdy}-t_{\rm brane}$ is still a direct line between $r_{\rm brane}$ and $r_{\rm bdy}$, and is given by eq.~\eqref{eq:Delta t marginal} which we copy here for convenience
\begin{equation}\label{eq:Delta t marginal0}
   \frac{ t_{\rm bdy}-t_{\rm brane}}{L^2}=-\frac{1}{r_0} \coth^{-1}\left(\sqrt{2}\right)-\frac{\log \left(\frac{r_{\rm brane}-r_{\infty}}{\sqrt{2} r_0}\right)}{\sqrt{2} r_0}\,.
\end{equation}
Inverting this equation to find $\delta P$ and plugging in the asymptotic behaviour of $t_{\rm brane}$ and $r_{\rm brane}$ leads to 
\begin{equation}
     \dv{c_{{}_V}}{t_{\rm bdy}} = \frac{P_\infty}{G_N L}- \frac{4 r_0^2 }{L^3 G_N \left(1+\sin^2 y_{\rm brane}\right)^2 }\left(3-2 \sqrt{2}\right)^{\sqrt{2}} D\left(y_{\rm brane}\right)  
 e^{-\frac{2\sqrt{2}r_0}{L^2}t_{\rm bdy}}\,,
\end{equation}
where
\begin{equation}
    D\left(y_{\rm brane}\right)=\left(\frac{\sec(y_{\rm brane})}{\sqrt{6}}\left(\sqrt{4+2\cos^2(y_{\rm brane})}+ 2 \sin(y_{\rm brane})\right) \right)^{2\sqrt{2}}.
\end{equation}
Of course, in the tensionless limit ($y_{\rm brane} \to 0$), the coefficient $D(y_{\rm brane}) \to 1$ in agreement with the ($\alpha=0,~y_{\rm brane}=0$) case above.

\subsubsection{Lloyd bound respecting region}\label{sec:Lloyd app}

When the magnitude of the JT gravity coupling $\alpha$ is smaller than the critical value given by eq.~\eqref{eq:alpha condition}, the volume complexity satisfies the Lloyd bound, in the sense that its time derivative reaches its asymptotic value from below. As before, we first explore a simple example, namely when there is no JT gravity coupling but the tension on the brane is nonzero, after which we generalize our findings.

\paragraph{No JT gravity, nonzero tension} In the case with no JT gravity coupling, $\alpha=0$, and nonzero tension, $y_{\rm brane} \neq 0$, the quadratic equation in~\eqref{eq:quartic} is
\begin{equation}
    P_v^2 - \frac{r_0^4}{L^4} \frac{r^2_{\rm brane}}{r^2_0} + \frac{r_0^4}{L^4}  \frac{r^4_{\rm brane}}{r_0^4} \cos^2(y_{\rm brane})=0\,.
\end{equation}
To solve it in the late time limit, we substitute $P_v^2 = P_\infty^2 - \delta P^2$ and solve it perturbatively in $\delta P$. The relevant solution is\footnote{In terms of tension, this solution reads
\begin{equation}
   \frac{r^2_{\rm brane}}{r^2_\infty} = \frac{1}{1-T_0} + \frac{2L^4\delta P^2}{r_0^4 T_0} +{\cal O}\left(\delta P^4\right)\,.
\end{equation}}
\begin{equation}\label{eq:simple tension App}
   \frac{r^2_{\rm brane}}{r^2_\infty} =\frac{1}{1+\sin(y_{\rm brane})} - \frac{2L^4\delta P^2}{r_0^4 \sin(y_{\rm brane})} +{\cal O}\left(\delta P^4\right)\,.
\end{equation}
Keeping in mind that $0>\sin(y_{\rm brane})>-1$, we see that $r_{\rm brane} > r_\infty$, and it reaches its final value from above. Note that the late time limit does not commute with the tensionless limit. Moreover, the deviation of $r_{\rm brane}$ from its final value is quadratic in $\delta P$, instead of linear. 

The time of intersection $t_{\rm brane}$ is given by\footnote{We implicitly assume that the intersection of brane with the maximal volume surface is inside the horizon. However, this intersection may lie outside the horizon. In this case, one should use 
\begin{equation}
    \frac{r_{\rm brane}^2}{r_0^2} = -\frac{1-\sin^2(y_{\rm brane}) \,{\rm tanh}^2\left(\frac{r_0 t_{\rm brane}}{L^2}\right)}{\cos^2(y_{\rm brane})}~.
\end{equation}
The results below should be modified accordingly.} combining eqs. \eqref{eq:embedding-y} and~\eqref{eq:simple tension App}, leading to
\begin{equation}\label{blaa}
    \coth^2\left(\frac{r_0 t_{\rm brane}}{L^2}\right) = \frac{1}{2}\csc^2(y_{\rm brane})\left(3-\sin(y_{\rm brane})\right) + {\cal O} \left(\delta P\right)\,.
\end{equation}
The evaluation of eq.~\eqref{eq:main integral} as in section~\ref{app:late time} and combining it with eqs.~\eqref{eq:simple tension App} and \eqref{blaa} then gives
\begin{equation}\label{eq:qqq App}
     \dv{c_{{}_V}}{t_{\rm bdy}} = \frac{P_\infty}{G_N L} - \frac{4 r_0^2}{G_NL^3} \left(3-2 \sqrt{2} \right)^{\sqrt{2}} B(y_{\rm brane}) D(y_{\rm brane}) e^{-\frac{\sqrt{2} r_0t_{\rm bdy}}{L^2}}\,,
\end{equation}
where
\begin{align}
     B(y_{\rm brane}) &= \left(2 \csc(y_{\rm brane}) \left(\sqrt{1+\sin(y_{\rm brane})}-1\right)-1\right)\times\nonumber\\
     &\quad\left(\frac{3+2\sin(y_{\rm brane}) + 2\sqrt{2 + 2 \sin(y_{\rm brane})}}{|1+2 \sin(y_{\rm brane})|}\right)^{\frac{1}{\sqrt{2}}}\,,\\
    D(y_{\rm brane}) &= e^{\sqrt{2} \coth^{-1} \left(-\csc(y_{\rm brane}) \sqrt{\frac{3-\sin(y_{\rm brane})}{2}} \right)}\,.\nonumber
\end{align}

Note that $\lim_{y_{\rm brane}\to 0}B(y_{\rm brane}) \propto - y_{\rm brane}$ which vanishes in the tensionless limit. In this case, the subleading corrections of order $\mathcal{O}\left(\exp\left(- \frac{2\sqrt{2}r_0}{L^2}t_{\rm bdy}\right)\right)$ in \eqref{eq:qqq App} become relevant, which is in agreement with the exponential behaviour found in the previous subsection. 

\paragraph{Near the boundary in parameter space} Next, we consider small deviations from the boundary into the Lloyd bound respecting region in parameter space. For concreteness we focus on the $\alpha>0$ when $t_{\rm bdy} \to \infty$, but similar results follow by instead using $\alpha<0$ and $t_{\rm bdy} \to -\infty$. First,  we find the location of the brane with
\begin{equation}
    \alpha = \frac{-\sin(y_{\rm brane})}{\cos^3(y_{\rm brane})} \frac{r_0}{L} - \delta \alpha \,,
\end{equation}
and $\delta \alpha>0$ small. Since we are in the Lloyd bound respecting region in parameter space, we substitute $P_v^2=P_\infty^2-\delta P^2$ into eq.~\eqref{eq:quartic} and solve perturbatively in $\delta P$ and $\delta \alpha$. However, the $\delta \alpha \to 0$ limit does not commute with the late time limit, $\delta P \to 0$. 

Solving eq.~\eqref{eq:quartic} perturbatively in $\delta \alpha$ and $\delta P$, while taking $\delta \alpha/\delta P \to 0$ gives
\begin{equation}
\begin{aligned}
    \frac{r_{\rm brane}^2}{r_\infty^2}&= 1 - \frac{L}{r_0} \cos^3(y_{\rm brane}) \sin(y_{\rm brane}) \delta \alpha \\
    &+ \frac{2L^2}{r_0^2}\sqrt{1+\sin^2(y_{\rm brane})}\delta P + {\cal O} \left(\delta P^2,~\delta \alpha \delta P,~ \delta \alpha^2 \right)\,.
\end{aligned}
\end{equation}
The intersection of the brane and the maximal volume surface is bigger than $r_\infty$ and approaches its asymptotic value from above. This result agrees with eq.~\eqref{eq:rbrane-boundaryapp} for the boundary in parameter space. Notice that since $\delta \alpha \ll \delta P$, the location of the brane is asymptotically close to $r_\infty$ resulting in the contour integral determining $t_{\rm bdy} - t_{\rm brane}$ to be similar to the boundary case of section~\ref{sec:boundary-app}, and the corresponding results will agree with that case in the $\delta \alpha\to 0$ limit. 

For late enough times, when $\delta P \ll \delta \alpha$, the situation is more similar to the ``no JT gravity but finite tension'' case discussed above. Solving eq.~\eqref{eq:quartic}  perturbatively in $\delta P$ and $\delta \alpha$, while taking $\delta P/\delta \alpha \to 0$ gives
\begin{equation}
\begin{aligned}
    \frac{r_{\rm brane}^2}{r_\infty^2} & = 1 + \frac{L}{r_0} \cos^3(y_{\rm brane}) \left(\sqrt{1+\sin^2(y_{\rm brane})}-\sin(y_{\rm brane})\right) \delta \alpha \\
    & \quad + \frac{2L^3}{r_0^3 \delta \alpha}\frac{\sqrt{1+\sin^2(y_{\rm brane})}}{\cos^3(y_{\rm brane})} \delta P^2 + {\cal O} \left(\delta P^4,\delta \alpha \delta P^2, \delta \alpha^2 \right)\,,
\end{aligned}
\end{equation}
which is consistent with eq.~\eqref{eq:simple tension App} in the limit $\alpha \to 0$, which implies $y_{\rm brane}$ is ${\cal O}(\delta \alpha)$. The intersection of the maximal volume surface and the brane remains above $r_\infty$ and approaches its asymptotic value from above as a function of time, that is, with decreasing $\delta P$. Notice that the asymptotic value of $r_{\rm brane}$ approaches $r_\infty$ from above as $\delta \alpha$ becomes smaller. As we will see shortly when we cross to the Lloyd bound violating region, the maximal volume surface starts probing regions behind $r_\infty$.

The time of intersection, $t_{\rm brane}$, is given by
\begin{equation}
\begin{split}
    \coth^2 &\left(\frac{r_0 t_{\rm brane}}{L^2}\right) = \frac{2+ \cos^2(y_{\rm brane})}{2\sin^2(y_{\rm brane})} \\
    &+ \frac{L}{2r_0} \frac{\cos^5(y_{\rm brane})}{\sin^2(y_{\rm brane})}\left(\sqrt{1+\sin^2(y_{\rm brane})} -\sin(y_{\rm brane}) \right)\delta \alpha + {\cal O}\left( \delta P\right)\,,
\end{split}
\end{equation}
which can be used in the equation above to find the asymptotic behaviour of $\delta P$.

The integration of~\eqref{eq:main integral} leads to

\begin{equation}\label{eq:qqqq App}
    \dv{c_{{}_V}}{t_{\rm bdy}} = \frac{P_\infty}{G_N L} - \frac{4 r_0^2}{G_NL^3} \left(3-2 \sqrt{2} \right)^{\sqrt{2}} B(y_{\rm brane},\delta \alpha) D(y_{\rm brane},\delta \alpha) e^{-\frac{\sqrt{2} r_0t_{\rm bdy}}{L^2}}\,,
\end{equation}
where
\begin{align}
    &\log B(y_{\rm brane},\delta \alpha) = \nonumber\\
    &\quad\sqrt{2} \tanh^{-1}\left(\frac{1}{\sqrt{2}} + \frac{1}{\sqrt{2}} \frac{L}{r_0} \cos^3(y_{\rm brane})\left(\sqrt{1+\sin^2(y_{\rm brane})}-\sin(y_{\rm brane})\right)\delta\alpha\right)\nonumber\\
    &\quad -2 \coth^{-1}\left(1 + \frac{L}{r_0} \cos^3(y_{\rm brane})\left(\sqrt{1+\sin^2(y_{\rm brane})}-\sin(y_{\rm brane})\right)\delta\alpha\right),\\
    &\log D(y_{\rm brane},\delta\alpha) = \nonumber\\
    &\quad -\sqrt{2} \, {\rm coth}^{-1} \sqrt{\frac{2+ \cos^2(y_{\rm brane})}{2\sin^2(y_{\rm brane})} + \frac{L}{2r_0} \frac{\cos^5(y_{\rm brane})\left(\sqrt{1+\sin^2(y_{\rm brane})} -\sin(y_{\rm brane}) \right)\delta \alpha}{\sin^2(y_{\rm brane})}}.\nonumber 
\end{align}

Importantly, close to the boundary in parameter space, when $\delta \alpha \to 0$ and $r_{\rm brane}$ approaches $r_\infty$, the coefficient in front of the exponential in eq.~\eqref{eq:qqqq App} vanishes because $B(y_{\rm brane},\delta \alpha)  \sim \delta \alpha$. When this happens, the subleading corrections would become important as was the case in section~\ref{sec:boundary-app}. This is consistent with our findings for the simple example of ($\alpha=0,y_{\rm brane} \neq 0$) discussed above. There the asymptotic value of $r_{\rm brane}$ approached $r_\infty$ for $y_{\rm brane}\to 0$ which led to a vanishing leading exponential correction to $\delta P$ and $dc_V/dt$.

\subsubsection{Lloyd bound violating region}\label{sec:no Lloyd app}

When the magnitude of the JT gravity coupling is bigger than the critical value given in eq.~\eqref{eq:alpha condition}, the volume complexity violates the Lloyd bound and ${\rm d} c_{{}_V}/{\rm d} {t_{\rm bdy}}$ reaches its asymptotic value from above. The simplest example of this class is when there is a nonzero JT gravity coupling, and the cosmological constant of the brane matches the bulk cosmological constant so that the brane is positioned at $y_{\rm brane}=0$. 

\paragraph{Nonzero JT gravity coupling, equal cosmological constants.} We begin with the case $\Lambda^{\rm brane}=\Lambda^{\rm bulk}$, so the location of the brane is set to $y_{\rm brane}=0$. For concreteness, we focus on the positive $\alpha$ case in the late time limit, but a similar analysis holds for negative $\alpha$ at early times ($t_{\rm bdy} \to -\infty$). In this limit, the equation determining the intersection of the maximal volume surface with the brane~\eqref{eq:quartic} is
\begin{equation}
    P_v^2 - \frac{r_0^4 + L^2 r_0^2 \alpha^2}{L^4} \frac{r_{\rm brane}^2}{r_0^2}+\frac{r_0^4 + L^2 r_0^2 \alpha^2}{L^4} \frac{r_{\rm brane}^4}{r_0^4}=0\,.
\end{equation}
We solve it in the late time limit once again, now with $P_v^2 = P_\infty^2 + \delta P^2$. We find
\begin{equation}\label{eq:simple JTapp}
    \frac{r_{\rm brane}^2}{r_\infty^2} = 1-\frac{L|\alpha|}{\sqrt{r_0^2+L^2\alpha^2}} + \frac{L^3 \delta P^2}{r_0^2|\alpha| \sqrt{r_0+L^2\alpha^2}} + {\cal O} \left( \delta P^4\right)\,.
\end{equation}

This time the point of intersection reaches deeper towards the singularity $r_{\rm brane}< r_\infty$ due to the contact term associated with the JT gravity action term pulling it inwards. The late time value is reached from above, and the deviation is quadratic in $\delta P$.

The time of intersection $t_{\rm brane}$ is
\begin{equation}\label{eq:cos ybrane}
    \coth^2 \left( \frac{r_0 t_{\rm brane}}{L^2}\right) = \csc^2(y_{\rm brane}) + \frac{1}{2} \cot^2(y_{\rm brane})\left(1- \frac{L|\alpha|}{\sqrt{r_0^2+L^2\alpha^2}} \right) + {\cal O}\left( \delta P\right)\,,
\end{equation}
which goes to zero in the $y_{\rm brane}\to 0$ limit. 

Putting all these expressions together, we find
\begin{equation}
    \dv{c_{_V}}{t_{\rm bdy}} =  \frac{P_\infty}{G_N L} + \frac{4 r_0^2}{G_NL^3} \left(3-2 \sqrt{2} \right)^{\sqrt{2}} B(\alpha) e^{-\frac{\sqrt{2} r_0t_{\rm bdy}}{ L^2}}\,,
\end{equation}
where
\begin{equation}
    B(\alpha) = \left(2\beta-2\sqrt{\beta^2-\beta}-1\right)\left(\frac{3\beta+2 \sqrt{2}\sqrt{\beta^2-\beta}-1}{1+\beta}\right)^{\frac{1}{\sqrt{2}}}\,\,
\end{equation}
with $\beta^2=1+\frac{r_0^2}{L^2\alpha^2}$. Notice that for $\alpha \to 0$, the coefficient $B(\alpha) \sim \alpha$ vanishes as expected because the asymptotic value of $r_{\rm brane}$ approaches $r_\infty$ for small $\alpha$.

\paragraph{Near the boundary in parameter space} As explained in the main text, one can generalize the results in a similar manner to section~\ref{sec:Lloyd app} using a perturbative expansion for the parameter
\begin{equation}
    \alpha = \frac{-\sin(y_{\rm brane})}{\cos^3(y_{\rm brane})} \frac{r_0}{L} + \delta \alpha \,,
\end{equation}
to locate the boundary in parameter space where there occurs a Lloyd bound violation via eq.~\eqref{eq:quartic} with $\delta \alpha>0$ small. Similar to section~\ref{sec:Lloyd app}, the $\delta \alpha \to 0$ limit does not commute with the late time limit ($\delta P \to 0$). In this case, the difference signals the onset of the violation of the Lloyd bound.

Concretely, $\delta \alpha \ll \delta P$ corresponds to times that are not late enough for the Lloyd bound to be violated. In this case, substituting $P_v^2=P_\infty^2-\delta P^2$ and solving eq.~\eqref{eq:quartic} perturbatively in $\delta \alpha$ and $\delta P$, while taking $\delta \alpha/\delta P \to 0$ gives
\begin{equation}
    \frac{r_{\rm brane}^2}{r_\infty^2}= 1 + \frac{L}{r_0} \cos^3(y_{\rm brane}) \sin(y_{\rm brane}) \delta \alpha + \frac{2L^2}{r_0^2}\sqrt{1+\sin^2(y_{\rm brane})}\delta P + {\cal O} \left(\delta P^2,\delta \alpha \, \delta P, \delta \alpha^2 \right).
\end{equation}
Note that because $\delta P \gg \delta \alpha$, the intersection of the brane and the maximal volume surface is still bigger than $r_\infty$ even though $\delta \alpha >0$ and $y_{\rm brane}\leq 0$. This is related to the fact that it is not late enough for the Lloyd bound to be violated despite being in the Lloyd bound violating region of parameter space. The location agrees with the results for the boundary in parameter space~\eqref{eq:rbrane-boundaryapp} for $\delta \alpha \to 0$.

For late enough times, $\delta P \ll \delta \alpha$, the Lloyd bound is indeed violated. Substituting $P_v^2 = P_\infty^2 + \delta P^2$ into eq.~\eqref{eq:quartic} and solving perturbatively in $\delta P$ and $\delta \alpha$, while taking $\delta P/\delta \alpha \to 0$ gives
\begin{equation}
\begin{aligned}
    \frac{r_{\rm brane}^2}{r_\infty^2} = 1 &- \frac{L}{r_0} \cos^3(y_{\rm brane}) \left(\sqrt{1+\sin^2(y_{\rm brane})}-\sin(y_{\rm brane})\right) \delta \alpha \\
    &+ \frac{2L^3}{r_0^3 \delta \alpha}\frac{\sqrt{1+\sin^2(y_{\rm brane})}}{\cos^3(y_{\rm brane})} \delta P^2 + {\cal O} \left(\delta P^4,\delta \alpha \delta P^2, \delta \alpha^2 \right).
\end{aligned}
\end{equation}
which is consistent with eq.~\eqref{eq:simple JTapp} for $y_{\rm brane} = 0$ and small $\alpha=\delta \alpha$. The asymptotic value of $r_{\rm brane}$ is smaller than $r_\infty$ for nonzero $\delta \alpha$ and it is reached from above as $\delta P$ goes to zero.

The time of intersection $t_{\rm brane}$ is given by
\begin{equation}
\begin{split}
    \coth^2 &\left( \frac{r_0 t_{\rm brane}}{L^2}\right) = \frac{2+ \cos^2(y_{\rm brane})}{2\sin^2(y_{\rm brane})} \\
    &+ \frac{L}{2r_0} \frac{\cos^5(y_{\rm brane})}{\sin^2(y_{\rm brane})}\left(\sqrt{1+\sin^2(y_{\rm brane})} -\sin(y_{\rm brane}) \right)\delta \alpha + {\cal O}\left( \delta P\right).
\end{split}
\end{equation}
Integrating \eqref{eq:main integral} and collecting all these results gives
\begin{equation}
    \dv{c_{_V}}{t_{\rm bdy}} =  \frac{P_\infty}{G_N L} + \frac{4 r_0^2}{G_NL^3} \left(3-2 \sqrt{2} \right)^{\sqrt{2}} B(y_{\rm brane},\delta \alpha) D(y_{\rm brane},\delta\alpha) e^{-\frac{\sqrt{2} r_0t_{\rm bdy}}{ L^2}}\,,
\end{equation}
where
\begin{align}
    &\log B(y_{\rm brane},\delta \alpha) = \nonumber\\
    &\quad \sqrt{2} \tanh^{-1}\left(\frac{1}{\sqrt{2}} - \frac{1}{\sqrt{2}} \frac{L}{r_0} \cos^3(y_{\rm brane})\left(\sqrt{1+\sin^2(y_{\rm brane})}-\sin(y_{\rm brane})\right)\delta\alpha\right)\nonumber\\
    &\quad  -2 \tanh^{-1}\left(1- \frac{L}{r_0} \cos^3(y_{\rm brane})\left(\sqrt{1+\sin^2(y_{\rm brane})}-\sin(y_{\rm brane})\right)\delta\alpha\right),\\
    &\log D(y_{\rm brane},\delta\alpha) = \nonumber\\
    &\quad -\sqrt{2} \, {\rm coth}^{-1} \sqrt{\frac{2+ \cos^2(y_{\rm brane})}{2\sin^2(y_{\rm brane})} + \frac{L}{2r_0} \frac{\cos^5(y_{\rm brane})}{\sin^2(y_{\rm brane})}\left(\sqrt{1+\sin^2(y_{\rm brane})} -\sin(y_{\rm brane}) \right)\delta \alpha}.\nonumber
\end{align}
Note that in the limit $\delta \alpha \to 0$, the coefficient $B(y_{\rm brane},\delta \alpha)\sim \delta \alpha$ vanishes, and the subleading corrections similar to the ones of section~\ref{sec:boundary-app} become important for the late time dependence of $dc_V/dt$.

\subsection{Late time expansion of \texorpdfstring{$t_{\rm brane}-t_{\rm bdy}$}{}}\label{app:late time}

To compute the late time expansion of $t_{\rm brane}-t_{\rm bdy}$, we need to evaluate eq.~\eqref{eq:main integral}, which we copy here
\begin{equation}\label{eq:main integral-app}
    I=-\int\frac{P_\infty~d r}{L^{-2}(r^2-r_0^2)\sqrt{L^{-4}(r^2-r_\infty^2)^2\pm\delta P^2}}+\mathcal{O}(\delta P^2)
\end{equation}
for different contours of integration which will depend on which region of parameter space we are in.

While the exact evaluation of the integral involves the elliptic integral of the third kind, for our purposes, we focus on obtaining an approximate expression suitable for the late time regime. To achieve this, we divide the range of the integral into different regions based on the values of $r$
\begin{equation}\label{eq:Pinfty pm delta}
    r=r_\infty+\delta r,\qquad P_v^2=P_\infty^2\pm\delta P^2~.
\end{equation}
We expand around $r=r_\infty$ and introduce the locations $(r_{\infty}\pm\Delta r)$ and $(r_{\infty}\pm\Delta r')$ to label the intermediate regions in $\qty[r_{\rm brane},r_{\text{bdy}}]$ where we evaluate the integral. We take the following order of limits: $\Delta r'/L \ll \delta P \ll \Delta r/L \ll r_0/L$.

We study each of the signs in eq.~(\ref{eq:Pinfty pm delta}), corresponding to violation or agreement with the Lloyd bound, separately below, as well as the boundary between the two regions, in which the Lloyd bound is saturated. 

\subsubsection{Scattering off the potential implies satisfying the Lloyd bound}
\label{app:no violation}
In this case, we take the negative sign in eq.~(\ref{eq:Pinfty pm delta}) so that the particle scatters off the potential $U_{\rm eff}(r)$. One can split the integrals in eq.~(\ref{eq:main integral-app}) in such a way that the relevant contributions are:
\begin{itemize}
    \item $\delta P\ll \delta r/L$: This corresponds to the range $r\in\qty[(r_{\infty}+\Delta r),r_{\rm brane}]\cup\qty[(r_{\infty}+\Delta r),r_{\text{bdy}}]$. The contribution to the integral, denoted as $I_1$, is approximately given by
\begin{equation}
\begin{aligned}
    I_1&\approx -\int
   \frac{P_{\infty }L^4}{\left(r^2-r_0^2\right) \left|r^2-r_{\infty }^2\right| } \, d r+\mathcal{O}\qty( \frac{L^2\delta P^2}{\Delta r^2})\\
   &=-\frac{L^4 P_{\infty } \left(r_{\infty } \tanh
   ^{-1}\left(\text{min}\left(\frac{r}{r_0},\frac{r_0}{r}\right)\right)-r_0 \coth
   ^{-1}\left(\frac{r}{r_{\infty
   }}\right)\right)}{r_0 r_{\infty }
   \left(r_{\infty }^2-r_0^2\right)}+\mathcal{O}\qty( \frac{L^2\delta P^2}{\Delta r^2})~.
\end{aligned}
\end{equation}

    \item $\delta P\sim\delta r$: This corresponds to $r\in\left[r_\infty+ \Delta r',r_\infty+\Delta r\right]$. +The contribution to the integral, denoted as $I_2$, is approximately given by
    \begin{equation}
    \begin{aligned}
        I_2&\approx-\int \frac{L^4 P_{\infty }}{\left(r_{\infty
   }^2-r_0^2\right) \sqrt{-\delta  P^2L^4+4 \delta r^2
   r_{\infty }^2}} \, d\delta r+\mathcal{O}\left(\frac{\Delta r}{L}\right)\\
   &=-\frac{L^4 P_{\infty }}{2 r_{\infty } \left(r_{\infty
   }^2-r_0^2\right)}\coth ^{-1}\left(\frac{2
   r_{\infty}\,\delta r}{\sqrt{-\delta  P^2L^4+4 \delta r^2 r_{\infty }^2}}\right)+\mathcal{O}\left(\frac{\Delta r}{L}\right)
   \end{aligned}
    \end{equation}
\end{itemize}

\begin{figure}[t]
\centering
\includegraphics[scale=1.7]{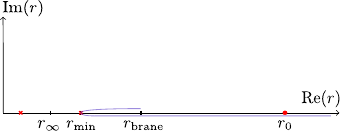}
\caption{Integration contour for eq.~\eqref{eq: decomposition-satisfied}, spanning from $r_{\rm brane}$ to $\infty$. The figure highlights the positions of significant features within the contour. The two branch points of the integrand are marked by red crosses, situated at $r_\infty \pm \frac{L^2}{2r_\infty}\delta P$. The red dot represents the pole of the integrand located at $r_0$. The turning point $r_{\rm min}$ coincides with one of the branch points.
Deep in this region of parameter space, $r_{\rm brane}-r_\infty\gg \mathcal{O}\left(\delta P\right)$.}
\label{fig:contourA1}
\end{figure}

The relevant case needed for the calculations in this work is for $r_{\rm brane}$ at a finite distance from $r_{\infty}$, then
\begin{equation}
t_{\rm bdy}-t_{\rm brane}=\eval{I_1}_{r_{\infty}+\Delta r}^{r_{\rm brane}}+2\eval{I_2}_{r_{\rm min}-r_{\infty}}^{\Delta r}+\eval{I_1}_{r_{\infty}+\Delta r}^{r_{\rm bdy}}~.
    \label{eq: decomposition-satisfied}
\end{equation}
The explicit evaluation leads to
\begin{equation}
    \frac{t_{\rm bdy}-t_{\rm brane}}{L^2}=-A(r_{\rm brane})-\frac{\sqrt{2} \log \left(\frac{L^2\delta P}{2 r_0^2}\right)}{r_0}\,,
\end{equation}
where
\begin{equation}
    A(r_{\rm brane})=\frac{-\tanh ^{-1}\left(\frac{r_{\rm brane}}{r_0}\right)+\sqrt{2} \coth ^{-1}\left(\frac{\sqrt{2} r_{\rm brane}}{r_0}\right)+2 \coth ^{-1}\left(\sqrt{2}\right)}{r_0}~,
\end{equation}
and we used that $r_{\rm min}-r_{\infty}=\frac{L^2\delta P}{\sqrt{2}r_0}$.

Notice that the final expression is again independent of the exact gluing location $r\pm \Delta r$. The respective late time growth of $P_v$ is then given by
\begin{equation}
    P_v=P_\infty-4 r_0^2 L^{-2} e^{-\sqrt{2} r_0 \left( A(r_{\rm brane})+L^{-2}\left(t_{\rm bdy}-t_{\rm brane}\right)\right)}~.
\end{equation}
We can rewrite this result as
\begin{equation}
    P_v=P_\infty-4 r_0^2 L^{-2} B(r_{\rm brane})e^{-\sqrt{2} r_0 L^{-2}\left(t_{\rm bdy}-t_{\rm brane}\right)}
\end{equation}
with
\begin{equation}
    B(r_{\rm brane})=\frac{\left(3-2 \sqrt{2}\right)^{\sqrt{2}} \left(1-\frac{r_0}{r_{\rm brane}}\right)^{-\frac{1}{\sqrt{2}}} \left(\frac{r_0+r_{\rm brane}}{r_{\rm brane}}\right)^{\frac{1}{\sqrt{2}}} \left(2 r_{\rm brane}-\sqrt{2} r_0\right)}{\sqrt{2} r_0+2 r_{\rm brane}}~.
\end{equation}

\subsubsection{No scattering off the potential implies violation of the Lloyd bound}\label{app:violation}
Now, we consider the positive sign case in eq.~(\ref{eq:Pinfty pm delta}). In this case, the particle has effective energy high enough to fall past $r_{\text{min}}$ towards $r_{\rm brane}$ without scattering off the potential. We decompose the integral~(\ref{eq:main integral-app}) into the following regions:
\begin{itemize}
    \item $\delta P\ll \delta r/L$: This corresponds to the range $r\in\qty[r_{\rm brane},(r_{\infty}-\Delta r)]\cup\qty[(r_{\infty}+\Delta r),r_{\text{bdy}}]$. The contribution to the integral, denoted as $I_1$, is approximately given by
\begin{equation}
\begin{aligned}
    I_1&\approx -\int
   \frac{P_{\infty }L^4}{\left(r^2-r_0^2\right) \left|r^2-r_{\infty }^2\right| } \, d r+\mathcal{O}\qty( \frac{L^2\delta P^2}{\Delta r^2})\\
   &=- \, {\rm sign} \left(r-r_\infty\right)\frac{L^4 P_{\infty } \left(r_{\infty } \tanh
   ^{-1}\left(\text{min}\left(\frac{r}{r_0},\frac{r_0}{r}\right)\right)-r_0 \tanh
   ^{-1}\left(\text{min}\left(\frac{r}{r_{\infty
   }},\frac{r_{\infty
   }}{r}\right)\right)\right)}{r_0 r_{\infty }
   \left(r_{\infty }^2-r_0^2\right)}\\
   &\quad\,+\mathcal{O}\qty( \frac{L^2\delta P^2}{\Delta r^2})~.
\end{aligned}
\end{equation}
    
\item $\delta P\sim\delta r/L$: This corresponds to $r\in\qty[(r_{\infty}-\Delta r),\,(r_{\infty}-\Delta r')] \cup \qty[(r_{\infty}+\Delta r'),\,(r_{\infty}+\Delta r)]$. The contribution to the integral, denoted as $I_2$, is approximately given by
    \begin{equation}
    \begin{aligned}
        I_2&\approx-\int \frac{L^4 P_{\infty }}{\left(r_{\infty
   }^2-r_0^2\right) \sqrt{\delta  P^2L^4+4 \delta r^2
   r_{\infty }^2}} \, d\delta r+\mathcal{O}\left(\frac{\Delta  r}{L}\right)\\
 &=-\frac{L^4 P_{\infty }}{2 r_{\infty } \left(r_{\infty
   }^2-r_0^2\right)}\tanh ^{-1}\left(\frac{2
   r_{\infty}\,\delta r}{\sqrt{\delta  P^2L^4+4 \delta r^2 r_{\infty }^2}}\right)+\mathcal{O}\left(\frac{\Delta  r}{L}\right)
   \end{aligned}
    \end{equation}
    
    \item $\delta P\gg \delta r/L$: This corresponds to $r\in\qty[(r_{\infty}-\Delta r'),(r_{\infty}+\Delta r')]$. The contribution to the integral, denoted as $I_3$, is approximately given by
    \begin{equation}    
    \begin{aligned}
        I_3&\approx -\int \frac{L^2 P_\infty}{(r_\infty^2-r_0^2)\delta P}d \delta r + {\cal O}\left( \frac{\Delta r'^2}{L^2}\right) \,,\\
        &= -\frac{ L^2 P_{\infty }\delta r}{\delta  P (r_{\infty }^2-r_0^2)}+\mathcal{O}\left( \frac{\Delta r'^2}{L^2}\right) 
        ~.
        \end{aligned}
    \end{equation}
    However, this contribution is subleading compared to the others and thus will not be necessary for our analysis.
\end{itemize}

\begin{figure}[t]
\centering
\includegraphics[scale=1.7]{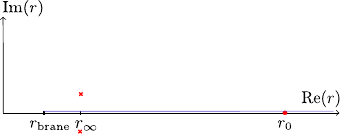}
\caption{Integration contour for eq.~\eqref{eq:Delta t to be inverted}, spanning from $r_{\rm brane}$ to $\infty$. The figure highlights the positions of significant features within the contour. The two branch points of the integrand are marked by red crosses, situated at $r_\infty \pm i\frac{L^2}{2r_\infty}\delta P$. The red dot represents the pole of the integrand located at $r_0$. Deep in this region of parameter space, $r_\infty-r_{\rm brane}\gg\mathcal{O}\left(\delta P\right)$.}
\label{fig:contourA2}
\end{figure}

By combining these different contributions, we find
\begin{equation}\label{eq:Delta t to be inverted}
    t_{\rm bdy}-t_{\rm brane}=\eval{I_1}_{r_{\rm brane}}^{r_{\infty}-\mu\Delta r}+\eval{I_2}_{-\Delta r}^{\Delta r}+\eval{I_1}_{r_{\infty}+\mu\Delta r}^{r_{\rm bdy}}.
\end{equation}
This results in
\begin{equation}\label{eq:Delta t to be inverted final1}
    \frac{t_{\rm bdy}-t_{\rm brane}}{L^2}=-A(r_{\rm brane})-\frac{\sqrt{2}}{r_0}\log\left(\frac{L^2\delta P}{4r_0^2}\right)\,,
\end{equation}
where the function $A(r_{\rm brane})$ is defined as
\begin{equation}\label{eq:Delta t to be inverted final2}
    A(r_{\rm brane})=\frac{1}{2r_0}\left(4\coth^{-1}\left(\sqrt{2}\right)-2\tanh^{-1}\left(\frac{r_{\rm brane}}{r_0}\right)+2\sqrt{2}\tanh^{-1}\left(\frac{r_{\rm brane}}{r_\infty}\right)\right).
\end{equation}
As before, the final expression is independent of the exact gluing location $r\pm \Delta r$.

Finally, by inverting eq.~(\ref{eq:Delta t to be inverted}) for $P_v$, we obtain the late time behaviour of the canonical momentum in the case of the Lloyd bound violation, given by
\begin{equation}\label{eq:Pv-violated}
    P_v=P_\infty+4^2r_0^2 L^{-2}\rme^{-\sqrt{2}r_0\left(A(r_{\rm brane})+ L^{-2}\left(t_{\rm bdy}-t_{\rm brane}\right)\right)}.
\end{equation}
This expression can also be written as
\begin{equation}\label{eq:Pv violated-2}
    P_v=P_\infty+4^2r_0^2L^{-2} B(r_{\rm brane}) e^{-\sqrt{2}r_0L^{-2}(t_{\rm bdy}-t_{\rm brane})}\,,
\end{equation}
where the function $B(r_{\rm brane})$ is defined as
\begin{equation}
    B(r_{\rm brane})=\frac{\left(3-2 \sqrt{2}\right)^{\sqrt{2}} \left(1-\frac{r_{\rm brane}}{r_0}\right){}^{-\frac{1}{\sqrt{2}}} \left(\frac{r_0+r_{\rm brane}}{r_0}\right){}^{\frac{1}{\sqrt{2}}} \left(r_0-\sqrt{2} r_{\rm brane}\right)}{r_0+\sqrt{2} r_{\rm brane}}~.
\end{equation}

\subsubsection{Critical case: saturating the Lloyd bound}
\label{app: boundary}
Finally, we turn our attention to the limiting case in phase space where the Lloyd bound can be violated or satisfied. We proceed in a similar manner to the above cases by assuming that the particle does scatter off the effective potential, but with the crucial difference that the turning point $r_{\text{min}}$ is now very close to the edge of the integration region determined by $r_{\rm brane}$. That means that we should take the negative sign case in eq.~(\ref{eq:Pinfty pm delta}). The decomposition of the integral is then similar to eq.~(\ref{eq: decomposition-satisfied}) but with the first term absent. We write \begin{equation}
    t_{\rm bdy}-t_{\rm brane}=\eval{I_2}_{r_{\rm brane}-r_{\infty}}^{\Delta r}+\eval{I_1}_{r_{\infty}+\mu\Delta r}^{r_{\rm bdy}}~.
\end{equation}

\begin{figure}[t]
\centering
\includegraphics[scale=1.7]{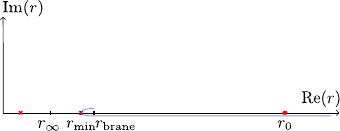}
\caption{Integration contour for eq.~\eqref{eq:Delta t marginal}, spanning from $r_{\rm brane}$ to $\infty$. The figure highlights the positions of significant features within the contour. The two branch points of the integrand are marked by red crosses, situated at $r_\infty \pm \frac{L^2}{2r_\infty}\delta P$. The red dot represents the pole of the integrand located at $r_0$. The turning point $r_{\rm min}$ coincides with one of the branch points, specifically $r_{\infty}+\frac{L^2}{2r_\infty}\delta P$.
At the boundary of the parameter space, $r_{\rm brane}$ is close to $r_{\rm min}$, and the two coincide when $y_{\rm brane}=0$.}
\label{fig:contourA3}
\end{figure}

We find that 
\begin{equation}\label{eq:Delta t marginal}
   \frac{ t_{\rm bdy}-t_{\rm brane}}{L^2}=-A-\frac{\log \left(\frac{r_{\rm brane}-r_{\infty}}{\sqrt{2} r_0}\right)}{\sqrt{2} r_0}\,,
\end{equation}
where
\begin{equation}
    A=\frac{1}{r_0}\coth ^{-1}\left(\sqrt{2}\right).
\end{equation}
Notice that the final expression is independent of the exact gluing location $r\pm \Delta r$.

We can use that $r_{\rm brane}-r_{\infty}=\frac{L^2\sqrt{1+\sin^2{y_{\rm brane}}}\delta P}{\sqrt{2}r_0}$ which follows from eq.~\eqref{eq:rbrane-boundary}. The asymptotic expansion of eq.~(\ref{eq:Pinfty pm delta}) can be then expressed as
\begin{equation}\label{eq:violation dCVdt}
    P_v=P_\infty-\frac{4 r_0^2}{L^2 \left(1+\sin^2 y_{\rm brane}\right)}  e^{-2 \sqrt{2} r_0  \left(A+L^{-2}\left(t_{\rm bdy}-t_{\rm brane}\right)\right)}~.
\end{equation}
Importantly, the exponential correction to volume growth contains a power $2\sqrt{2}r_0$.


\section{Extrinsic curvature of the shell worldvolume in the wormhole picture}
\label{app:ext_curve}
This appendix is devoted to the derivation of the equation (\ref{eq:Kij explicitly}) of the extrinsic curvature of the worldvolume of the shell given by the profile (\ref{eq:shell_profile}). For concreteness, we focus on the right part of the long wormhole geometry of section~\ref{sec:Engelhardt-Folkestad}, but a similar calculation follows for the left side. The worldvolume of the shell is embedded into the right half of the wormhole spacetime by $w^\mu=(\tau(\eta),~y(\eta),~x)$. We can describe a basis of tangential vectors to the worldvolume surface of the shell as
\begin{equation}
    \rme^\mu_1=\partial_\eta w^\mu=(\dot{\tau}(\eta),~{\dot{y}(\eta)},~0)~,\quad \rme^\mu_2=(0,~0,~1)~\,,
\end{equation}
where the dot denotes the derivative with respect to $\eta$. Then, any normal vector $n^\mu$ with respect to the shell must satisfy $n_\mu \rme^\mu_{a}=0$ ($a=1,~2$), which can be found as
\begin{equation}
    n^\mu=-\frac{\cos y}{L\sqrt{\dot{\tau}^2-\dot{y}^2}}(\dot{y},~\dot{\tau},~0)~.
\end{equation}
We are choosing the orientation of $n^\mu$ such that it is pointing outwards. Here the normalization is chosen as $g_{\mu\nu}n^\mu n^\nu=1$ with the metric in (\ref{MetricGlobal}). This allows us to evaluate the extrinsic curvature:
\begin{equation}
    K_{ab}=\frac{1}{2}n^\mu\partial_{\mu}h_{ab}=-\frac{\cos y}{2L\sqrt{\dot{\tau}^2-\dot{y}^2}}\qty(\dot{y}\partial_{\tau}+\dot{\tau}\partial_{y})h_{ab}~,
\end{equation}
where $h_{ab}$ is given by
\begin{equation}\label{eq:hij}
    ds^2_{\rm ind}=h_{ab}dz^a dz^b=\frac{1}{\cos^2y}\qty((\dot{y}^2-\dot{\tau}^2)L^2d\eta^2+\frac{r_0^2}{L^2}\cos^2\tau d x^2)~,
\end{equation}
and $z^a=(\eta,x)$ are coordinates for the shell worldvolume whose indices are denoted by Latin letters. 

Explicitly, we have:
\begin{align}
    K_{\eta\eta}&=L \dot{\tau} \frac{\tan y}{\cos y}
   \sqrt{\dot{\tau}^2-\dot{y}^2}~,\\
    K_{xx}&=-\frac{r_0^2 \cos \tau 
   \left(\dot{\tau} \cos \tau \tan
   y -\dot{y}\sin \tau \right)}{L^3\cos y \sqrt{\dot{\tau}^2-\dot{y}^2}}~,\\
    K&\equiv h^{ab}K_{ab}=-\frac{2 \dot{\tau} \sin y -\dot{y} \tan \tau \cos y}{L \sqrt{\dot{\tau}^2-\dot{y}^2}}~.
\end{align}
So far, there have been no approximations, however, we are interested in the profile given eq.~\eqref{eq:y new}:
\begin{equation}
    \tau(\eta)=\eta~,\quad y(\eta)=y_0-\varepsilon \cos^2y_0 \tan \eta~.
\end{equation}
In that case
\begin{equation}\label{eq:approx hab}
h_{ab}~dz^adz^b=\qty({\sec^2y_0}-2 \epsilon \tan\left(\tau_{\rm brane}\right) \tan y_0)\qty(-L^2d\eta^2+\frac{r_0^2}{L^2}\cos^2\left(\tau_{\rm brane}\right)~{d x^2})+\mathcal{O}(\varepsilon^2)~.    
\end{equation}
Now we can write the extrinsic curvature to order $\mathcal{O}(\epsilon)$ in perturbation theory:
\begin{align}
    K_{ab}&=-\frac{\sin y_0}{L} h_{ab}-\varepsilon\qty(\frac{r_0^2\cos y_0\sin^3\left(\tau_{\rm brane}\right)}{L^3\cos\left(\tau_{\rm brane}\right)}\delta^x_a\delta^x_b+L\cos y_0\tan\left(\tau_{\rm brane}\right)\delta_a^\eta\delta_b^\eta)+\mathcal{O}(\varepsilon^2)~,\\
    &=-\frac{\sin y_0}{L} h_{ab}-\varepsilon \frac{\cos y_0  \tan \left(\tau_{\rm brane}\right)}{L} \left(\frac{r_0^2}{L^2}\delta^x_a\delta^x_b-\cos^2 y_0 h_{ab}\right)+\mathcal{O}(\varepsilon^2)~,\label{eq:Kij varepsilon}\\
    K&=-\frac{2}{L}\sin y_0 - \varepsilon\frac{\cos^3 y_0}{L} \tan \left(\tau_{\rm brane}\right)\qty(\tan^2 \left(\tau_{\rm brane}\right)-1)+ {\cal O}(\varepsilon^2)~.
\end{align}

\bibliographystyle{JHEP}
\bibliography{refs}

\end{document}